\documentclass[structabstract]{aa}  
\usepackage{graphicx}
\usepackage{txfonts}
\usepackage{pdfpages}
\usepackage{multirow}
\usepackage{natbib}

\newcommand{\htwos}{H$_{2}$S}

\newcommand{\sotwo}{SO$_{2}$}

\newcommand{\htwoclplus}{H$_{2}$Cl$^{+}$}
\newcommand{\htwohclplus}{H$_{2}^{37}$Cl$^{+}$}

\newcommand{\chplus}{CH$^{+}$}
\newcommand{\shplus}{SH$^{+}$}
\newcommand{\ohplus}{OH$^{+}$}
\newcommand{\htwooplus}{H$_{2}$O$^{+}$}
\newcommand{\htwoo}{H$_{2}$O}

\newcommand{\htwoeo}{H$_{2}^{18}$O}
\newcommand{\hcoplus}{HCO$^{+}$}
\newcommand{\ntwohplus}{N$_{2}$H$^{+}$}

\newcommand{\nhthree}{NH$_{3}$}

\newcommand{\meth}{CH$_{3}$OH}
\newcommand{\form}{H$_{2}$CO}

\newcommand{\oursource}{OMC-2~FIR~4}
\newcommand{\numberoflines}{719}
\newcommand{\numberofmolecules}{26}
\newcommand{\numberofisotopologs}{14}

\newcommand{\vlsr}{v$_{\textrm{lsr}}$}
\newcommand{\coolco}{$60$}		
\newcommand{\coolwater}{$13$}	
\newcommand{\coolmeth}{$9$}	

\newcommand{\compA}{\textit{Quiescent gas}}
\newcommand{\compB}{\textit{Wings}}
\newcommand{\compC}{\textit{Broad blue}}
\newcommand{\compD}{\textit{Foreground slab}}
\newcommand{\compE}{\textit{Other}}
\newcommand{\compa}{\textit{quiescent gas}}
\newcommand{\compb}{\textit{wings}}
\newcommand{\compc}{\textit{broad blue}}
\newcommand{\compd}{\textit{foreground slab}}
\newcommand{\compe}{\textit{other}}

\newcommand{\msol}{M$_{\odot}$}
\newcommand{\lsol}{L$_{\odot}$}
\newcommand{\tkin}{$T_{\textrm{kin}}$}

\newcommand{\herschel}{\emph{Herschel}}
\newcommand{\hso}{\emph{Herschel} Space Observatory}
\newcommand{\herhifi}{\emph{Herschel}/HIFI}

\newcommand{\changesone}[1]{#1}
\newcommand{\changestwo}[1]{#1}
\newcommand{\resubmit}[1]{#1}
\newcommand{\retwo}[1]{#1}
\newcommand{\rethree}[1]{#1}

\begin{document}
   \title{The \emph{Herschel}/HIFI spectral survey of OMC-2 FIR~4 (CHESS)}
   
   \titlerunning{HIFI spectral survey of OMC-2 FIR 4 (CHESS)}

   \subtitle{An overview of the 480 to 1902~GHz range}

   \author{M.~Kama\inst{1}, A.~L\'{o}pez-Sepulcre\inst{2}, C.~Dominik\inst{1,3}, C.~Ceccarelli\inst{2}, A.~Fuente\inst{4}, E.~Caux\inst{5,6},  R.~Higgins\inst{7}, A.G.G.M.~Tielens\inst{8}, and T.~Alonso-Albi\inst{4}
          }

   \authorrunning{Kama et al.}

\institute{Astronomical Institute `Anton Pannekoek', University of Amsterdam, Amsterdam, The Netherlands, \email{M.Kama@uva.nl}
        \and
	UJF-Grenoble 1 / CNRS-INSU, Institut de Plan\'{e}tologie et d\textquoteright Astrophysique de Grenoble (IPAG) UMR 5274, Grenoble, F-38041, France
        \and
	Department of Astrophysics/IMAPP, Radboud University Nijmegen, Nijmegen, The Netherlands
        \and
	Observatorio Astron\'omico Nacional, P.O. Box 112, 28803 Alcal\'a de Henares, Madrid, Spain
        \and
	Universit\'e de Toulouse, UPS-OMP, IRAP, Toulouse, France
        \and
	CNRS, IRAP, 9 Av. colonel Roche, BP 44346, 31028 Toulouse Cedex 4, France
	\and
	KOSMA, I. Physik. Institut, Universit\"{a}t zu K\"{o}ln, Z\"{u}lpicher Str. 77, 50937 K\"{o}ln, Germany
	\and
	Leiden Observatory, P.O. Box 9513, NL-2300 RA, Leiden, The Netherlands
            }

   \date{}

 
  \abstract
   {Broadband spectral surveys of protostars offer a rich view of the physical, chemical and dynamical structure and evolution of star-forming regions. The \hso\ opened up the terahertz regime to such surveys, giving access to the fundamental transitions of many hydrides and to the high-energy transitions of many other species.}
   {A comparative analysis of the chemical inventories and physical processes and properties of protostars of various masses and evolutionary states is the goal of the Herschel CHEmical Surveys of Star forming regions (CHESS) key program. This paper focusses on the intermediate-mass protostar, OMC-2 FIR 4.}
   {We obtained a spectrum of OMC-2 FIR 4 in the 480 to 1902 GHz range with the HIFI spectrometer onboard \herschel\, and carried out the reduction, line identification, and a broad analysis of the \resubmit{line profile} components, excitation, and cooling.}
   {\resubmit{We detect \numberoflines\ spectral lines from 40 species and isotopologs. The line flux is dominated by CO, \htwoo, and \meth. The line profiles are complex and vary with species and upper level energy, but clearly contain signatures from quiescent gas, a broad component likely due to an outflow, and \retwo{a foreground cloud}.}}
   {\resubmit{We find abundant evidence for warm, dense gas, as well as for an outflow in the field of view. Line flux represents 2\% of the $7~L_{\odot}$ luminosity detected with HIFI in the 480 to 1250~GHz range. Of the total line flux, \coolco\% is from CO, \coolwater\% from \htwoo\ and \coolmeth\% from \meth. A comparison with similar HIFI spectra of other sources is set to provide much new insight into star formation regions, a case in point being a difference of two orders of magnitude in the relative contribution of sulphur oxides to the line cooling of Orion KL and OMC-2 FIR 4.}}


   \maketitle

\section{Introduction}\label{sec:intro}

\changesone{With the proliferation of high-sensitivity broadband receivers, wide frequency coverage spectral surveys covering $>10$~GHz are becoming the \changestwo{norm \citep[e.g.][]{Johanssonetal1985, Blakeetal1986, Cernicharoetal1996, Schilkeetal1997, Cernicharoetal2000, Cauxetal2011}.} Such surveys provide comprehensive probes of the chemical inventory, excitation conditions and kinematics of sources such as protostars. Here, we present the first \emph{Herschel}/HIFI spectral survey of an intermediate-mass protostellar core, \oursource\ in the Orion~A molecular cloud, covering 480 to 1902~GHz.}

\changesone{The chemical composition of a protostar is linked to its evolutionary state and history, for example the relative abundances of the sulphur-bearing species in a protostellar core depend on the gas temperature and density, as well as the composition of the ices formed during the prestellar core phase \citep[e.g.][]{Wakelametal2005}. The chemical makeup of the gas also plays a role in the physical evolution of the protostar, for example by coupling with the magnetic field or by its role in the cooling of the gas \citep[e.g.][]{Goldsmith2001}. The large number of spectral lines captured by a survey places strong constraints on the excitation conditions and even spatially unresolved physical structure.}

\changesone{The HIFI spectrometer \citep{deGraauwetal2010} onboard the \hso\ \citep{Pilbrattetal2010} made a large part of the far-infrared or terahertz-frequency regime accessible to spectral surveys \citep{Berginetal2010, Ceccarellietal2010, Crockettetal2010, Zernickeletal2012, Neilletal2012, vanderWieletal2013}. Previous studies at these frequencies have been mostly limited to small frequency windows on the ground or, for space missions such as ISO, orders of magnitude lower spectral resolution and sensitivity than HIFI. Many hydrides and high-excitation lines of key molecules such as CO and \htwoo\ are routinely observable while Herschel is operational. Spectral surveys of star forming regions with HIFI are the focus of the CHESS\footnote{\texttt{http://www-laog.obs.ujf-grenoble.fr/heberges/hs3f/}; PI Cecilia Ceccarelli.} \citep{Ceccarellietal2010} and HEXOS\footnote{\texttt{http://www.hexos.org}, PI Edwin A. Bergin} \citep{Berginetal2010} key programs. A comparison of low- to high-mass protostars, a key goal of CHESS, offers insight into the physics and chemistry of star formation through the entire stellar mass range.} \changestwo{The intermediate-mass protostar in the CHESS sample is \oursource\ in Orion.}

\changesone{\resubmit{Orion is a giant molecular cloud complex at a distance of $\sim 420$~pc (\citealt{Mentenetal2007} found $414\pm7$~pc to the Orion Nebula Cluster and \citealt{Hirotaetal2007} $437\pm19$~pc to Orion~KL). Various stages of star formation are represented}: Orion Ia and Ib are $10$~Myr old clusters, while Class 0 protostars are still abundant in the OMC-1, -2 and -3 subclouds. \changestwo{The OMC-2 cloud core contains a number of protostars, including \oursource\ as the dominant Class~0 object. It is among the closest intermediate-mass protostellar cores and possibly an example of triggered star formation \citep{Shimajirietal2008}.}}

\retwo{While earlier studies attributed a luminosity of $400$ or $1000$~\lsol\ to \oursource\ \citep{Mezgeretal1990, Crimieretal2009}, recent \herschel\ and SOFIA observations found $30$ to $50$~\lsol\ \citep{Adamsetal2012}. \citet{Adamsetal2012} further found} an envelope mass of 10~\msol\ for FIR~4, while previous authors found it to be $\sim 30$~\msol\ \citep{Mezgeretal1990, Crimieretal2009} and continuum interferometry has yielded even larger estimates, $60$~\msol\ \citep{Shimajirietal2008}. \rethree{The factor of 20 difference in luminosity is apparently related to the improved spatial resolution of the Herschel and SOFIA data used by \citet{Adamsetal2012} compared to that used in earlier work, as well as to differences in the SED integration annuli, with one focusing on the mid-infrared peak and the other covering the whole millimetre source, as discussed elsewhere \citep{LopezSepulcreetalInterferometry}. Our results do not depend on the exact value of the luminosity, although eventually this issue will require a dedicated analysis to facilitate a proper classification of \oursource.}

This paper is structured as follows: the observations, their calibration and reduction are discussed in Sect.~\ref{sec:obs}; we summarize the quality and molecular inventory of the data and present a rotational diagram analysis in Sect.~\ref{sec:results}; line profile components and energetics are discussed in Sect.~\ref{sec:discussion}; and the conclusions are summarized in Sect.~\ref{sec:conclusions}.

\retwo{The reduced data and the list of line detections presented in this paper will be available on the CHESS key program website\footnote{\texttt{http://www-laog.obs.ujf-grenoble.fr/heberges/hs3f/}}. The data can also be downloaded via the Herschel Science Archive\footnote{\texttt{http://herschel.esac.esa.int/Science\_Archive.shtml}}.}

   \begin{figure*}[!ht]
   \centering
   \includegraphics[clip=,width=0.98\textwidth]{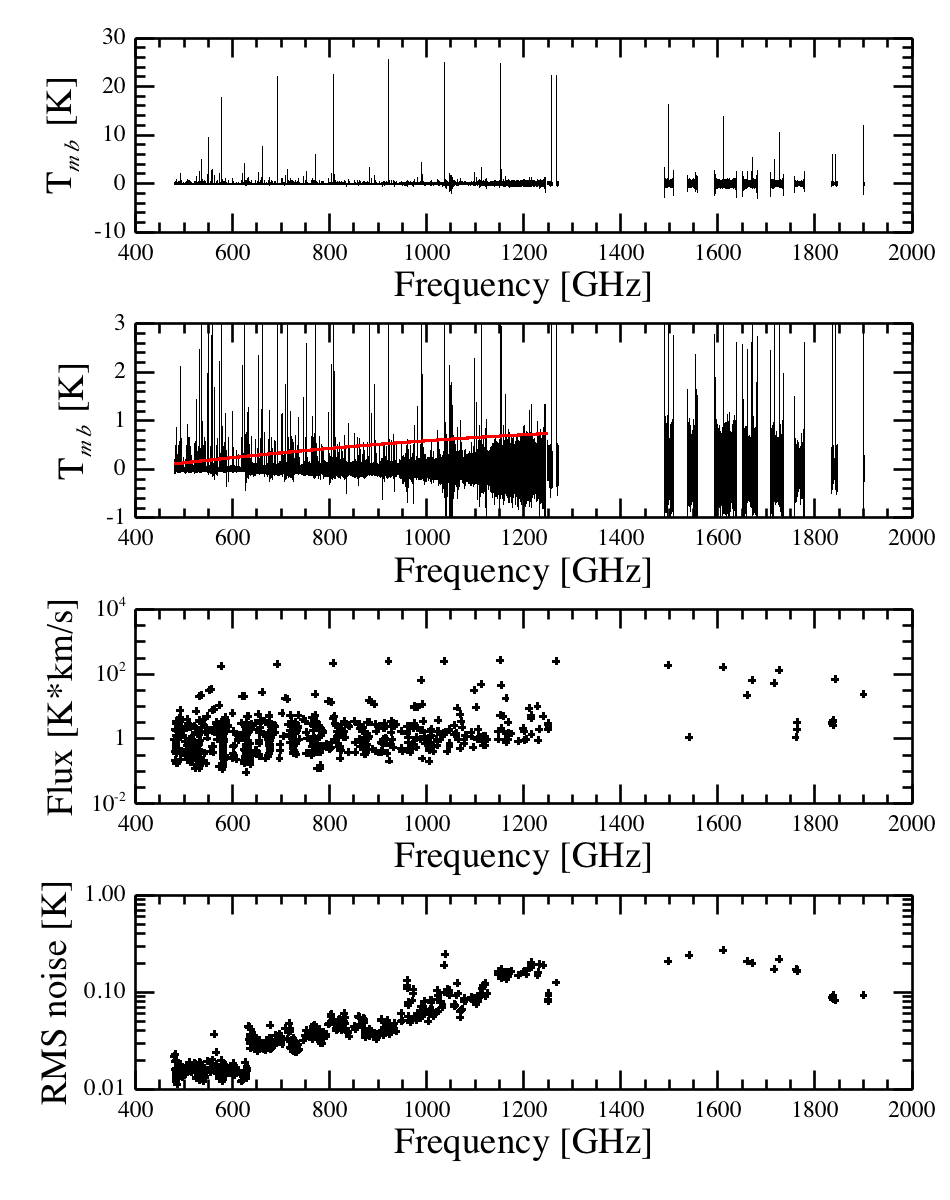}
      \caption{\textbf{Upper panel: }Full baseline-subtracted spectral survey (black line) at 1.1~MHz resolution. The set of bright lines towering above the rest is CO, the feature at 1901~GHz is CII. \textbf{Second panel: }\rethree{Full baseline-subtracted spectral survey (black line) on a blown-up $y$-scale to emphasize weak lines and a second-order polynomial fit (red) to the subtracted continuum in bands 1a through 5a (red).} \textbf{Third panel: }$T_{\textrm{mb}}$ scale integrated intensity of each detected transition. \textbf{Lower panel: } The local RMS noise around each detected transition.}
         \label{fig:fullsurvey}
   \end{figure*}

\section{Observations and data reduction}\label{sec:obs}

The data, presented in Fig.~\ref{fig:fullsurvey}, were obtained with the HIFI spectrometer on the \hso\ in 2010 and 2011, as part of the \herhifi\ guaranteed-time key program CHESS \citep{Ceccarellietal2010}. The spectral scan observations were carried out in dual beam switch (DBS) mode, using the Wide Band Spectrometer (WBS) with a native resolution of 1.1~MHz (0.7~to~0.2~km/s). \changestwo{The data were downloaded from the \herschel\ Science Archive, re-pipelined, reduced, and then deconvolved with the HIPE software \citep[][version 6.2.0 for the SIS bands and 8.0.1 for the HEB bands]{ott2010}.} \resubmit{Kelvin-to-Jansky conversions were carried out with the factors given by \citet{Roelfsemaetal2012}, which is also the standard reference for other instrumental parameters.}

\changesone{The spectrum on which line identification was carried out was obtained by stitching together the \changestwo{deconvolved} spectral scans and single-setting observations. In case of band overlap, the spectra were cut and stitched at the central frequency of the overlap. In principle, a lower RMS noise could be obtained \resubmit{for the band overlap regions by combining the data from adjacent bands}, but this requires corrections for sideband gain ratio variations, which are still being characterized (see also Sect.~\ref{sec:overlaps}).}

\subsection{Data quality after reduction}

After default pipelining, the data quality is already very high. However, in several bands unflagged spurious features (\textit{spurs}) prevent the deconvolution algorithm from converging. This problem is resolved by manually flagging the spurs missed by the pipeline. The baseline level in all bands is mostly gently sloping, but has occasional noticeable ripples. The data quality after spur flagging and baseline subtraction is excellent, as seen in Fig.~\ref{fig:fullsurvey}. Updated reductions will be provided on the CHESS KP and \herschel\ Science Center websites.

In bands 1 through 5, an important concern is ghosts from bright ($\rm T_{a} \geq 3~K$) lines \citep{ComitoSchilke2002}. The effect of ghosts on the overall noise properties is negligible, but they may locally imitate or damage true lines. To check this, we performed a separate reduction where bright lines are masked out and not used in the deconvolution. In bands 6 and 7, we have no ghost problems, as the signal to noise ratio of even the strongest lines is small, and ghosts are typically at the scale of double sideband intensity variations, which are $\leq 10$\% of the peak intensity. Table~\ref{tab:bandnoise} lists the measured root-mean-square (RMS) noise for the central part of each full band at 1.1~MHz resolution. Bands 6 and 7 were observed only partially, their RMS noise values can be seen in the bottom panel of Fig.~\ref{fig:fullsurvey}.

\begin{table}
	\caption{Summary of the full-band HIFI observations and of fluxes at selected standard wavelengths.}
	\label{tab:bandnoise}
	\begin{tabular}{ c r c c c c c }	
	Band	&	$\rm\nu_{band}^{\textrm{a}}$	& HPBW$^{\textrm{b}}$		& RMS$\rm_{obs}^{\textrm{c}}$		& Lines			& Flux$^{\textrm{d}}$		& F$\rm_{\nu}^{\textrm{e}}$ \\
			&	GHz				& ''			&		mK				& GHz$^{-1}$	& $\rm W\cdot m^{-2}$	& Jy		\\
	\hline
	\hline
	1a				&	520		& 	41	&	$16$	&	$1.9$	& $5.2(-14)$	&	63	\\
	1b				&	595		& 	36	&	$16$	&	$1.4$	& $7.1(-14)$	&	84	\\
	2a				&	675		& 	31	&	$30$	&	$1.2$	& $1.1(-13)$	&	106	\\
	2b				&	757		& 	28	&	$34$	&	$1.1$	& $1.3(-13)$	&	145	\\
	3a				&	830		& 	26	&	$45$	&	$0.7$	& $9.3(-14)$	&	134	\\
	3b				&	909		& 	23	&	$40$	&	$0.7$	& $2.0(-13)$	&	156	\\
	4a				&	1005	& 	21	&	$70$	&	$0.7$	& $2.5(-13)$	&	196	\\
	4b				&	1084	& 	20	&	$85$	&	$0.4$	& $1.7(-13)$	&	217	\\
	5a				&	1176	& 	18	&	$158$	&	$0.3$	& $3.4(-13)$	&	272	\\
	\hline
	158~$\mu$m		&	1902	&	11	&			&			& --			&	153	\\
	194~$\mu$m		&	1545	&	14	&			&			& --			&	256	\\
	350~$\mu$m		&	857		&	25	&			&			& --			&	173 \\
	450~$\mu$m		&	666		&	32	&			&			& --			&	107	\\
	\hline
	1a to 5a			&			&		&			&			& $1.3(-12)$	&		\\ 
	\end{tabular}
	\textbf{Notes. }$^{\mathrm{a}}$ The central frequency of the band. $^{\mathrm{b}}$ Beam size around the band center. $^{\mathrm{c}}$ RMS noise around the band center.	$^{\mathrm{d}}$ Band-integrated flux, in $\rm W\cdot m^{-2}$. The notation is $f(g) = f\cdot 10^{g}$. Due to overlap between the bands, the sum of band-integrated fluxes exceeds the total at the bottom. $^{\mathrm{e}}$ Flux at the band center, in Jy.
\end{table}

\subsection{Line identification}

All the lines detected in the survey are summarized in Fig.~\ref{fig:fullsurvey}, where we show the full HIFI spectrum, and the integrated flux and corresponding RMS noise level of each line. \changesone{As the main detection criterion, we adopted a limit of $S \geq 5$ for the signal to noise (significance) of the integrated line flux. We use a local definition of the signal to noise ratio:}

\begin{equation}
S = \left| \frac{ \int_{i=1}^{N}{\textrm{T}_{\textrm{mb},i}}dv }{ \textrm{RMS} \cdot dv \cdot \sqrt{N} } \right|,
\label{eq:fluxsigma}
\end{equation}

where $S$ is the significance, the integral gives the integrated intensity, $i$ is the channel index, $\textrm{T}_{\textrm{mb},i}$ is the main beam temperature of channel $i$ and $dv$ the channel width, $N$ is the number of channels covered by the line, and RMS is the local RMS noise around the line, measured at a resolution of 1.1~MHz. \changestwo{Given $\sim 10^{6}$ channels in the data and a typical extent of $10\ldots100$ channels per line, the total number of false-positives for a flux detection limit of $S=5$ in the entire survey is negligible.}

\begin{figure}[!h]
  \centering
  \includegraphics[width=1.0\columnwidth]{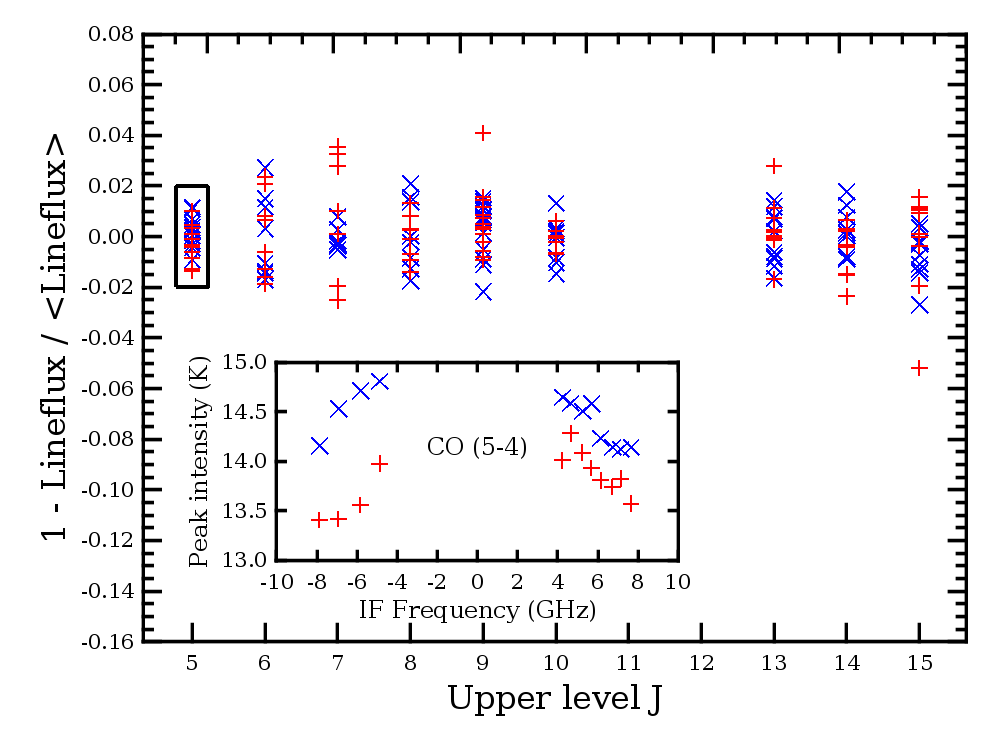}
  \caption{\changestwo{Fractional difference of the double-sideband CO line fluxes from their mean for each rotational line. The H polarization is shown in blue (x) and V in red ($+$). The inset shows the peak line intensity of the CO (5-4) transition, highlighted in the main plot with a box, versus intermediate frequency (IF) position. Negative IF frequency denotes lower side band (LSB).}}
        \label{fig:intensity_scatter}
\end{figure}

Lines in the survey were mostly identified using the \resubmit{JPL\footnote{\texttt{http://spec.jpl.nasa.gov/}} \citep{Pickettetal1998} and CDMS\footnote{\texttt{http://www.astro.uni-koeln.de/cdms/}} \citep{Mulleretal2005} catalogs. The literature was consulted for \htwooplus. The line identification was carried out in two phases and relied on the fact that the \oursource\ spectrum, while relatively rich in lines, is sufficiently sparse at our sensitivity that most lines can be individually and unambiguously identified.}

In phase 1, we employed the CASSIS\footnote{CASSIS has been developed by IRAP-UPS/CNRS, see \texttt{http://cassis.irap.omp.eu}.} software to look for transitions of more than 80 molecules or isotopologs that had been previously reported in spectral surveys or were expected to be seen in the HIFI data. Reasonable cutoffs were applied to limit the number of transitions investigated. For example, for CH$_{3}$OH, we set $\rm E_{u} < 10^{3}$~K, A$\rm_{ul} > 10^{-4} s^{-1}$ and $|K_{u}|$~$\leq 5$, yielding $>2000$ transitions in the HIFI range. \resubmit{The location of each of these transitions was then visually inspected in the survey data. \retwo{Once marked as a potential detection, a feature was not excluded from re-examination as a candidate for another species, with the intention of producing a conservative list of blend candidates.} For suspected detections, the line flux was measured in a range covering the line and accommodating potential weak wings (typically $\sim20$~km/s in total). The local RMS noise was determined from two line-free nearby regions for each line, and the line flux signal-to-noise ratio was calculated from Eq.~\ref{eq:fluxsigma}. The majority of the investigated transitions were not deemed even candidate detections, and $\sim10$\% of all chosen candidates turned out to be below the $S=5$ limit. The identification process was greatly sped up by the use of the CASSIS software as well as custom HIPE scripts. Nearly all the lines reported in this paper were identified in phase 1.}

\resubmit{In phase 2, we performed an unbiased visual inspection of all the data, looking for features significantly exceeding the local RMS noise level and not yet marked as detections in Phase 1. This process was aided by an automated line-finder that uses a sliding box to identify features exceeding $S=3$ and $5$ in the spectral data and also labels all previously identified lines. Phase 2 yielded a large number of candidates, of which only \retwo{a subset were confirmed as line detections, while the rest failed our signal-to-noise test.}}

\resubmit{\retwo{A summary of the detections is given in Sect.~\ref{sec:detections}.} The bottom two panels of Fig.~\ref{fig:fullsurvey} summarize the absolute line fluxes and RMS noise values of the detected lines.}

\subsection{Flux calibration accuracy}\label{sec:overlaps}

\changestwo{For HIFI, instrumental effects such as the sideband gain ratios and standing waves play a dominant role in the calibration accuracy and precision \citep{Roelfsemaetal2012}. Standing waves arise from internal reflections in the instrument and, to first order, contribute a constant $\lesssim 4$\% to the flux calibration uncertainty across each band. The sideband gain ratio (SBR) characterizes the fraction of the total double-sideband intensity that comes from the upper or lower side band (USB, LSB). In an ideal mixer, the USB and LSB contributions are equal, however in reality this is not exactly the case, particularly toward the edges of the receiver bands.  Here, we discuss the flux calibration uncertainty in the data, using the overlaps of adjacent HIFI bands, as well as an analysis of the CO line properties at the double-sideband stage.}

\begin{figure*}[!ht]
  \centering
  \includegraphics[clip=,width=1.0\linewidth]{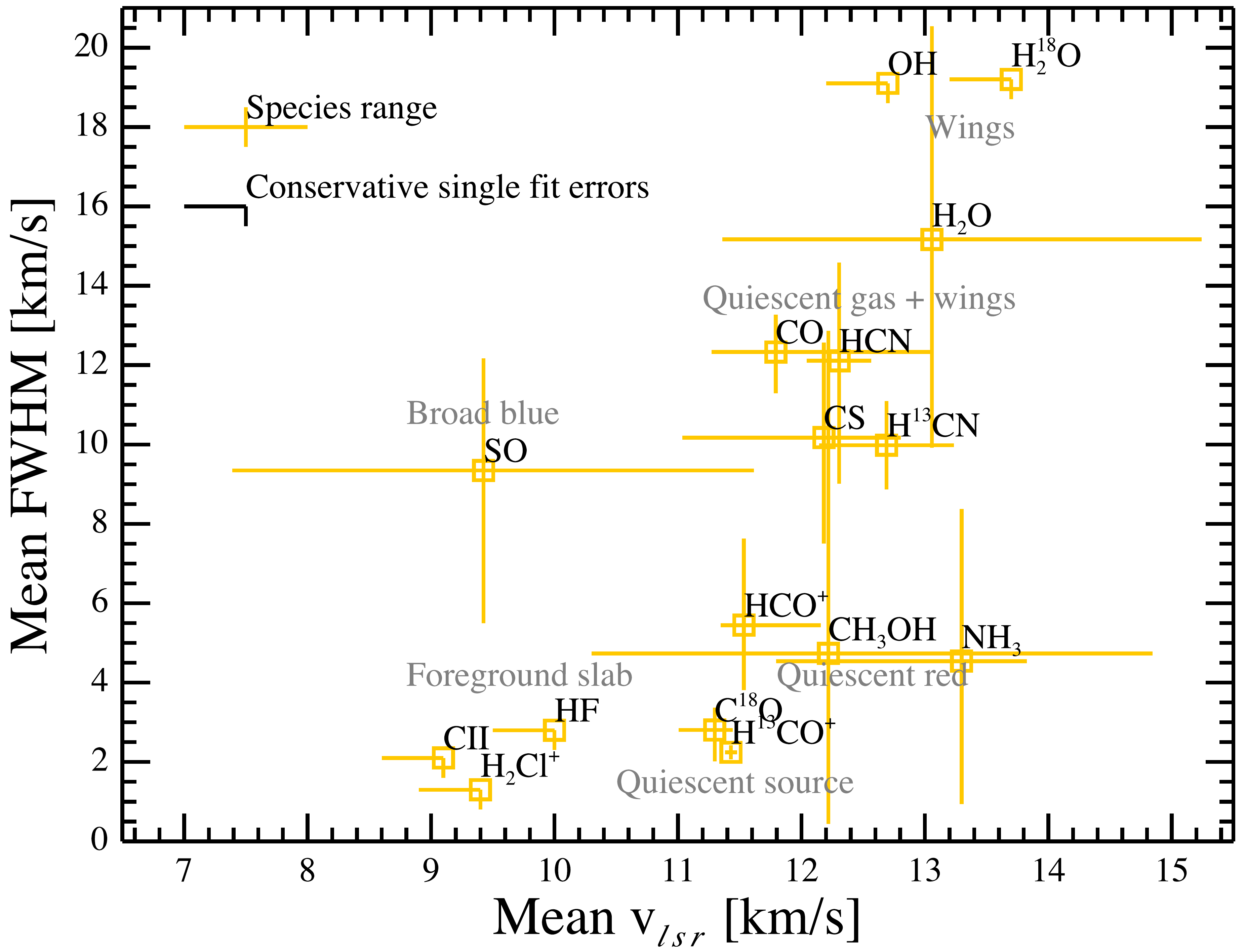}
     \caption{\retwo{Mean velocities and linewidths from Gaussian fits to lines of representative species. Several groups emerge, these are labeled in gray and referred to in the text. The orange bars give the range of fit parameters of the full set of lines of each species. The black bars, \rethree{at top left}, show the conservative Gaussian parameter fit uncertainties (see also Sect.~3.2).  Where the orange bars are one-sided, showing the conservative fit errors, only a single line was detected or simultaneous fitting of multiple lines forced the species to appear at a single \vlsr. For a discussion, see Sect.~\ref{sec:kinematicstory}.}}
        \label{fig:velowidth}
\end{figure*}

\begin{figure}[!ht]
  \centering
  \includegraphics[clip=,width=1.0\linewidth]{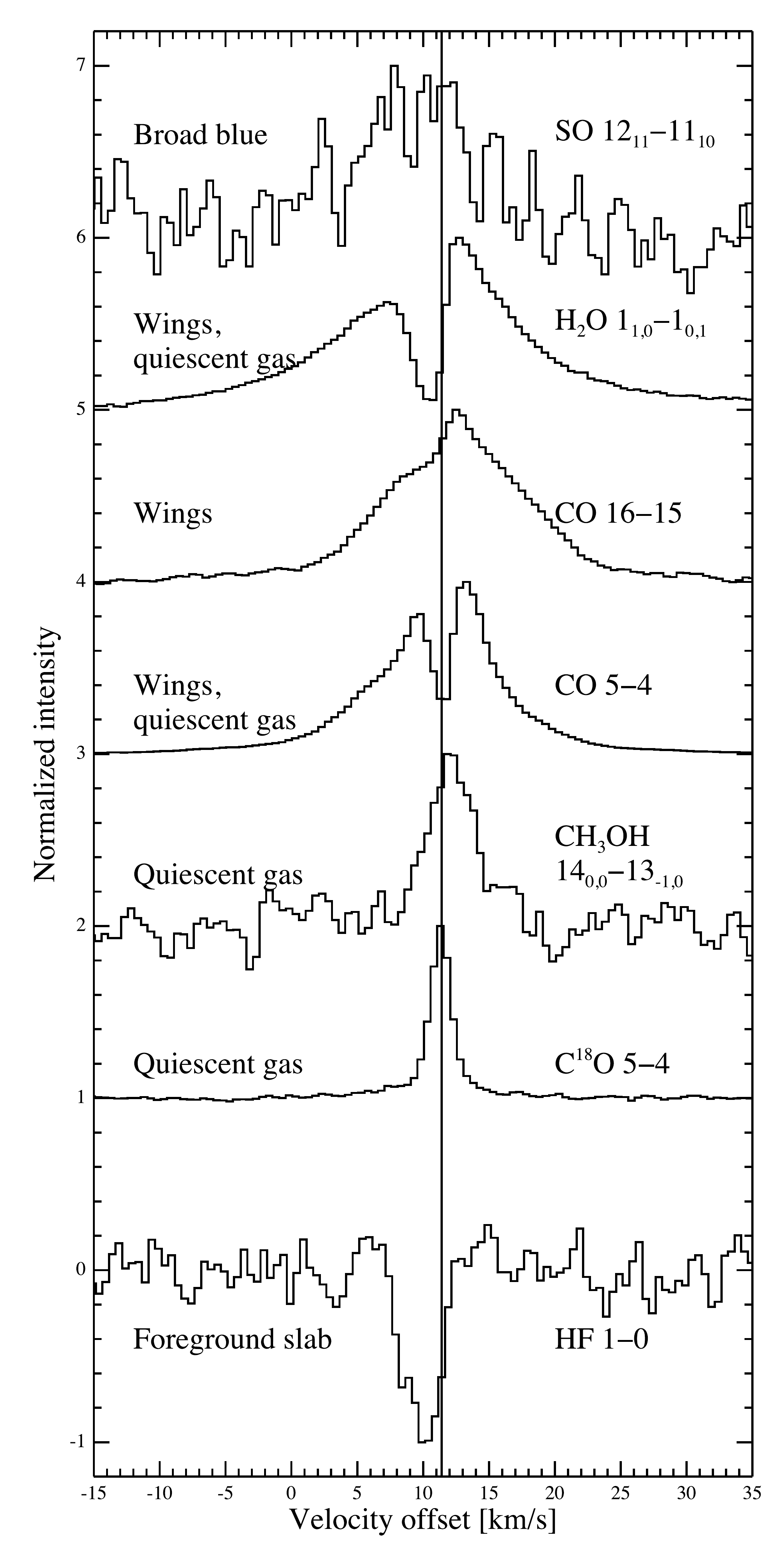}
     \caption{\resubmit{Main components of the line profiles. The solid vertical line shows the source velocity, $11.4$~km/s. The two velocity regimes of \compa\ are illustrated by C$^{18}$O and \meth. \retwo{The deep, narrow absorption feature in CO is due to emission in the reference beams, while the absorption in \htwoo\ appears to be \rethree{source-related}.} The \compc\ component is represented by SO. The HF line traces foreground material in the \compd, at 9~km/s, with another contribution from the \compa.}}
        \label{fig:linecomponents}
\end{figure}

\changestwo{Of all the lines in our survey, 10\% are in overlapping sections of HIFI bands, which we use to obtain a repeat measurement of their fluxes after deconvolution, and thus an estimation of the calibration and processing uncertainties. Focussing on the overlap of bands 1b and 2a, where the SBR variations are known to be large \citep{higginsphd2011}, we find that the line flux uncertainty is at the $\sim10$\% level, although this may depend to a substantial degree on how the data reduction is carried out.}

\changestwo{To estimate the SBR impact in the centers of the HIFI bands, we analyze the fluxes of $^{12}$CO lines in the double sideband data.} In Fig.~\ref{fig:intensity_scatter}, we show the fractional difference from the mean for the integrated intensity of each CO line observed in Spectral Scan mode, i.e. with multiple LO settings. H and V polarization are shown in blue (x) and red (+), respectively. The inset in Fig.~\ref{fig:intensity_scatter} shows the variations of the CO~($5-4$) line peak intensity with LO setting, revealing a correlation with the distance of the line from the center of the intermediate frequency (IF) range, consistent with a SBR variation across the band. This correlation is similar for the other CO lines. There is also a systematic offset between the intensities in the H and V polarizations, which we do not consider. We find the relative variations from uncorrected SBRs within a band to be $\lesssim 4$\%\changestwo{, although other factors may contribute to the total flux uncertainty budget.}

\changestwo{Corrections for the SBR variations are already in the pipeline for band 2a and are being characterized for all bands \citep{higgins2009, higginsphd2011}. A more thorough analysis of the impact of SBRs as well as operations such as baselining and deconvolution on the line fluxes is needed to exploit the full potential of HIFI. \resubmit{This is particularly important for absorption and weak emission lines.}}


\begin{table*}
	\centering
	\caption{Summary of the detected species.}
	\label{tab:species}
	\begin{tabular}{l c c c c c c l}
	Species	& \#	& $\rm E_{u}$ range	& $\rm\overline{v_{lsr}}$	& $\rm\overline{FWHM}$	& $\int T_{mb}dv$	& Flux	& Line components	\\
		&	&		K	&	km/s	& km/s	&	$\rm K\cdot km/s$	&	$\rm W\cdot m^{-2}$	& \\
	\hline
CO$^{s1}$ 			& 11 & 83 \ldots 752 & 11.8 & 12.3 & 2.2(3) & 2.9(-14) 				& \compA, \compb. \\
$^{13}$CO$^{s2}$	 & 8 & 79 \ldots 719 & 11.9 & 4.7 & 1.3(2) & 1.2(-15)			& \compA, \compb.\\ 
C$^{18}$O$^{s3}$	 & 5 & 79 \ldots 237 & 11.3 & 2.8 & 1.3(1) & 1.1(-16)			& \compA. \\ 
C$^{17}$O$^{s4}$	 & 3 & 81 \ldots 151 & 10.8 & 3.2 & 1.6(0) & 1.0(-17)				& \compA. \\ 
	\hline
H$_{2}$O$^{s5}$		 & 11 & 53 \ldots 305 & 13.1 & 15.2 & 4.4(2) & 6.2(-15)			& \compA, \compb. \\ 
\htwoeo$^{s6}$	 & 1 & 61 & 13.7 & 19.2 & 1.1(0) & 6.9(-18)					& \compB.\\ 
OH$^{s7}$ 			 & 6 & 270 & 12.7 & 19.1 & 8.9(0) & 1.9(-16) 					& \compB.\\ 
OH$^{+}$$^{s8}$ 		 & 8 & 44 \ldots 50 & -- & -- & -6.2(0) & -7.2(-17)					& \compD.  \\ 
\htwooplus$^{s9}$ 	& 1 & 54	& 8.4	& 2.5	& -8.9(-1)	& -1.2(-17)							& \compD. \\ 
	\hline
CH$_{3}$OH$^{s10}$ 	 & 431 & 33 \ldots 659 & 12.2 & 4.7 & 5.1(2) & 4.5(-15)		& \compA. \\ 
H$_{2}$CO$^{s11}$ 	 & 74 & 97 \ldots 732 & 11.9 & 4.7 & 9.5(1) & 7.0(-16)			& \compA. \\ 
	\hline
HCO$^{+}$$^{s12}$ 	 & 8 & 90 \ldots 389 & 11.5 & 5.4 & 1.1(2) & 9.3(-16)			& \compA, \compb(?). \\ 
H$^{13}$CO$^{+}$$^{s12}$  & 2 & 87 \ldots 117 & 11.4 & 2.2 & 7.0(-1) & 4.4(-18)		& \compA. \\ 
N$_{2}$H$^{+}$$^{s13}$  & 7 & 94 \ldots 349 & 11.7 & 3.0 & 2.6(1) & 2.2(-16)		& \compA. \\ 
	\hline
CI$^{s14}$ 			 & 2 & 24 \ldots 63 & 11.9 & 1.8 & 9.6(0) & 7.3(-16)						& \compA. \\ 
CII$^{s15}$ 			& 1 & 91 & 9.1 & 2.1 & 2.4(1) & 5.4(-16)								& \compD. \\ 
CH$^{+}$$^{s16}$ 		& 1 & 40 & 9.8 & 6.0 & -2.8(0) & -2.8(-17) 							& \compD. \\ 
CH$^{s17}$			& 3 & 26 & 12.7 & 2.5 & 4.4(-1) & 3.6(-18)							& \compA. \\ 
CCH$^{s18}$			& 20 & 88 \ldots 327 & -- & -- & 1.1(1) & 9.0(-17)						& \compA, \compb. \\ 
	\hline
HCN	$^{s19}$		& 9 & 89 \ldots 447 & 12.3 & 12.1 & 1.1(2) & 9.8(-16)$^\textrm{a}$			& \compA, \compb.  \\ 
H$^{13}$CN$^{s20}$	 & 2 & 87 \ldots 116 & 12.7 & 10.0 & 1.6(0) & 1.0(-17)				& \compA, \compb. \\ 
HNC	$^{s21}$		 & 2 & 91 \ldots 122 & 11.6 & 2.6 & 2.0(0) & 1.3(-17)					& \compA. \\ 
CN$^{s22}$			 & 20 & 82 \ldots 196 & 12.5 & 8.1 & 1.0(1) & 7.5(-17)				& \compA, \compb. \\ 
NH$^{s23}$			 & 5 & 47 & -- & -- & --$^\textrm{a}$ & --$^\textrm{a}$								& \compA? \\ 
\nhthree$^{s25}$		& 7  & 28\ldots286	& 13.3	& 4.5	& 2.6(1)	& 3.2(-16)				& \compA, \compb. \\ 
$^{15}$NH$_{3}$$^{s26}$ & 1 & 28 & 11.3 & 5.8 & 1.4(-1) & 1.0(-18)						& \compA. \\ 
	\hline
CS$^{s27}$			 & 12 & 129 \ldots 543 & 12.2 & 10.3 & 2.0(1) & 1.5(-16)				& \compA, \compb. \\ 
C$^{34}$S$^{s27}$	 & 1 & 127 & 10.0 & 1.7 & 2.1(-1) & 1.2(-18)								& \compA? \\ 
H$_{2}$S$^{s28}$	 & 6 & 55 \ldots 103 & 11.6 & 5.0 & 1.4(1) & 1.4(-16)						& \compA, \compb? \\ 
SO$^{s29}$			 & 12 & 166 \ldots 321 & 9.4 & 9.3 & 5.5(0) & 3.7(-17)				& \compC. \\ 
SO$_{2}$$^{s30}$		 & 2 & 65 \ldots 379 & 11.1 & 10.0 & 2.7(-1) & 1.7(-18)				& \compC, \compa, \compb? \\ 
\shplus$^{s31}$		& 2 & 25 & 12.6 & 2.8 & 2.0(-1) & 1.2(-18)							 & \compA, \compb? \\ 
	\hline
HCl$^{s32}$			 & 10 & 30 \ldots 90 & 11.4 & -- & 4.9(0)	& 9.8(-17)					& \compA, \compb. \\ 
H$^{37}$Cl$^{s32}$	 & 10 & 30 \ldots 90 & 11.4 & -- & 9.5(-1) & 8.6(-18)$^\textrm{b}$			& \compA, \compb. \\ 
\htwoclplus$^{s33}$	&	5	& 23 \ldots 58	& 9.4	& 1.3	& -8.2(-1)	& 9.3(-18)		& \compD. \\ 
\htwohclplus$^{s33}$	&	1	& 58	& 9.4	& 1.3	& -3.6(-1)	& -4.4(-18)				& \compD. \\ 
	\hline
HDO$^{s6}$		 & 3 & 43 \ldots 95 & 12.7 & 3.8 & 6.5(-1) & 6.0(-18)					& \compA, \compb? \\ 
DCN	$^{s34}$		 & 2 & 97 \ldots 125 & 12.0 & 4.9 & 3.6(-1) & 2.2(-18)					& \compA. \\ 
ND$^{s35}$			 & 1 & 25 & 11.2 & 2.5 & 2.6(-1) & 1.6(-18)									& \compA? \\ 
NH$_{2}$D$^{s36}$ 	& 2 & 24 & 11.3 & 2.6 & 6.2(-1) & 3.6(-18)						& \compA. \\ 
	\hline
HF$^{s32}$			 & 1 & 59 & 10.0 & 2.8 & -2.0(0) & -2.9(-17)							& \compD, \compa. \\ 
	\hline
	\hline
All$^\textrm{c}$		& \numberoflines	& 23 \ldots 752	& 12.0	& 5.4	& 4.0(3) 	& 4.8(-14)	& \\
\hline
	\end{tabular}

	\raggedright{
	\textbf{Notes. }The table gives the number of detected transitions for each species, the range of upper level energies, the typical $\rm v_{lsr}$ and FWHM, the total line flux in $\rm K\cdot km/s$ and in $\rm W\cdot m^{-2}$, \changesone{where the exponential is given in brackets}, and a note on the dominant line profile components. The line counts include hyperfine components and blends. \resubmit{Blends between species are excluded from the per-species flux sums, but included in the total line flux measured in the survey.} The species are grouped similarly to Section~\ref{sec:species}. \changesone{Dashes represent cases where a good single Gaussian fit was not obtained, mostly due to blending.} \resubmit{$^\textrm{a}$NH is unambiguously detected in absorption on the HCN~$11-10$ line.}; $^\textrm{b}$\changesone{due to a blend with \meth\ on the $2-1$ line, only the H$^{37}$Cl~$1-0$ flux is given}; $^\textrm{c}$ Excluding some blended lines.\\
	\textbf{References. } $^{s1}$\citet{Winnewisseretal1997}; $^{s2}$\citet{Cazzolietal2004}; $^{s3}$\citet{Klapperetal2001}; $^{s4}$\citet{Klapperetal2003}; $^{s5}$\citet{Pickettetal2005}; $^{s6}$\citet{Johns1985}; $^{s7}$\citet{BlakeFarhoomandPickett1986}; $^{s8}$\citet{Mulleretal2005}; $^{s9}$\citet{Murtzetal1998}; $^{s10}$\citet{Mulleretal2004}; $^{s11}$\citet{Mulleretal2000a}; $^{s12}$\citet{Lattanzietal2007}; $^{s13}$\citet{Paganietal2009}; $^{s14}$\citet{Cooksyetal1986a}; $^{s15}$\citet{Cooksyetal1986b}; $^{s16}$\citet{Muller2010}; $^{s17}$\citet{McCarthyetal2006}; $^{s18}$\citet{Padovanietal2009}; $^{s19}$\citet{Thorwirthetal2003}; $^{s20}$\citet{CazzoliPuzzarini2005}; $^{s21}$\citet{Thorwirthetal2000}; $^{s22}$\citet{Klischetal1995}; $^{s23}$\citet{FloresMijangosetal2004}; $^{s24}$\citet{Mulleretal1999}; $^{s25}$\citet{Yuetal2010}; $^{s26}$\citet{Huangetal2011}; $^{s27}$\citet{KimYamamoto2003}; $^{s28}$\citet{Belov1995}; $^{s29}$\citet{Bogeyetal1997}; $^{s30}$\citet{Mulleretal2000b}; $^{s31}$\citet{BrownMuller2009}; $^{s32}$\citet{Noltetal1987}; $^{s33}$\citet{Arakietal2001}; $^{s34}$\citet{Brunkenetal2004}; $^{s35}$\citet{Takanoetal1998}; $^{s36}$\citet{Fusinaetal1988}.}
\end{table*}

\section{Overview and results}\label{sec:results}

\subsection{Detected lines and species}\label{sec:detections}

We found and identified a total of \numberoflines\ lines from \numberofmolecules\ molecular and atomic species and \numberofisotopologs\ secondary isotopologs at or above a flux signal to noise level of 5. \retwo{All the detected features were identified, i.e. no unidentified lines remained.} The detections are summarized in Table~\ref{tab:species}, and a full list of lines\retwo{, including a description of potential blends,} is given in Table~A1, in Appendix~\ref{apx:detections}. Of the detected lines, 431 or 60\% belong to \meth. Another 74 or 10\% belong to \form. In comparison, from SO and SO$_{2}$ we detect only twelve and two transitions, respectively. Four deuterated isotopologs are detected: HDO, DCN, NH$_{2}$D and ND. The molecular ions include \hcoplus, \ntwohplus, \chplus, \shplus, \htwoclplus, \ohplus\ and \htwooplus.

The upper level energies of the detected transitions range from 24 to 752~K and the typical value is $\rm E_{u} \sim 100$~K, indicating that much of the emitting gas in the beam is warm or hot. \changesone{Many of the transitions have very high critical densities, $\rm n_{cr} > 10^{8} cm^{-3}$, suggesting that much of the emission originates in dense gas, probably a compact region. Self-absorption in CO, \htwoo\ and \nhthree\ indicates that foreground material is present at the source velocity, while blueshifted continuum absorption in \ohplus, \chplus, HF and other species points to another foreground component.}

In terms of integrated line intensity, CO dominates with \coolco\% of the total line flux, \htwoo\ is second with \coolwater\% and \meth\ third with \coolmeth\%. The line and continuum cooling are discussed in more detail in Sect.~\ref{sec:cooling}.

\begin{table*}
	\caption{\retwo{Results of rotational diagram analyses for isotopologs of CO} in two upper level energy ranges, $E_{\textrm{u}} < 200$~K and $E_{\textrm{u}} > 200$~K.}
	\label{tab:rotdiag}
	\centering
	\begin{tabular}{ l c c c c c l }
	Species	& Size	& \multicolumn{2}{c}{$E_{\textrm{u}} < 200$~K}	& \multicolumn{2}{c}{$E_{\textrm{u}} > 200$~K} & Notes	\\
	\hline
			&	& $T_{\textrm{rot}}$~[K]	& $N$~[cm$^{-2}$]	& $T_{\textrm{rot}}$~[K]	& $N$~[cm$^{-2}$]	&	\\
	\hline
	\hline
	CO			& beam-filling & $208$			& $6.0\cdot10^{16}$			& $226$			& $6.6\cdot10^{16}$			& \retwo{Only wings, $|v-v_{\textrm{lsr}}|\geq 2.5$~km/s.}	\\
	CO			& beam-filling & $67$			& $5.1\cdot10^{17}$			& $149$			& $2.4\cdot10^{17}$			& \retwo{Only wings, optical depth corrected.}	\\
	$^{13}$CO	& beam-filling & $68$			& $1.3\cdot10^{16}$			& $152$			& $4.8\cdot10^{15}$			& 	\\
	C$^{18}$O	& beam-filling & $52$			& $2.2\cdot10^{15}$			& --				& --							&	\\
	C$^{17}$O	& beam-filling & $36$			& $1.0\cdot10^{15}$			& --				& --							& Two lines ($7-6$ is excluded). \\
	\hline
	CO			& 15''	& $89$			& $3.4\cdot10^{17}$			& $177$			& $2.1\cdot10^{17}$			& \retwo{Only wings, $|v-v_{\textrm{lsr}}|\geq 2.5$~km/s.}	\\
		CO			& 15'' & $45$			& $5.0\cdot10^{18}$			& $124$			& $8.7\cdot10^{17}$			& \retwo{Only wings, optical depth corrected.}	\\
	$^{13}$CO	& 15'' 	& $46$			& $1.3\cdot10^{17}$			& $126$			& $1.8\cdot10^{16}$			& 	\\
	C$^{18}$O	& 15'' 	& $38$			& $2.3\cdot10^{16}$			& --				& --							&	\\
	C$^{17}$O	& 15'' 	& $27$			& $1.4\cdot10^{16}$			& --				& --							& Two lines ($7-6$ is excluded).
	\end{tabular}

	\raggedright{
	\textbf{Notes. }Uncertainties for the parameters were calculated including the effects of RMS noise and the relative flux calibration errors (10\%), and were found to always be $<~10$\%.
	}
\end{table*}

\subsection{Line profiles}\label{sec:components}

There is a wealth of information in the line profiles of the detected species. Two striking examples are the different \resubmit{line profile} components revealed by the different molecules and substantial changes in line parameters such as $\rm v_{lsr}$ and full-width half maximum (FWHM) with changing upper level energy or frequency for some molecules.

\retwo{The statistical significance of the flux of each detected line, by definition, is at least $5\sigma$. For some lines, this makes a visual confirmation unintuitive, however for clarity we prefer the formal signal-to-noise cutoff. As an example, the weakest CS lines are consistent with the intensity as modeled based on the stronger lines in the rotational ladder. The fluxes in Appendix~\ref{apx:detections} originate in channel-by-channel integration, while the velocity is given both from the first moment as well as a Gaussian fit to the profile, and the FWHM is given only from Gaussian fitting.}

\retwo{The parameters and uncertainties for each transition are given in Appendix~\ref{apx:detections}. We give here typical uncertainties of the Gaussian parameters for the 479 unblended lines. For \vlsr, a cumulative 82\% of the lines have uncertainties below $0.1$~km/s, 98\% fall below $0.5$~km/s and only one line has $\delta$\vlsr$~>~1$~km/s. This is an \htwohclplus\ line, which represents a weak set of blended hyperfine components. For FWHM, a cumulative 46\% of lines have uncertainties below $0.1$~km/s and 91\% are below $0.5$~km/s. Fourteen lines have $\delta$FWHM$~>~1$~km/s, but none of these have FWHM/$\delta$FWHM$~<5$, except for one weak \meth\ line and the \htwohclplus\ feature mentioned above.}

\resubmit{For the purposes of this paper, we distinguish the following line profile components, as illustrated in Figs.~\ref{fig:linecomponents}~and~\ref{fig:velowidth}: the \compa, \compb, \compc, \compd, and \compe. Their nature is discussed in Sect.~\ref{sec:kinematicstory}. We emphasize that this is first and foremost a morphological classification, and that the underlying spatial source structure may be more complex than is immediately apparent from the emergent line profiles.}

\resubmit{The \compa\ refers to \rethree{relatively} narrow lines, $FWHM~\sim~2\ldots6$~km/s. \rethree{While it is not clear that $6$~km/s really originates in quiescent material, currently we lump these velocities together to denote material likely related to various parts of the envelope, to distinguish them from broad lines tracing outflowing gas.} \compA\ may have at least two subcomponents, one at the source velocity of $11.4$~km/s and another at $12.2$~km/s. The \compb\ refers to a broad component, difficult to fully disentangle but apparently centered around $\sim~13$~km/s and with $FWHM~\geq~10$~km/s. The \compc\ is traced by SO and is unique for its combination of blueshifted velocity and large linewidth, both $\sim~9$~km/s. Other species may have contributions from this component. The \compd\ refers to narrow lines at $\sim9$~km/s, most of them in absorption. We use the term \compe\ for line profile features which do not fall in the above categories. The velocities of all components shift around by $\sim1$~km/s with species and excitation energy, pointing to further substructure in the emitting regions, although part of this variation is certainly due to measurement uncertainties and could also be due to uncertainties of order 1~MHz in the database frequencies of some species.}

\retwo{Using the \texttt{HIPI}\footnote{\texttt{nhscsci.ipac.caltech.edu/sc/index.php/Hifi/HIPI}} plugin for \texttt{HIPE}, we inspected several species for contaminating emission in the reference spectra of the dual beam switch observations. The species were those with the strongest lines or where contamination might be suspected. For CO, CI and CII strong emission in one or, in the case of CO, both of the HIFI dual beam switch reference positions creates artificial absorption-like features where emission is subtracted out of the on-source signal. This is evident for CO as deep, narrow dips in the line profiles, and makes determining the \compa\ contribution to these lines very inaccurate. In the case of CII, the line itself peaks to the blue of the reference position emission and we judge the problem to be less severe, similarly to CI where only one reference position appears to be substantially affected. The deep self-absorption in several \htwoo\ lines appears to be related to the source. Extended weak \htwoo~$1_{1,0}-1_{0,1}$ emission is present on $\sim5$' scales in OMC-2 \citep{Snelletal2000}, but this emission seems to peak strongly on OMC-2. Previous observations have demonstrated the large extent of uniform CO~$1-0$ \citep{Shimajirietal2011} and CII emission \citep{Herrmannetal1997}, consistent with the contamination seen in the HIFI spectra.}

\subsection{Line density}

In Table~\ref{tab:bandnoise}, we give the line density per GHz measured in each band. \resubmit{The typical value is 1~line~GHz$^{-1}$. \rethree{At 500~GHz, with RMS noise levels of $16$~mK, the line density is $\rm 1.9~GHz^{-1}$, and at 1~THz, with RMS noise levels of $70$~mK, it is $\rm 0.7~GHz^{-1}$.  At 1200~GHz, with and RMS noise of $158$~mK, the line density is only $\rm 0.3~GHz^{-1}$. The frequency resolution is $1.1$~MHz in all cases.} Clearly, the line density decreases markedly with frequency. This is due to} the decreasing sensitivity of our survey and the increasing demands on temperature and density to excite high-lying rotational levels. A similar decrease for Orion~KL was discussed by \citet{Crockettetal2010}. As the line detections cover only 7\% of all frequency channels at 500~GHz, a range where the highest number of transitions from \meth, \sotwo\ and other ``weed'' molecules is expected, we conclude that our survey is far from the line confusion limit.

\subsection{The continuum}

\resubmit{\retwo{While continuum emission studies are not the main goal of our HIFI survey, the high quality of the spectra allows a continuum level to be determined for use in line modeling. For example, local wiggles in the baseline can mimick a continuum and distort the absorption line to continuum ratio. We provide here a global second-order polynomial fit to the continuum in bands 1a through 5a}, where the data quality is highest and the frequency coverage is complete. We stitched the spectra with baselines intact and sampled every 10th channel to reduce the data volume. All spectral regions containing line detections were excluded from the fit. For a polynomial of the form}

\begin{equation}
T_{\textrm{mb}} \textrm{[K]} = a + b\cdot\nu\textrm{[GHz]} + c\cdot(\nu\textrm{[GHz]})^{2},
\end{equation}

\resubmit{we find the parameters to be $a~=~-0.51979$, $b~=~0.0015261$ and $c~=~-4.1104\cdot10^{-7}$. This fit is displayed in the second panel of Fig.~\ref{fig:fullsurvey} and is valid in the range 480 to 1250~GHz.}


\subsection{Rotational diagrams for CO isotopologs}

\retwo{We performed a basic rotational diagram \citep{GoldsmithLanger1999} analysis for CO. This is more straightforward than for many other species due to the detection of several isotopologs as well as the low critical density. The results are summarized in Table~\ref{tab:rotdiag}, with rotational excitation temperatures and local thermodynamic equilibrium (LTE) column densities. The rotational temperature, $T_{\textrm{rot}}$, equals the kinetic one if the emitting medium is in LTE, and if optical depth effects and the source size are accounted for. For subthermal excitation, if the source size is known, $T_{\textrm{rot}}$ gives an upper limit on \tkin, and the obtained column density is a lower limit on the true value. We provide results for two source sizes: beam-filling and 15''. The latter is consistent with the dense core \citep[in press]{Shimajirietal2008, LopezSepulcreetalInterferometry}.}

\retwo{For $^{12}$CO, we used only the flux more than $2.5$~km/s away from the line centre, to avoid the reference contamination that affects the lower lines. Thus, effectively the $^{12}$CO diagram results are for the line wings. In each velocity bin, the $^{12}$CO lines were corrected for optical depth using $^{13}$CO, yielding results that match those of $^{13}$CO very well. Assuming an isotopologue ratio of 63, the $^{12}$CO optical depth away from the line centre is found to be $\leq2$ and the value decreases with increasing $J_{\textrm{u}}$. We expect the C$^{18}$O and C$^{17}$O isotopologs to be optically thin, and a comparison of their column densities with that of $^{13}$CO shows that the latter is only slightly optically thick \rethree{at the line center, which dominates the flux of this species}. For $^{13}$CO and C$^{18}$O, we find a column density ratio of 6, while 7 is expected.}

\retwo{To look for changes with increasing $E_{\textrm{u}}$, we perform the analysis in two excitation regimes: $E_{\textrm{u}} < 200$~K and $E_{\textrm{u}} > 200$~K. We find that $T_{\textrm{rot}}$ increases by a factor of $2-3$ with $J_{\textrm{u}}$, in other words \rethree{higher-lying transitions have a higher excitation temperature}. Thus, while the results in Table~\ref{tab:rotdiag} show that a somewhat higher $T_{\textrm{rot}}$ is found for C$^{18}$O than for C$^{17}$O, more lines are detected for the former, and fitting the same transitions for both species gives a better match.}

\retwo{Due to its low critical density ($n~\lesssim~10^{6}$~cm$^{-3}$ up to $J_{\textrm{u}}~=~16$), CO is easily thermalized at densities typical of protostellar cores, $\sim10^{6}$~cm$^{-3}$. Such densities may not exist in the regions that emit in the line wings. Therefore, the temperature values in Table~\ref{tab:rotdiag} are upper limits on the kinetic temperature.}

\begin{figure}[!ht]
  \centering
  \includegraphics[clip=,width=1.0\columnwidth]{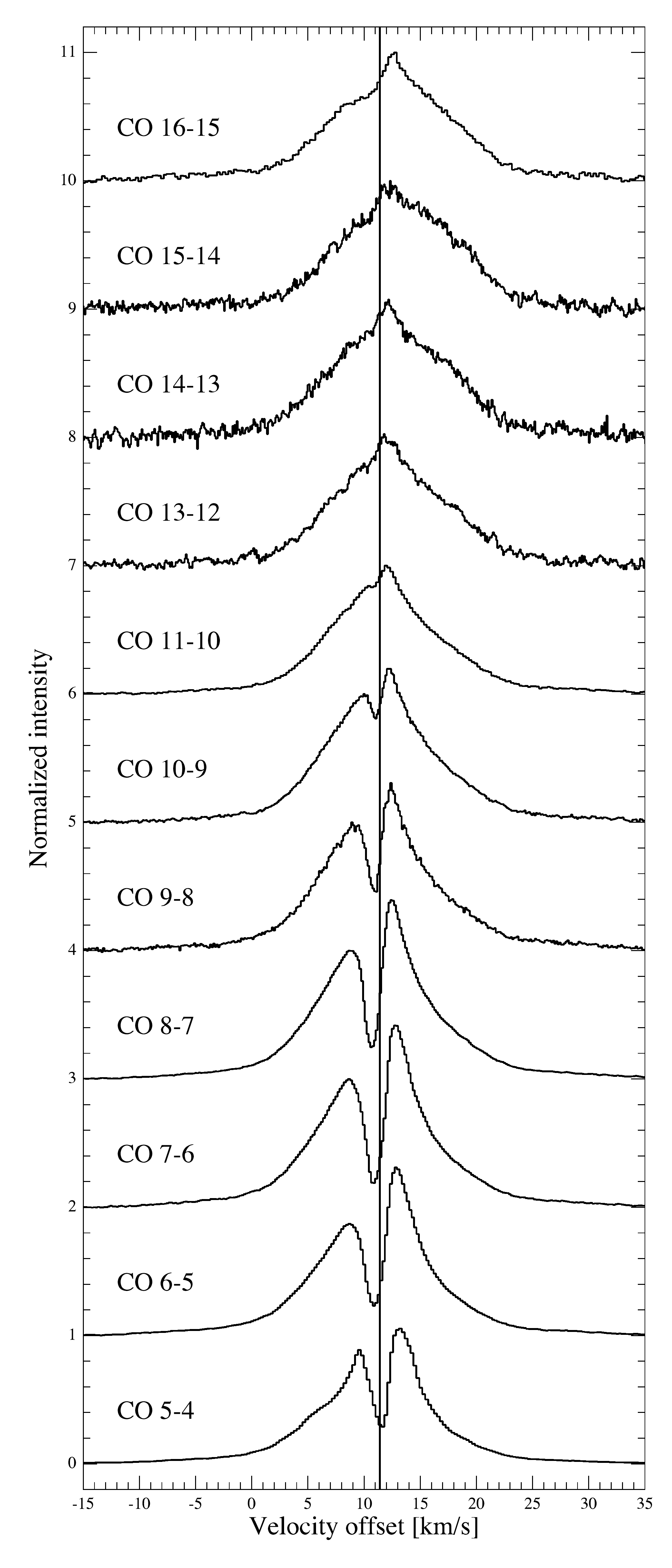}
     \caption{Normalized profiles of the CO lines detected in the survey, displayed with a vertical offset of unity between each profile. The solid vertical line marks the source velocity of 11.4~km/s. The \compb\ are increasingly important toward high rotational levels. \retwo{Up to $J_{\textrm{u}}~=~11$, the \compa\ is masked by contamination in the reference beams, which causes absorption-like dips in the profiles.}}
        \label{fig:COlines}
\end{figure}

\begin{figure}[!ht]
  \centering
  \includegraphics[clip=,width=1.0\columnwidth]{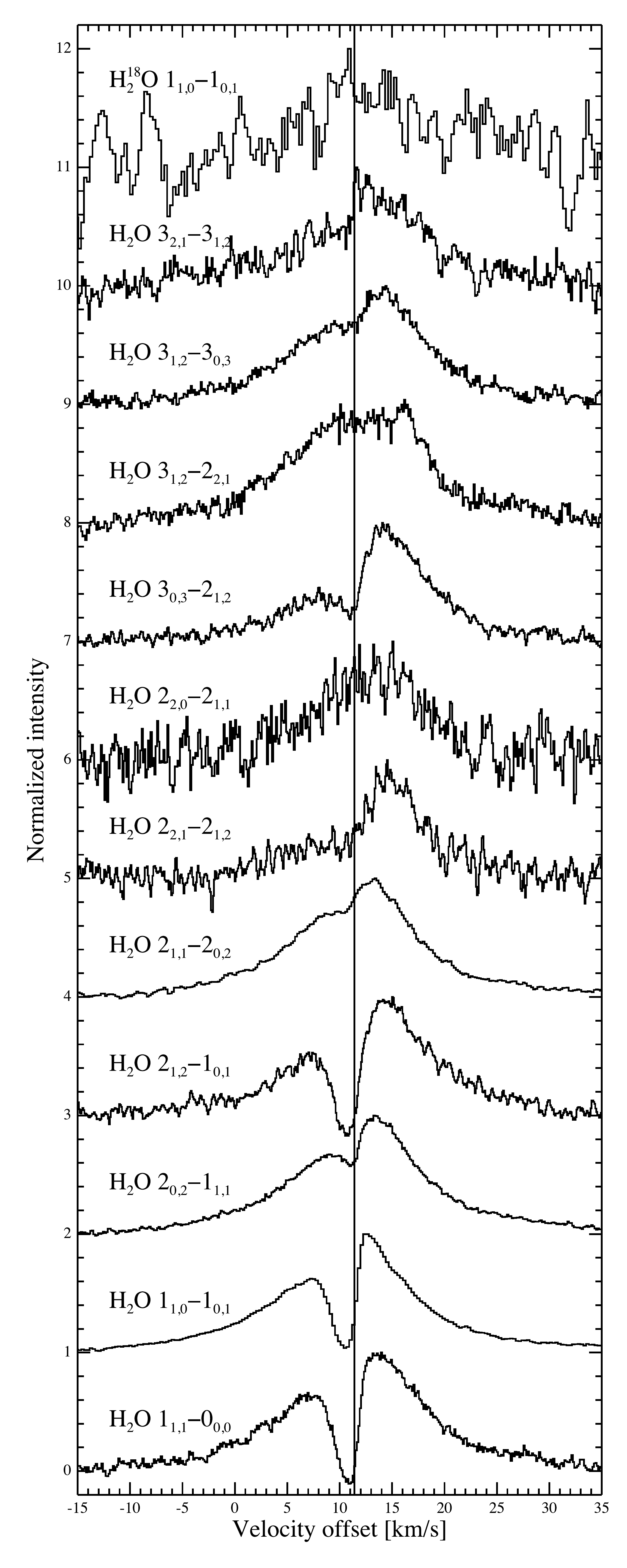}
     \caption{Normalized profiles of the \htwoo\ lines detected in the survey, displayed with a vertical offset of 1 between each profile. The solid vertical line marks the source velocity of 11.4~km/s. \retwo{The narrow dips are dominated by self-absorption.}}
        \label{fig:H2Olines}
\end{figure}

\retwo{Analyses of ground-based observations of C$^{18}$O in \oursource\ are consistent with our results. The column density we find assuming a beam-filling source, $N~=~2.2\cdot10^{15}$~cm$^{-2}$, is almost exactly the same as that found by \citet{CastetsLanger1995} from the $1-0$ and $2-1$ lines with with the 15~m SEST telescope, $N~=~2.5\cdot10^{15}$~cm$^{-2}$. Using the IRAM 30~m telescope and the same transitions, \citet{AlonsoAlbietal2010} found $T_{\textrm{rot}}~=~22$~K and $N~=~4.8\cdot10^{15}$~cm$^{-2}$. For a centrally concentrated source, such an increase of average column density with decreasing beam size ($\theta_{\textrm{SEST}}~\approx~2\cdot\theta_{\textrm{IRAM}}$) is expected, although given that the true uncertainties on any column density determination are likely around a factor of two, the difference may be insignificant. The different  C$^{18}$O rotational temperatures, 52~K from HIFI and 22~K from IRAM, again assuming beam-filling, indicate that the high-$J$ lines preferentially probe warmer gas, \resubmit{consistent with our rotational diagram results in the low and high $E_{\textrm{u}}$ regimes.}}

\subsection{Comments on individual species}\label{sec:species}

Here, we comment on each detected species. All quoted line parameters are from single-Gaussian fits, unless explicitly stated otherwise.

\subsubsection{CO, $^{13}$CO, C$^{18}$O, C$^{17}$O}\label{sec:CO}

The stitched spectrum of \oursource\ is dominated by the CO ladder, towering above the other lines in Fig.~\ref{fig:fullsurvey} and shown individually in Fig.~\ref{fig:COlines}. The peak $\rm T_{mb}$ values are in the range $4.4\ldots20.1$~K. \resubmit{\retwo{The lines are dominated by emission in the \compb\ and \compa\ velocities, but up to $J_{\textrm{u}}~=~11$, emission from the reference beams masks the narrower component and results in fake absorption features due to signal subtraction. } With increasing $J_{u}$, the Gaussian fit velocity shifts from 11.9 to 13.3~km/s.}

As the relative contribution of the wings increases with increasing $J_{\textrm{u}}$ level, or equivalently with decreasing beam size, the \textit{wing} component likely \resubmit{traces gas that is hotter than any other important CO-emitting region in \oursource.} \retwo{On the other hand, referring to the rotational temperatures in Table~\ref{tab:rotdiag}, we see that the optical depth corrected $^{12}$CO wing $T_{\textrm{rot}}$ matches that of the \emph{entire} $^{13}$CO line very well -- there is a correlation between the fluxes in the line wings and centres.}

\begin{figure}[!h]
  \centering
  \includegraphics[clip=,width=1.0\columnwidth]{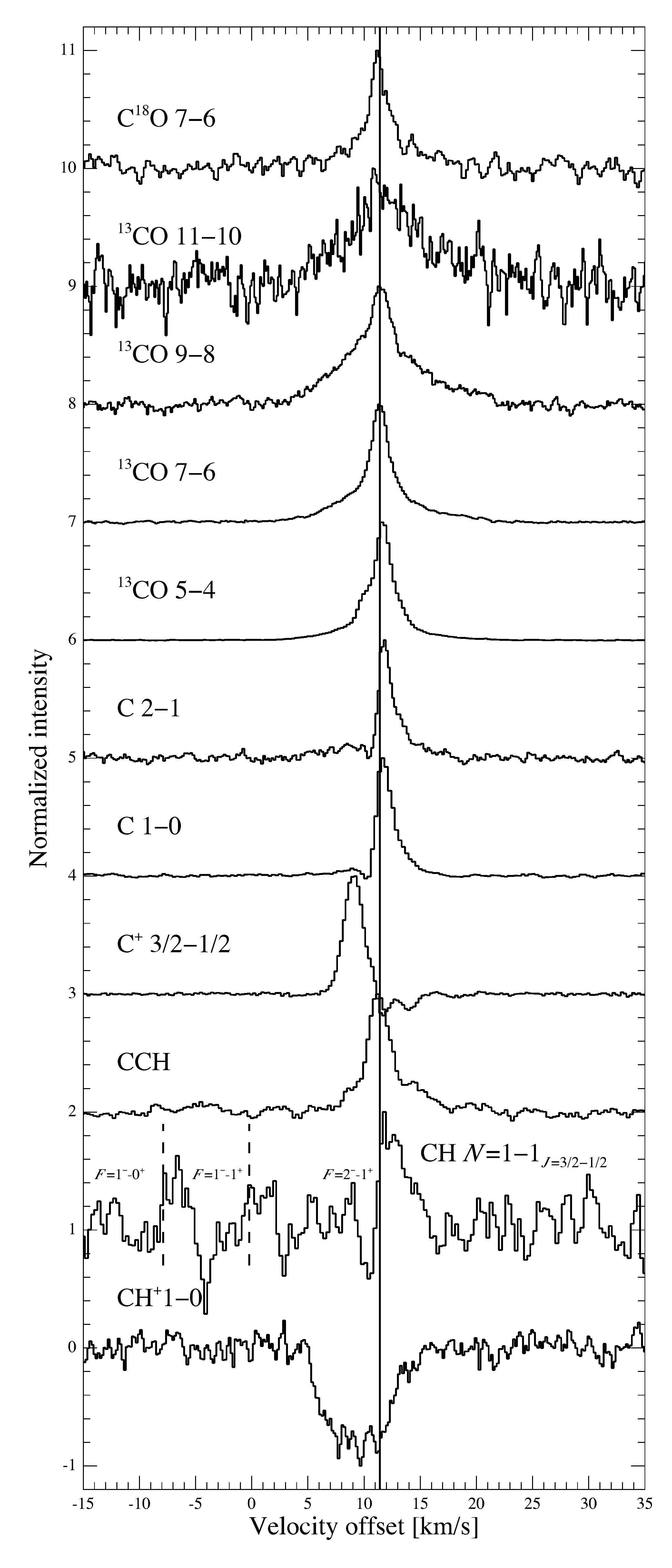}
     \caption{Normalized line profiles of carbon-bearing species, illustrating their different kinematics. The solid vertical line marks the source velocity of 11.4~km/s. The short vertical dashed lines show the CH hyperfine components except the one the $\rm v_{lsr}$ is centered on. \retwo{No CASSIS model was made for CH, see text.} The CII absorption at $11\ldots15$~km/s is an artefact due to contamination in the reference spectra. \retwo{The dip in CI at $\sim10$~km/s as also due to reference beam contamination.}}
        \label{fig:carboncomparison}
\end{figure}

We also detect several lines of the isotopologs $^{13}$CO, C$^{18}$O and C$^{17}$O. While C$^{18}$O and C$^{17}$O trace the \compa\ component, the $^{13}$CO lines contain hints of the \compb\, as seen in Fig.~\ref{fig:carboncomparison}. From $J_{u} = 5$ to 11, the Gaussian linewidth of $^{13}$CO increases near-linearly from 3~km/s to 9~km/s. With increasing $J_{u}$, the width of the C$^{18}$O lines changes from 2 to 3.4~km/s, a less pronounced change than $^{13}$CO but similar to \ntwohplus.

\resubmit{\retwo{We list the C$^{17}$O~$6-5$ line as} blended with \meth, but similar methanol transitions are not detected and the line is unusually narrow and redshifted to be a \meth\ detection, suggesting the flux originates purely in the CO isotopologue. The C$^{17}$O~$7-6$, however, has a significant flux contribution from a blended \form\ line, as evidenced by the detection of formaldehyde transitions similar to the blended one.}

\subsubsection{Water: \htwoo, OH, \ohplus, \htwooplus}\label{sec:water}

\begin{figure}[!h]
  \centering
  \includegraphics[clip=,width=1.0\columnwidth]{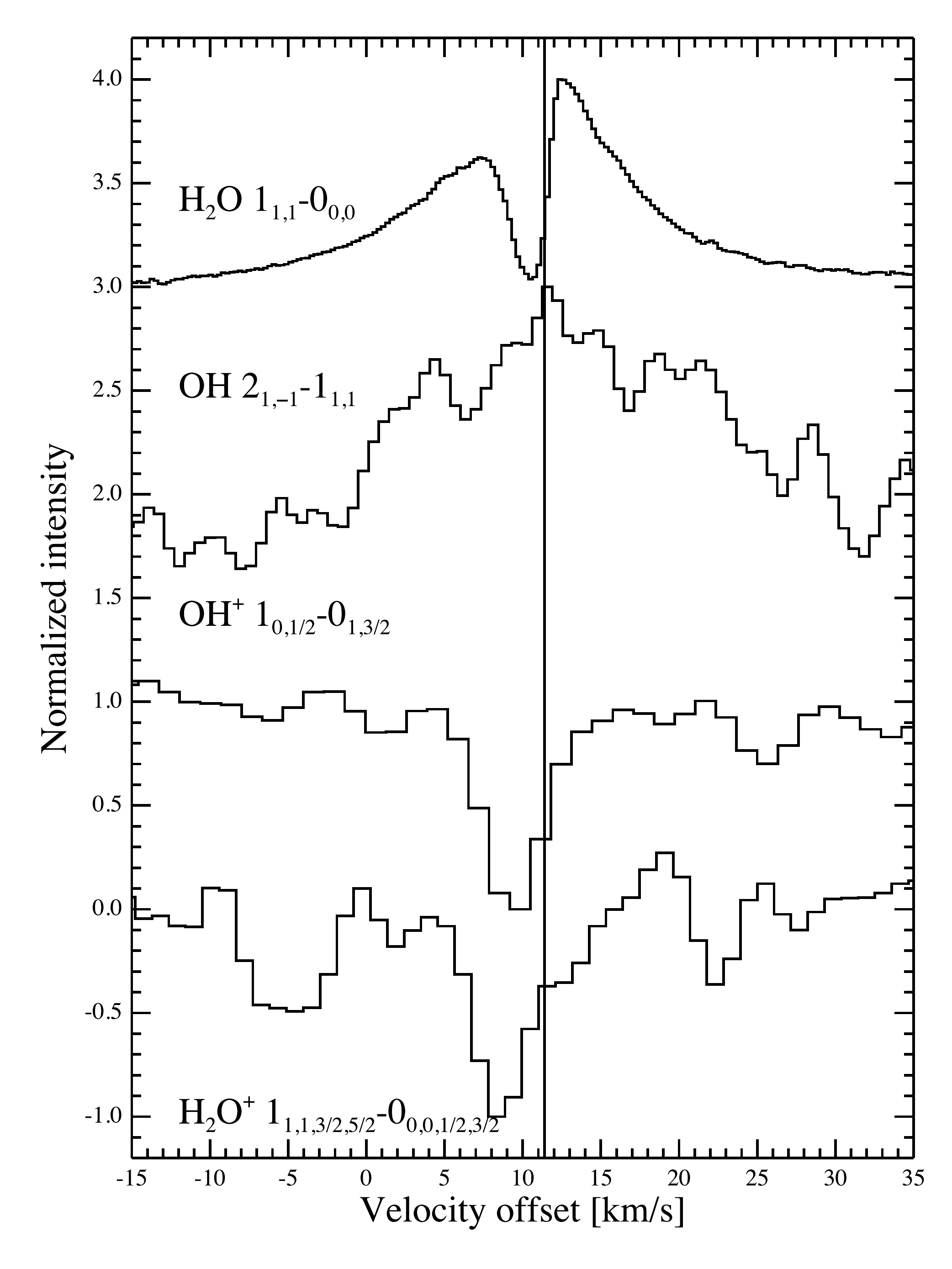}
     \caption{Normalized line profiles of some oxygen-bearing species, in particular relating to \htwoo\ chemistry. The solid vertical line marks the source velocity of 11.4~km/s. \retwo{The OH, \ohplus\ and \htwooplus\ spectra have been gaussian-smoothed to 4~MHz for clarity.}}
        \label{fig:hydro}
\end{figure}

One of the key molecules observable with HIFI, water, is well detected in \oursource, as seen in Figs.~\ref{fig:H2Olines}~and~\ref{fig:hydro}. Similarly to CO, the \htwoo\ lines are self-absorbed within a $\sim 0.5$~km/s blueshift from the source velocity, \resubmit{corresponding to the \compa\ but in absorption. They also clearly display the \compb\ component, Fig.~\ref{fig:velowidth} shows the \htwoo\ wings are broader than those of CO and the lines are typically centered near 13~km/s.} The peak intensity of the lines does not exceed $\sim 4.5$~K. The single-Gaussian fit linewidths vary considerably, from 9.9 to 20.5~km/s, although it must be kept in mind the line profiles are complex. The isotopologue \htwoeo, weakly detected, is centered at $\rm v_{lsr} = 13.7$~km/s \resubmit{ and 19.2~km/s wide. As seen in the top panel of Fig.~\ref{fig:H2Olines}, the isotopologue profile is flat and weak, and thus difficult to interpret, but it appears to be as broad as \htwoo\ itself.} The \htwoo\ lines point to a complex underlying velocity and excitation structure within the envelope.

\resubmit{The $3/2-1/2$ transition of OH, comprizing six hyperfine components, is detected in emission and one set of hyperfine components is shown in Fig.~\ref{fig:hydro}. An LTE fit with CASSIS, including the hyperfine structure, yields $\rm v_{lsr} = 12.7$ and $\rm FWHM = 19.1$~km/. This is consistent with the \compb\ component, \resubmit{of which OH may be the best tracer in terms of line profile complexity}. The fit is mostly useful for constraining the kinematic parameters of this species -- the source size, column density and excitation temperature are not well constrained. In the best-fit model, the hyperfine components are optically thin ($\tau~\lesssim~0.01$). The OH line parameters are very similar to those typical for \htwoo\ and for high-$J$ CO lines.}

Multiple \ohplus\ transitions are detected in absorption at 9.3~km/s, part of the \compd\ component. We also detect the \htwooplus\ $1_{1,1}-0_{0,0}$ line at 8.4~km/s. \resubmit{Selected lines are presented in Fig.~\ref{fig:hydro}. These ions and the tenuous gas they probe are analyzed in a companion paper \citep{LopezSepulcreetal2013}.}

\begin{figure}[!h]
  \centering
  \includegraphics[clip=,width=1.0\columnwidth]{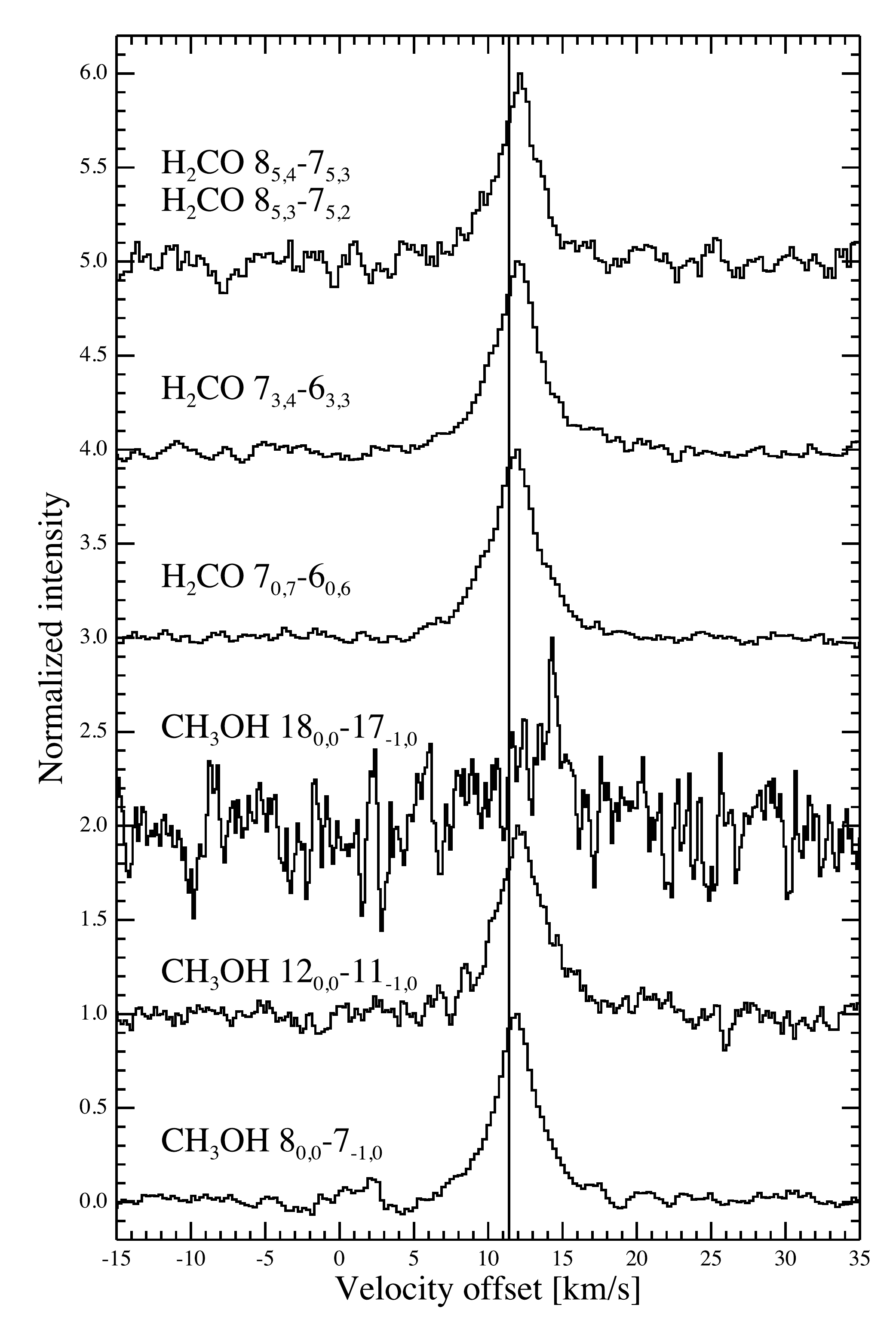}
     \caption{Normalized line profiles of some \meth\ and \form\ lines, representing, from bottom to top, upper level energies $E_{\textrm{u}}~\approx~100$~K, 200~K and 400~K for both species. The solid vertical line marks the source velocity of 11.4~km/s.}
        \label{fig:methform}
\end{figure}

\subsubsection{\meth\ and \form}

\resubmit{Lines of \meth\ and \form\ are shown in Fig.~\ref{fig:methform}, although due to the enormous number of detections the only criterion for the displayed subset is to cover upper level energies from $\sim100$~K through 200~K to $400$~K for both species.}

\resubmit{\meth\ dominates the number of detected lines in \oursource\ with 431 lines, and \form\ is second with 74. While the median \meth\ velocity is \vlsr$=~12.2$~km/s and FWHM$~=~4.7$~km/s, the lines display a significant trend in $\rm v_{lsr}$ with upper level energy}: at $E_{\textrm{u}} = 50$~K, they peak at 11.8~km/s, while by $E_{\textrm{u}}~=~450$~K, the typical peak has shifted to 12.5~km/s. \resubmit{This suggests the low-excitation lines are dominated by the source velocity component of the \compa, while the redshifted ($\sim~12$~km/s) component becomes increasingly dominant at high excitation energies.} \changesone{This dominance of an underlying hot component is in line with the conclusion of our previous \meth\ analysis \citep{Kamaetal2010}, where a subset of the lines was presented, including line profiles and rotational diagrams.}

The \form\ lines range in upper level energy from 97~K to 732~K, and \resubmit{the median parameters, \vlsr$~=~11.9$~km/s and FWHM$~=~4.7$~km/s}, are statistically indistinguishable from those of \meth. The line of sight velocity trend of \form\ also resembles \resubmit{that of \meth}.

\resubmit{As no isotopologs of \meth\ and \form\ are detected and because their excitation will be analyzed in a separate paper, we do not give rotational diagram results for these species in Table~\ref{tab:rotdiag}. However, the shifting of the emission peaks with upper level energy and frequency suggests these species probe multiple regions of \oursource.}

\subsubsection{\hcoplus\ and \ntwohplus}\label{sec:bigions}

\begin{figure}[!h]
  \centering
  \includegraphics[clip=,width=1.0\columnwidth]{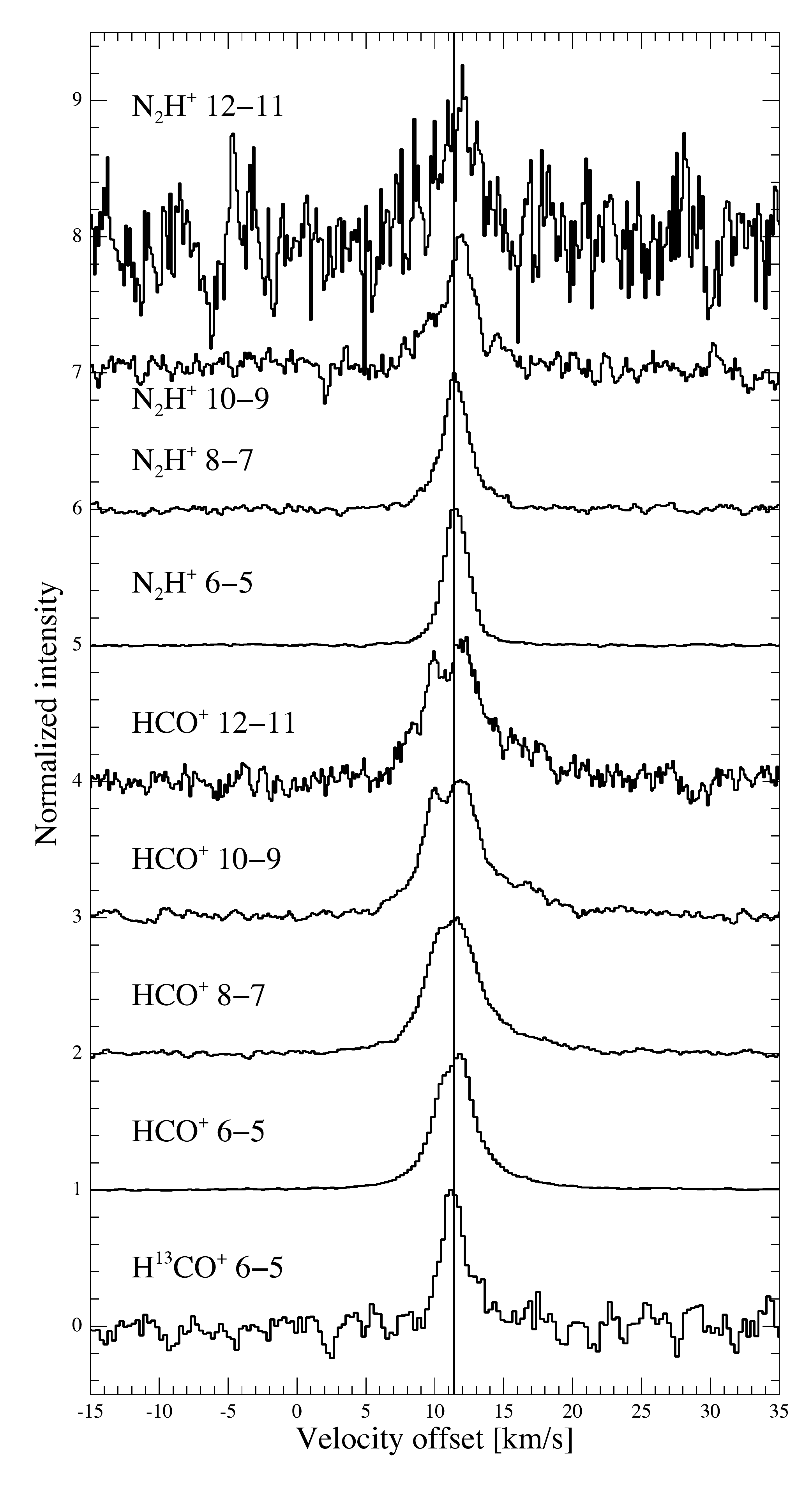}
     \caption{Selection of normalized line profiles of \hcoplus\ and \ntwohplus, illustrating trends in the line kinematics. The solid vertical line marks the source velocity of 11.4~km/s.}
        \label{fig:ioncomparison}
\end{figure}

The \hcoplus\ and \ntwohplus\ lines, \resubmit{an overview} of which can be seen in Fig.~\ref{fig:ioncomparison}, are dominated by \resubmit{the \compa\ component. The \compb\ component appears to contribute at a low level to \hcoplus\ emission.} Towards higher $J$ levels, the \hcoplus\ lines become double-peaked, with one component near 10~km/s and another at 12~km/s. An examination of Fig.~\ref{fig:ioncomparison} shows that the double peak is relevant mostly for the highest \hcoplus\ lines, for which the critical densities are $\sim 10^{8}$~cm$^{-3}$ and lower level energies $> 200$~K. \resubmit{The apparent double peak may be due to self-absorption at $\sim10$~km/s. \retwo{There is no indication in the reference spectra of contamination problems for \hcoplus.} H$^{13}$CO$^{+}$ is a prime example of the \compa\ component.}

The \ntwohplus\ stays single-peaked, but at $E_{\textrm{u}}~>~200$~K ($J_{\textrm{u}}~>~9$), the peak redshifts to 12~km/s. While the \ntwohplus\ lines are narrower than those of \hcoplus\ by roughly a factor of two, their widths and velocities are similar to the H$^{13}$CO$^{+}$, C$^{18}$O and C$^{17}$O lines.

\subsubsection{Carbon: CI, CII, CH, CCH, \chplus}

Aside from CO, a number of simple carbon-bearing species are detected. A comparison of some lines of $^{13}$CO, CI, CII, CCH, CH and \chplus\ is given in Fig.~\ref{fig:carboncomparison}, indicating substantial variations in the line profiles.

\resubmit{The CI and $^{13}$CO lines also correspond to the \compa.} The highest-$J$ $^{13}$CO lines show a broad base, likely the \compb. \retwo{The CI lines show an absorption dip at around 10~km/s, which is due to emission in the reference positions and may be the cause of the slight redshift observed for CI in Fig.~\ref{fig:velowidth}. The CI line fluxes from the bulk of \oursource\ are underestimated due to the self-absorption.}

\resubmit{The CII line peaks at 9.5~km/s and likely represents emission from the \compd.} The CII absorption in the $11\ldots15$~km/s velocity range is an artefact caused by emission in the reference position spectra. \changesone{To estimate the amount of flux lost due to this at the line peak, we reprocessed the data with one DBS reference spectrum at a time. We established that, while the $11\ldots15$~km/s absorption features vary strongly with the choice of reference spectrum, the 9.5~km/s peak varies only by $\sim 10$\%, suggesting that the amount of line flux lost in this component through reference subtraction is small or that the large-scale emission at this velocity around OMC-2 is remarkably uniform. \retwo{The latter is unlikely as the difference spectrum of the two reference positions shows a strong positive-negative residual, indicating a velocity offset between the components.} If the extended CII emission at 9.5~km/s in OMC-2 is of comparable intensity to the detected peak, any intensity fluctuations must be at the $\leq 10$\% level on a scale of 6', given a resolution of 12''.} Accounting for only the upper level population, we obtain a lower limit of $\rm 3.4\cdot10^{16}$~$\rm cm^{-2}$ on the CII column density toward FIR~4.

\begin{figure}[!h]
  \centering
  \includegraphics[clip=,width=1.0\columnwidth]{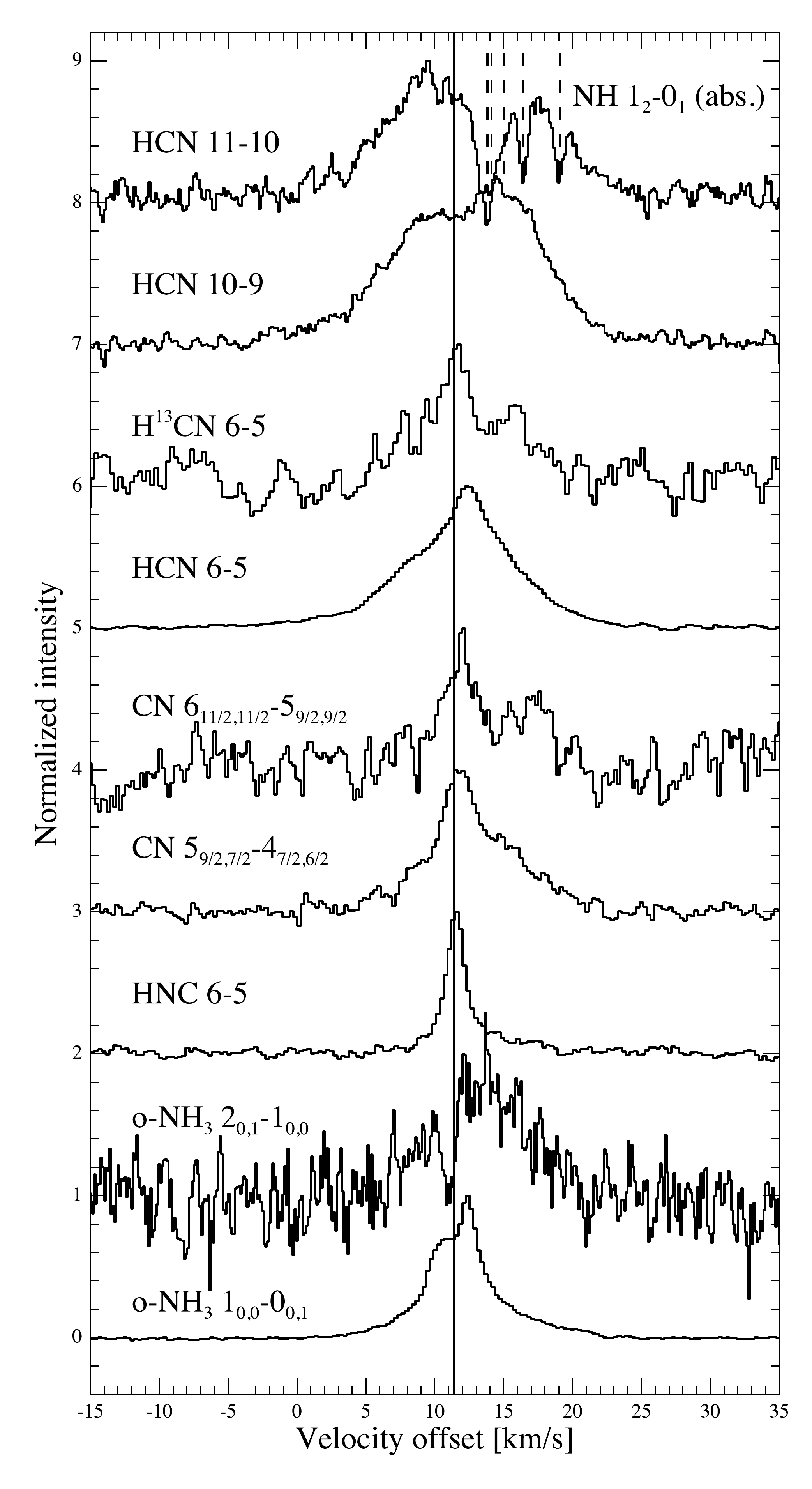}
     \caption{Examples of normalized line profiles of nitrogen-bearing species. The solid vertical line marks the source velocity of 11.4~km/s. \resubmit{The top line profile is centered on HCN and the dashed vertical lines mark the locations of the strongest NH hyperfine components.} }
        \label{fig:nitrogencomparison}
\end{figure}

CCH is dominated by the \compa\ component, but shows evidence for \compe\ components. \resubmit{The lines have a broad base which appears redshifted, perhaps indicating a contribution from the \compb. \retwo{The CH lines are slightly redshifted and correspond to the \compa\ emission. No CASSIS model fitting was done, the line parameters were obtained from a Gaussian fit to a detected isolated hyperfine component. The parameters are similar to those of \shplus\ and HDO.}} The CH$^{+}$ ion absorption line is strongly saturated, \resubmit{leading to the increased linewidth seen in Fig.~\ref{fig:velowidth}, but it is centered on the \compd\ component. We compare \chplus\ and \shplus\ in Sect.~\ref{sec:sulphur}.}

\subsubsection{Nitrogen-bearing molecules}

Aside form \ntwohplus, which is covered in Sect.~\ref{sec:bigions}, the main detected nitrogen-bearing species are HCN, CN, HNC and the nitrogen hydrides. There are substantial differences in the profiles with species as well as with upper level energy for the nitrogen-bearing species, similarly to carbon and sulphur. In Fig.~\ref{fig:nitrogencomparison}, lines of several representative nitrogen-bearing molecules are shown on a common velocity scale.

\begin{figure}[!h]
  \centering
  \includegraphics[clip=,width=1.0\columnwidth]{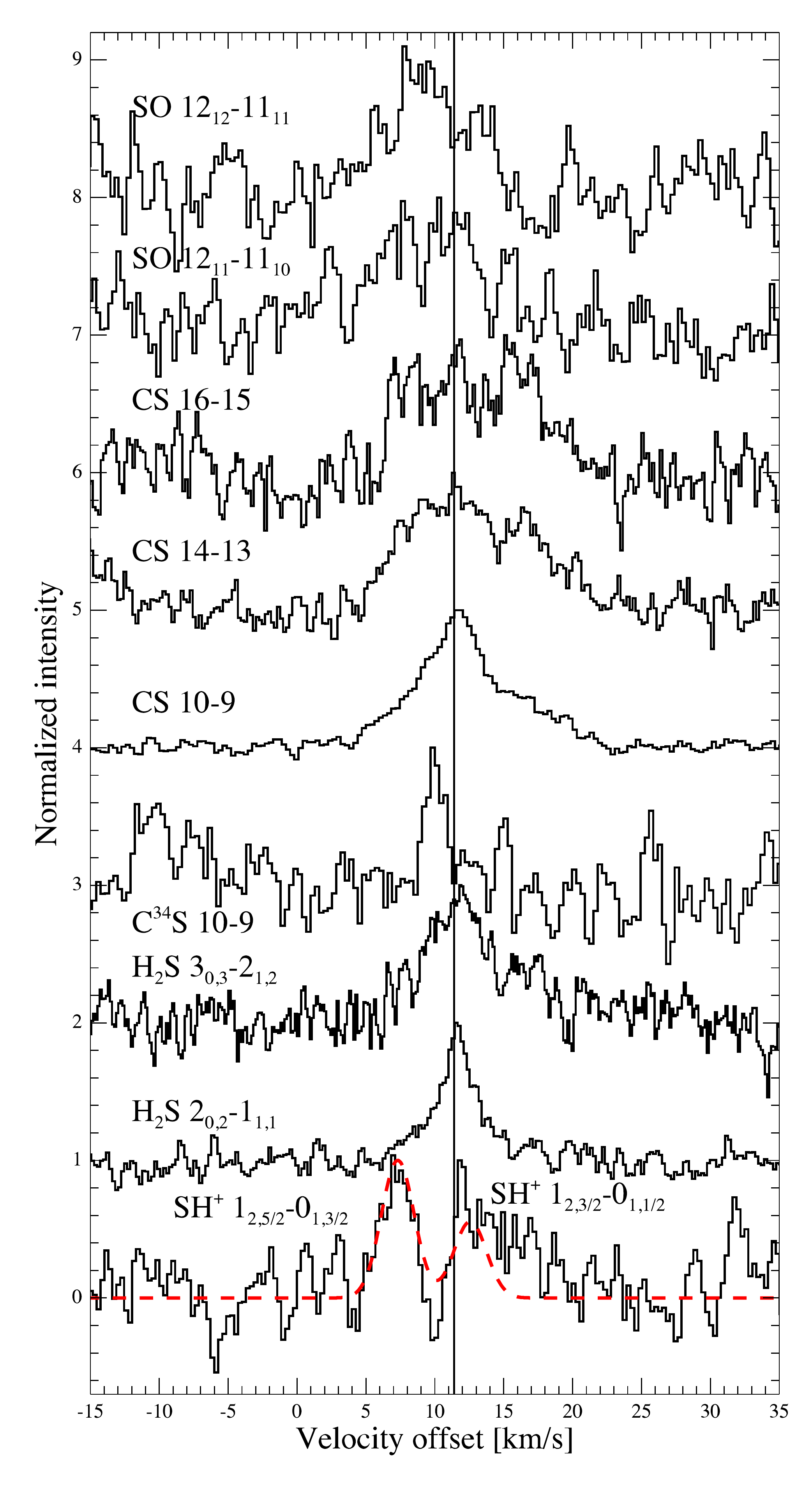}
     \caption{Examples of normalized line profiles of sulphur-bearing species. The solid vertical line marks the source velocity of 11.4~km/s. \retwo{The CASSIS model for \shplus, showing two hyperfine components, is given by the dashed red line.}}
        \label{fig:sulphurcomparison}
\end{figure}

\textit{\textbf{HCN, CN and HNC. }}Gaussian fits to the HCN lines peak at $12.1$~km/s. \resubmit{Their large linewidth, characteristic of the \compb\ component, puts them close to CO and \htwoo\ in Fig.~\ref{fig:velowidth}.} The upper level energies of the detected HCN lines range from 89 to 447~K. The high-lying lines, in particular, have critical densities around $10^{10}$~cm$^{-3}$, indicating that the emitting regions contain very dense and hot gas. \resubmit{The H$^{13}$CN profiles are also broad, although they show a peak corresponding to the \compa.}

\resubmit{The two detected HNC lines trace the \compa\ component. The CN line profiles are dominated by the \compa\ component, but there is also a contribution from the \compb, as seen in Fig.~\ref{fig:nitrogencomparison}.}

\textit{\textbf{NH and \nhthree. }} We detect some hyperfine components of the NH~$1_{2}-0_{1}$ transition in absorption on the HCN~$11-10$ emission line, \resubmit{as shown in Fig.~\ref{fig:nitrogencomparison}. The lines are close to the \compa\ component velocity.} The dominant hyperfine components are in absorption until below the continuum level. The ND~(1$_{2,3,4}$-0$_{1,2,3}$) transition is detected in emission \resubmit{and is shown in Fig.~\ref{fig:deuteratedcomparison}}.

Seven transitions of \nhthree\ are detected, covering 28 through 286~K in upper level energy. \resubmit{As seen in Fig.~\ref{fig:nitrogencomparison}, the lines have a broad base, consistent with the \compb, and they show narrow self-absorption near the \compa\ component. \retwo{We find no evidence for emission in the reference positions to be causing the absorption, but due to the relative weakness of the features this cannot be fully excluded at present.} While the lines have a similar appearance to CO, \htwoo\ and OH, and are centered at the same velocity, they are typically a factor of four narrower, as seen in Fig.~\ref{fig:velowidth}. We return to this point in Sect.~\ref{sec:kinematicstory}.} The fundamental ortho-\nhthree\ line has a particularly interesting profile, with a sharp peak at $12.5$~km/s and a plateau at $10\ldots12$~km/s.

\subsubsection{Sulphur-bearing molecules}\label{sec:sulphur}

The detected sulphur-bearing species are CS, C$^{34}$S, \htwos, SO, \sotwo\ and \shplus. As seen from the selection of lines in Fig.~\ref{fig:sulphurcomparison}, there are significant differences between their line profiles.

\resubmit{The detected CS lines cover an upper level energy range of $\rm 129~K \leq E_{u}~\leq~543~K$ and seem similar to CO in that they may have contributions from the \compa\ and the \compb.} The critical densities of the high-lying lines are $n~>~10^{8}$~cm$^{-3}$, indicating the presence of dense and hot gas in the emitting regions. \resubmit{The C$^{34}$S $10-9$ line is narrow and peaks at around 10~km/s, likely tracing a dense, warm clump with a high CS abundance, at that velocity.} \retwo{The clumpy distribution of C$^{34}$S in \oursource\ is explored further by \citet[in press]{LopezSepulcreetalInterferometry}.}

\resubmit{The \htwos\ lines are dominated by the \compa\ component, but the \compb\ or \compc\ components may be present at a low level.}

\resubmit{The SO lines are the sole clear tracer of the \compc\ component. They cover} an upper level energy range of 165 to 321~K and have critical densities of order $10^{8}$~cm$^{-3}$, suggesting dense and hot gas in the emission region. \resubmit{It is notable that the SO lines are $9.3$~km/s broad and peak at $9.4$~km/s, 2~km/s to the blue from the OMC-2 systemic velocity. We detect only two transitions of \sotwo, which are similar in their median properties to CO, perhaps having contributions from the \compc, \compa\ and \compb\ components.}


The detected \shplus\ hyperfine line emission is shown in the lowest part of Fig.~\ref{fig:sulphurcomparison}. The line profiles are very similar to CH, as seen in Fig.~\ref{fig:velowidth} and by comparing with the second-to-lowest panel in Fig.~\ref{fig:carboncomparison}. \retwo{The line parameters were obtained by fitting the hyperfine structure with the CASSIS software, and found to be \vlsr$~=~(12.6\pm0.3)$~km/s and FWHM$~=~(2.8\pm0.3)$~km/s.} \resubmit{The central velocity of \shplus\ matches that of the \compb\ component, but the linewidth resembles the \compa. It is interesting to note that we detect \chplus\ in absorption at 9.8~km/s and \shplus\ in emission around 12.2~km/s, contrary to the strong correlation between these species seen in a sample of galactic sight-lines by \cite{Godardetal2012}. The synthesis path of \shplus\ is $\rm S^{+}~+~H_{2}~\rightarrow~SH^{+}~+~H$, which has an activation barrier of 9860~K, more than twice the 4640~K barrier to the production of \chplus\ via an analogous path. This may offer an explanation to our nondetection of \shplus\ in the \compd\ component, but it is not immediately clear why we do not detect \chplus\ at the same velocity as \shplus\ -- excitation and chemistry may both play a role. In the models of \citet{Brudereretal2010}, there are regions near a protostellar outflow base where the \shplus\ abundance is elevated by several orders of magnitude with respect to that of \chplus\ due to sulphur evaporation from grains.}

\begin{figure}[!h]
  \centering
  \includegraphics[clip=,width=1.0\columnwidth]{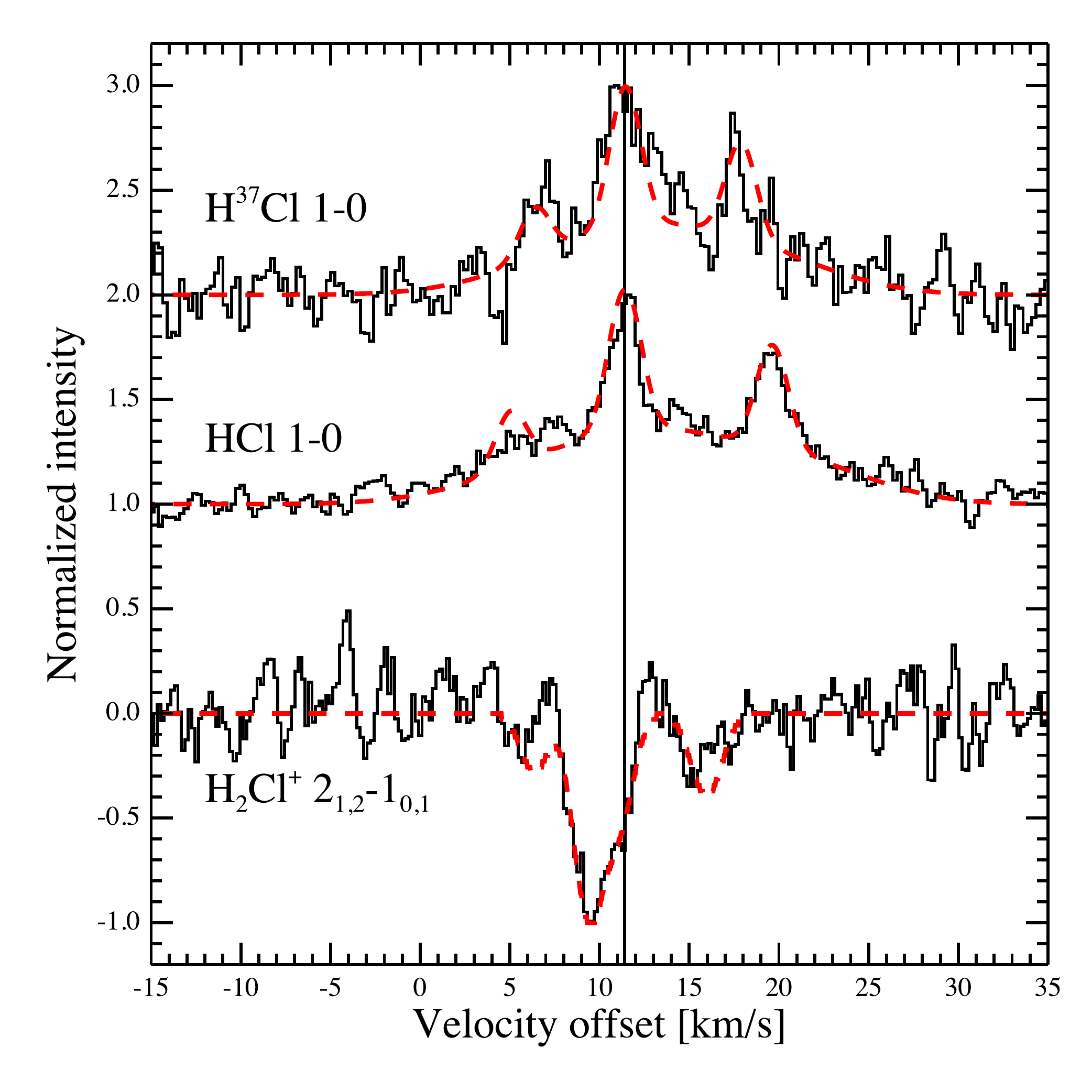}
     \caption{Normalized line profiles of the main detected HCl and \htwoclplus\ transitions. The solid vertical line marks the source velocity of 11.4~km/s. \retwo{The CASSIS models for each species, used to obtain the \vlsr\ and width of the lines, are shown with red dashed lines.}}
        \label{fig:chlorine}
\end{figure}

\subsubsection{Chlorine-bearing molecules}

Lines of isotopologs containing $^{35}$Cl and $^{37}$Cl of two of the main chlorine-bearing molecules are detected: \retwo{HCl~$1-0$ and $2-1$, and \htwoclplus~$1-0$ and $2-1$. For HCl, the line peak is at $\sim 11.4$~km/s, the \compa\ velocity.} \resubmit{There is a broad base which may be identified with the \compb\ component.} \retwo{For \htwoclplus, we fitted the hyperfine structure with CASSIS, obtaining \vlsr$~=~(9.4\pm0.2)$~km/s and FWHM$~=~(1.8\pm0.5)$~km/s. Several examples of transitions of chlorine-bearing species are shown in Fig.~\ref{fig:chlorine}.}

Hydrogen chloride is predicted to be the dominant gas-phase chlorine reservoir in high-extinction regions \citep{Blakeetal1986, NeufeldWolfire2009}, consistent with the expectation that the 11.4~km/s component traces the large-scale envelope. On the other hand, \htwoclplus\ is predicted to be a significant carrier in photon-dominated regions, and correspondingly \htwoclplus\ is detected in absorption in the blue-shifted \compd\ component. \changesone{The chlorine-bearing species will be the focus of an upcoming paper \citep{Kamaetal2012inprep}.}

\begin{figure}[!h]
  \centering
  \includegraphics[clip=,width=1.0\columnwidth]{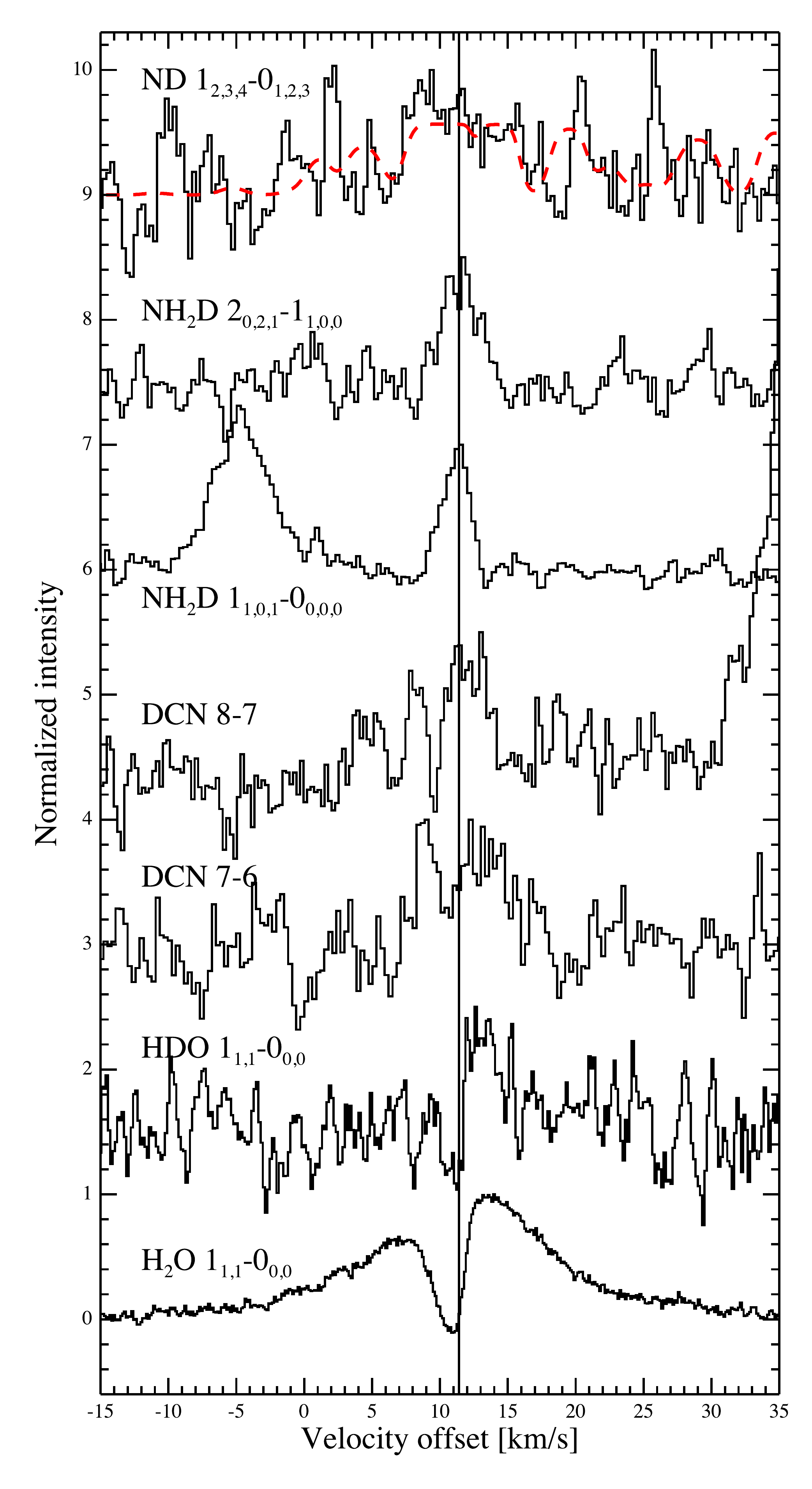}
     \caption{Selection of lines of deuterated species. A water line has been added for comparison with HDO. The solid vertical line marks the source velocity of 11.4~km/s. \retwo{The CASSIS model for ND is marked with a dashed red line.}}
        \label{fig:deuteratedcomparison}
\end{figure}

\subsubsection{Deuterated species}
   
Our HIFI survey is poor in deuterated species, the only detections are HDO, DCN, NH$_{2}$D and ND. \resubmit{The strongest lines are shown in Fig.~\ref{fig:deuteratedcomparison}. \retwo{The HDO lines are very weak, therefore firm conclusions cannot be drawn, but the $1_{1,1}-0_{0,0}$ line shown in Fig.~\ref{fig:deuteratedcomparison} suggests similarities with \htwoeo\ in velocity, and our tests with Gaussian decomposition of the \htwoo\ lines also suggest a similar emission peak location.} The DCN median \vlsr\ is 12~km/s, consistent within the errors with HCN but with the linewidth again much narrower and corresponds to the redshifted side of the \compa\ component, also traced by \meth\ and \form. Singly deuterated ammonia, NH$_{2}$D, peaks at v$\rm_{lsr} \approx 11.2$~km/s and corresponds well to other tracers of the \compa\ component. The ND line is centered near the \compa\ velocity but its width corresponds better to the \compb\ or \compc\ component. However, the line is weak and should be interpreted with caution.}

\subsubsection{HF}

The $J~=~1-0$ transition of hydrogen fluoride is detected in absorption at 10.0~km/s\resubmit{, and is shown in Fig.~\ref{fig:linecomponents}. The linewidth is 2.8~km/s and the estimated column density $(1.2\pm0.3)~\cdot~10^{13}$~cm$^{-2}$. Modeling by \citet{LopezSepulcreetal2013} suggests the presence of two different velocity components: one at 9~km/s, corresponding to the \compd, and a dominant component around 11~km/s, roughly the \compa\ velocity.}

\section{Discussion}\label{sec:discussion}

Much information about the kinematic, excitation and abundance structure of \oursource\ is encoded into the roughly 700 detected spectral lines. \changestwo{The source is kinematically and chemically complex.} Substantial amounts of molecular gas are present in the HIFI beam at all probed scales (40''\ldots 11'' or 17000~AU\ldots 5000~AU). We detect emission from high-$J$ transitions of CO, CS, HCN, \hcoplus\ and other species. The high critical densities (up to $n~\sim~10^{9}$~cm$^{-3}$) and \retwo{excitation energies} indicate that at least some of the emitting regions are very dense and hot or able to excite \retwo{some} transitions via infrared pumping. We focus below on the kinematical components and the energetics, deferring more detailed analyses to future papers.

\subsection{Interpretation of the kinematic components}\label{sec:kinematicstory}

\changestwo{We now discuss the characteristics and possible nature of the kinematic components defined in Sect.~\ref{sec:components}.} \resubmit{We emphasize again that the features discussed are predominantly line profile components and need not have a one-to-one correspondonence with actual source components. The diversity of line profile shapes is, in any case, a clear indication of the complexity of the underlying spatial variations of composition, density, temperature and velocity.}

\retwo{A correlation between the rotational temperature of the $^{12}$CO line wings and entire $^{13}$CO lines was pointed out in Sect.~\ref{sec:CO}. \rethree{This is based on a $^{12}$CO flux integration starting from $\pm2.5$~km/s from the line centre, and it is not clear if the same result would be found if the integration only included emission from much higher velocities, e.g. starting from $\pm10$~km/s. If the correlation holds}, it implies that there is a physical connection between the emission in the CO line centre and high-velocity wings and a Gaussian decomposition into a narrow and broad component may not be useful. This may contribute to a better understanding of the origin of the CO emission. The results of a detailed study of the CO and \htwoo\ line wings will be presented in a separate paper \citep{Kamaetal2013inprep}. For the moment, we proceed with describing the morphological components proposed earlier in this paper.}

\compA. \resubmit{This line profile component may have several subcomponents. One, centered around $11.0\ldots11.5$~km/s, likely corresponds to the large-scale ($\sim 10^{4}$~AU or 25'') envelope of \oursource\ and matches the velocity of the bulk OMC-2 cloud \citep{CastetsLanger1995, Asoetal2000}. This component is most clearly seen in C$^{18}$O, C$^{17}$O, \ntwohplus\ and NH$_{2}$D, and it \resubmit{appears to contribute} to the emission of most detected species. The other component is centered near 12.2~km/s and is best traced by \meth, \form\ and DCN. \rethree{We reiterate that some of the lines classified as \compa\ have substantial widths, $>5$~km/s, however we lump them together with other lines which are relatively narrow when compared to the broadest lines, which must trace outflowing rather than envelope material.}}

\retwo{The \htwoo, \nhthree\ and high-$J$ \hcoplus\ self-absorption (if \hcoplus\ is indeed self-absorbed) is blueshifted with respect to the source rest frame. One hypothesis for explaining this blueshifted self-absorption feature is a slowly expanding layer of warm gas in the inner envelope. Similar absorption features on broad \htwoo\ line profiles in a sample of protostars are discussed as originating in the inner envelope by \citet{Kristensenetal2012}, who successfully reproduced such line profiles with an adaptation of the \citet{Myersetal1996} model, incorporating outflow emission partially obscured by an expanding layer of less excited gas.}

\changesone{Our previous analysis of \meth\ lines in a subset of the present data revealed that the \compa\ component is dominated at high upper level energies by a compact, hot component, which we identified with a hot core \citep{Kamaetal2010}. \resubmit{The rotational diagram results for other species in Table~\ref{tab:rotdiag} provide further evidence for the importance of an underlying hot component.}}

\compB. \resubmit{The broad wings are seen most clearly in OH, \htwoo\ and CO, and likely trace at least one outflow. The existence of a broad component in the CO~$3-2$ line was noted already by \citet{Johnstoneetal2003}. An interpretation of this component is made difficult by the projected overlap of \oursource\ and one lobe of an outflow from the nearby source, FIR~3 \citep{Shimajirietal2008}. Given the prominence and symmetry of the wings in the high rotational lines of CO, and the broadness of the HCN and CS lines -- all indicative hot or dense emitting material -- the wing emission may originate in a compact outflow from FIR~4 itself \citep{Kamaetal2013inprep}.}

\resubmit{Broad lines of OH correlate well with water and are associated with outflow shocks \citep{Wampfleretal2010} and in \oursource, the large width of the OH lines (FWHM$~=~19.1$~km/s) is consistent with an outflow shock origin. Furthermore, our modeling finds the OH lines to be optically thin, so together with the highest-$J$ CO lines, OH may be the most straightforward tracer of this outflow.}

\resubmit{In Fig.~\ref{fig:velowidth}, \nhthree\ corresponds to the outflow tracers CO, \htwoo\ and OH in terms of \vlsr, but has a typical linewidth a factor of four smaller. In the L1157-B1 outflow context, similar observations have been explained with a model where \nhthree\ is destroyed in a shock at velocities $>15$~km/s, while \htwoo, for example, maintains its high abundance \citep{Vitietal2011}. Other species, such as \meth\ and \form, should also show emission at the outflow velocity, and a weak wing seems to indeed be present.}

\compC. \resubmit{This component only appears clearly in lines of SO, although other species such as CO and CS do have a blue wing component that may be related. Its nature is unclear.} \retwo{It may be related to the compact blueshifted spot seen in low-frequency interferometry of SiO and other species by \citet{Shimajirietal2008}. Indeed, the SiO lines in our follow-up survey with the IRAM 30~m telescope also show a blueshifted emission component.}

\compD. \resubmit{The slab component, at \vlsr$~\sim~9.5$~km/s, is traced almost exclusively by absorption lines of molecular ions associated with photon-dominated regions (PDRs), such as \ohplus, \htwooplus\ and \htwoclplus. The CII emission peak also corresponds to this component, as does part of the absorption in the fundamental line of HF. In addition to the set of species tracing the component, the fact that absorption is seen in low-lying lines suggests a very tenuous medium. In a companion paper \citep{LopezSepulcreetal2013}, we present a detailed analysis of this component, finding it to be a low-density and low-extinction ($A_{\textrm{v}}~\approx~1$~mag) PDR cloud in front of OMC-2, irradiated on one side by a heavily enhanced FUV field.}

\begin{figure}[!h]
  \centering
  \includegraphics[clip=,width=1.0\columnwidth]{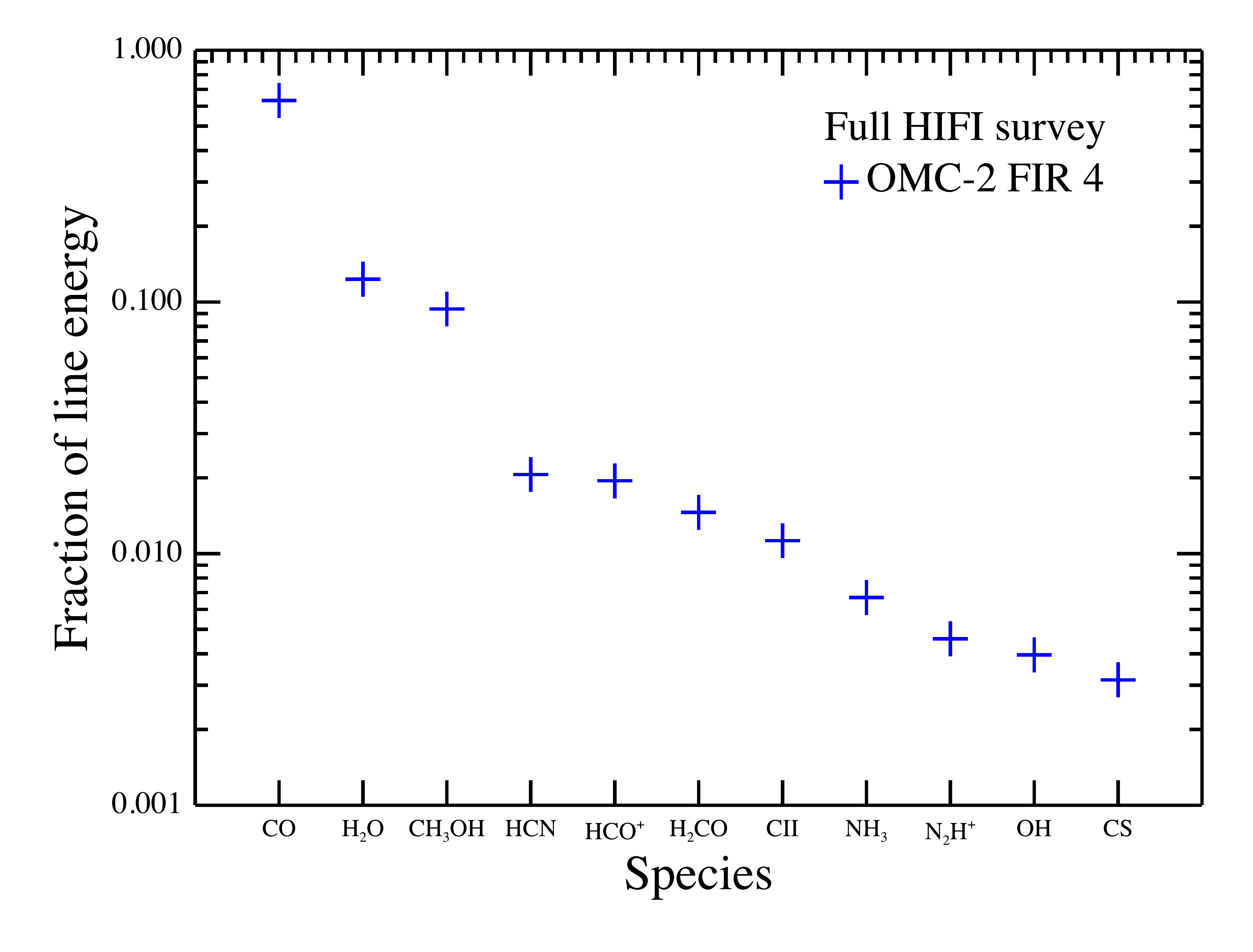}
     \caption{\changesone{Fractional contribution of various species to the line emission from \oursource. The 11 dominant cooling species, integrated across the entire 480 to 1902~GHz HIFI survey, are displayed. Note that the units are in percent of the total line flux in the HIFI survey, excluding continuum emission.}}
        \label{fig:coolingOMC2}
\end{figure}

\compE. \resubmit{Aside from the above four components, there is variation in line profiles between different species, upper level energies and observing frequencies. This includes the evolution of \ntwohplus\ and especially \hcoplus\ line profiles with increasing $J$ level (Fig.~\ref{fig:ioncomparison}) and the plateau-and-peak profile of the fundamental ortho-\nhthree\ line (Fig.~\ref{fig:nitrogencomparison}). Many of these line profile aspects may be subcomponents of the \compa\ and other aforementioned categories. A full interpretation of this variety of features, while important, is outside the scope of this paper.} \retwo{Pending further analyses of the HIFI data, we refer the reader to previous \citep{Shimajirietal2008} and upcoming \citep[in press]{LopezSepulcreetalInterferometry} papers for interferometric results on the small-scale structure of \oursource.}

\subsection{Line and continuum cooling}\label{sec:cooling}

We investigated the role of molecular line emission in gas cooling by summing the integrated intensities of the detected transitions of each species. \changesone{The results are given in the second-to-last column of Table~\ref{tab:species}, and in Fig.~\ref{fig:coolingOMC2}, we show the 11 dominant cooling molecules.} Within the frequency coverage of the survey, CO is the dominant molecular coolant, emitting \coolco\% of the total energy in the detected lines. The second most important is \htwoo\ with \coolwater\% and the third is \meth\ with \coolmeth\%. Two other notable coolants are \hcoplus\ and HCN, both contributing 2\% \changesone{of the total line flux}.

\retwo{Due to contamination in the reference positions, the total flux in CO lines is likely underestimated by a few tens of percent, as estimated from the lack of contamination in $J_{\textrm{u}}~>~11$ and a multicomponent analysis of the lower-$J$ lines. Thus, the true CO to total line flux ratio may be as high as $\sim70\ldots80$\%.}

To measure the relative importance of line and continuum cooling, we integrated the spectra with baselines intact in the fully frequency sampled region of the survey (bands 1a through 5a or 480 to 1250~GHz), obtaining a flux of $\rm 1.3\cdot 10^{-12}~W\cdot m^{-2}$. Of this, $\rm 2.7\cdot 10^{-14}~W\cdot m^{-2}$ or 2\% is in lines. In their 300~GHz range study of five low- to intermediate-luminosity protostars including \oursource, \citet{Johnstoneetal2003} found lines to contribute $<8$\% of the measured continuum in their entire sample, consistent with our terahertz-range result. Assuming a distance of 420~pc, the flux we measure with HIFI between 480 and 1250~GHz corresponds to a luminosity of $\sim7$~\lsol, of which $0.1$~\lsol\ is line emission.

It is remarkable that the sulphur oxides, SO and \sotwo, contribute negligibly to the line cooling in \oursource, in striking contrast to Orion~KL, as discussed in Sect.~\ref{sec:oricomp}.

\begin{figure}[!h]
  \centering
  \includegraphics[clip=,width=1.0\columnwidth]{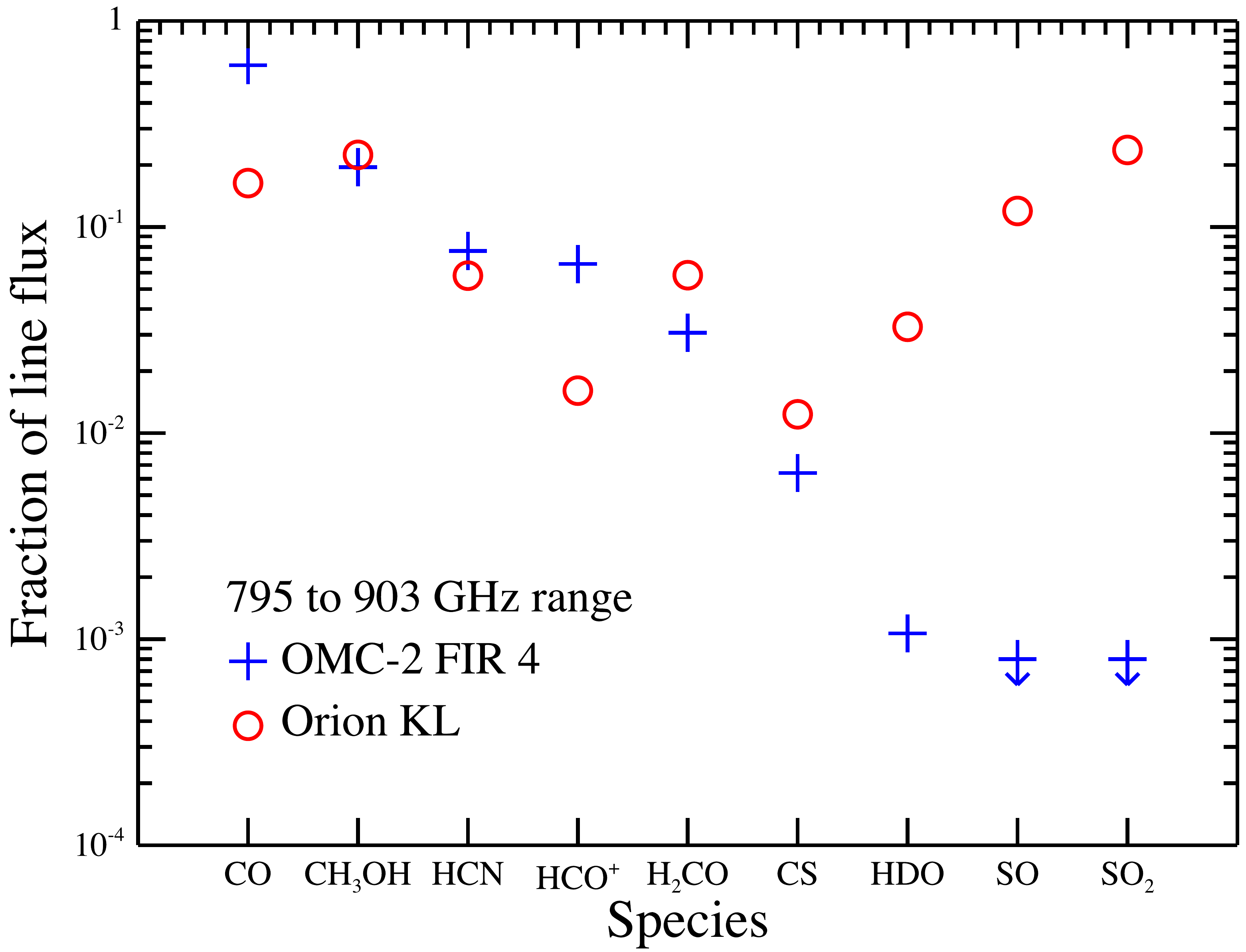}
     \caption{Line cooling in the 795 -- 903~GHz range in two sources in Orion. The symbols mark the percentage of line emission in each dominant cooling species in \oursource\ (crosses, this work) and in Orion~KL \citep[circles,][]{Comitoetal2005}. The upper limits are $5\sigma$. Note that the units are in percent of all line flux within the 795~to~903~GHz range, excluding continuum emission.}
        \label{fig:cooling}
\end{figure}

\resubmit{As the frequency coverage above 1250~GHz is not complete, the cooling contributions of some species in the HIFI range may differ from those given in Table~\ref{tab:species}, but for most species the total line fluxes should not differ much from their total 480 to 1902~GHz fluxes. Undetected weak lines contributing to the continuum may also introduce a small correction to the quoted numbers. The total line emission from \oursource\ may have a large contribution from CO, \htwoo\ and OI lines outside the high end of the HIFI frequency range, toward the mid- and near-infrared.} \retwo{A study of the spectrally unresolved CO lines seen toward \oursource\ with \emph{Herschel}/PACS by \citet{Manojetal2013} determined the CO luminosity between $J_{\textrm{u}}~=~14$ and $46$ to be $0.3$~\lsol. A comparison of this with our HIFI CO line luminosity of $0.1$~\lsol\ shows that transitions at frequencies within the HIFI range contribute roughly the same total flux as those at higher frequencies. As the HIFI and PACS ranges overlap, the total CO luminosity from this region cannot be much above $0.4$~\lsol\rethree{, which is between $0.04$\% and $0.8$\% of the total luminosity, depending on the source luminosity estimate ($50$ to $1000$~\lsol, as discussed in Sect.~\ref{sec:intro})}.}

\subsubsection{\resubmit{Comparison with other sources}}\label{sec:oricomp}

\resubmit{As a detailed comparison with other sources is outside the scope of this paper, we provide here an initial view.}

In Fig.~\ref{fig:cooling}, we compare the relative line cooling in the 795 through 903~GHz range in \oursource\ and Orion KL, a well-studied strong line and continuum emitter. The numbers here are not to be confused with the results for the entire survey given above\retwo{, furthermore the units here are K$\cdot$km/s, while the full-survey fluxes were compared on the W$\cdot$m$^{-2}$ scale}. The selected species contain the seven most important coolants for either \resubmit{source}. \htwoo\ is not considered in this comparison as it has no lines in the 795 to 903~GHz range. In \oursource\ in this range, CO emission contains 61\% of the line energy, followed by \meth\ with 20\%, HCN with 8\% and \hcoplus\ with 7\%. In Orion~KL, \sotwo\ dominates with 24\% of the energy, followed by \meth\ with 22\%, CO with 16\% and SO with 12\%.

\resubmit{While the sulphur oxides contribute at the $10\ldots20$\% level to line cooling in Orion~KL, in \oursource\ both contribute $\leq 0.1$\% in the 795 to 903~GHz range, a difference of at least two orders of magnitude. This may be due to an exceptional contribution of energetic shocks in Orion~KL. A similar scarcity of sulphur oxides in comparison to Orion~KL has been reported before in samples of low- as well as high-mass protostars \citep{Johnstoneetal2003, Schilkeetal2006}. In addition to shocks, factors such as the thermal evolution of the gas and dust may also play an important role in abundance variations of sulphur-bearing molecules \citep{Wakelametal2004, Wakelametal2005}. The difference of peak velocity we observe between SO and \meth, $\sim2$~km/s, suggests that SO emission in \oursource\ is dominated by a spatially distinct region, a result seen also in Sagittarius~B2 \citep{Nummelinetal2000}.}

The nature and heating processes of Orion~KL are currently under debate in the community and our comparison once again confirms the exceptional nature of this source. \changesone{It must be kept in mind that the quoted Orion~KL observations were carried out with the Caltech Submillimeter Observatory, with a beam roughly a third the size of that of \herschel\ at equivalent frequencies, therefore beam dilution effects may be important and the above comparison should be repeated with the HIFI survey of that source \citep{Crockettetalinprep}. \retwo{The accelerating publication of HIFI spectral surveys and line observations of a number of other protostars \citep[e.g.][see also other papers from the CHESS, HEXOS, and WISH key programs]{Zernickeletal2012, Neilletal2012, Cauxetalinprep} will soon allow more, and more detailed, comparative analyses to be made, across molecular species as well as protostellar properties.}} For example, the number of species found in the HIFI spectral survey of \oursource, \numberofmolecules\ excluding isotopologs, is smaller than that found in similar data for the high-mass star forming regions NGC~6334I \citep[46 species,][]{Zernickeletal2012} and Sagittarius~B2(N) \citep[$\geq40$ species,][]{Neilletal2012}. This may be related to the substantially lower luminosity of \oursource\ compared to the more massive sources.

\section{Conclusions}\label{sec:conclusions}

\begin{itemize}

\item{We present a \herhifi\ spectral survey of \oursource\ in the range 480 to 1901~GHz, one of the first spectral surveys of a protostar in the terahertz regime.}

\item{We find \numberoflines\ lines in the survey, originating from \numberofmolecules\ different molecular and atomic species and tracing a large range of excitation conditions, with $24~\leq~E_{\textrm{u}}~\leq~752$~K.}

\item{The line profiles have contributions from the large-scale envelope of \oursource, at least one outflow, a newly discovered foreground PDR and other components. The broad outflow emission has a redshifted offset of 1.5~km/s from the source velocity. Narrow, blue-shifted self-absorption on broad emission lines of \htwoo, \nhthree, and possibly \hcoplus, may originate in an expanding layer in the inner envelope. Broad, blue-shifted SO lines trace a new component of unclear nature.}

\item{The cooling budget of \oursource\ between 480 and 1250~GHz is dominated by continuum radiation, with lines contributing 2\%. The total flux received in this range is $\rm 1.3\cdot 10^{-12}~W\cdot m^{-2}$ or $\sim7$~\lsol\ at 420~pc.}

\item{Of all the detected line flux in W$\cdot$m$^{-2}$, \coolco\% is from $^{12}$CO, \coolwater\% from \htwoo\ and \coolmeth\% from \meth. Every other species contributes at the $\leq 2$\% level.}

\item{The dominant cooling molecules in \oursource\ and Orion~KL are similar, but the relative role of SO and \sotwo\ in the energy budget of \oursource\ is two orders of magnitude smaller than in Orion~KL, indicating a substantial difference in the role of shocks or in the thermal evolution.}

\item{In terms of composition and dominant chemical species, \oursource\ is well in line with results for other protostars from ground-based instruments. It is thus a nearby intermediate-mass star formation laboratory for which an exceptional spectral dataset is now available.}

\end{itemize}

\acknowledgements{\changesone{
The authors would like to thank Charlotte Vastel for help with the spectroscopic data; our CHESS, HEXOS and WISH colleagues for useful discussions; and the HIFI Instrument Control Center and \herschel\ Science Center teams for their efforts. M.K. gratefully acknowledges funding from the Netherlands Organisation for Scientific Research (NWO) Toptalent grant number 021.002.081, the Leids Kerkhoven-Bosscha Fonds and the COST Action on Astrochemistry. A.L.S. and C.C. acknowledge funding by the French Space Agency CNES and from ANR contract ANR-08-BLAN-022. A.F. acknowledges support from the CONSOLIDER INGENIO 2010 program, grant CSD2009-00038. \herschel\ is an ESA space observatory with science instruments provided by European-led Principal Investigator consortia and with important participation from NASA. This work is based on analysis carried out with the CASSIS software, developed by IRAP-UPS/CNRS (\texttt{http://cassis.irap.omp.eu}).
}}

\bibliographystyle{aa}
\bibliography{fir4}

\appendix

\section{Detected transitions}\label{apx:detections}

\resubmit{In Table~A1, we present a frequency-sorted table of the transitions detected in the survey. \retwo{In the first six columns, we give the database properties of each transition. Columns 7-10 give the Gaussian fit velocity and width, and their associated formal uncertainties. Columns 11-14 give the peak and integrated intensity, and associated uncertainty. Blending is indicated in column 15, by a \texttt{B:} followed by an identification of the blended line (only one \texttt{B:} line is given also for multiple overlapping blends). Line profile parameters (\vlsr, FWHM, $T_{\textrm{mb}}$) are not given for the lines which are blended.} For some species, such as OH and \shplus, the line parameters were determined by fitting models of the hyperfine line profile to the data, these results can be seen in Table~\ref{tab:species} and Fig.~\ref{fig:velowidth} \retwo{and plotted with red dashed lines in figures showing examples of the data.}}



\section*{Supplementary material}

Table A1: Transitions detected in the HIFI spectral survey of OMC-2 FIR 4.

| ν GHz (1) | Species (2) | Transition (3) | ν_rest GHz (4) | E_u K (5) | A_ul rad/s (6) | *Gaussian fit* v_lsr km/s (7) | δv_lsr km/s (8) | FWHM km/s (9) | δFWHM km/s (10) | T_mb K (11) | *from moments* RMS mK (12) | Flux K km/s (13) | δFlux K km/s (14) | Blend(B) / Notes (15) |
|---|---|---|---|---|---|---|---|---|---|---|---|---|---|---|
| 480.269 | $CH_3OH$ | $3_2 - 3_1$ | 480.269 | 51.64 | 4.89e-04 | 12.26 | 0.18 | 5.20 | 0.42 | 0.17 | 22 | 1.01 | 0.04 | |
| 481.504 | $CH_3OH$ | $10_0 - 9_0$ | 481.504 | 44.67 | 3.94e-04 | 11.36 | 0.14 | 3.64 | 0.34 | 0.14 | 23 | 0.42 | 0.04 | |
| 481.916 | $C^{34}S$ | $10_0 - 9_0$ | 481.916 | 127.23 | 2.38e-03 | 10.01 | 0.10 | 1.68 | 0.24 | 0.05 | 19 | 0.21 | 0.03 | |
| 482.282 | $CH_3OH$ | $10_0 - 9_0$ | 482.282 | 140.60 | 5.02e-04 | 11.80 | 0.06 | 3.61 | 0.13 | 0.37 | 16 | 1.93 | 0.03 | |
| 482.959 | $CH_3OH$ | $10_{-1} - 9_{-1}$ | 482.959 | 133.15 | 5.00e-04 | 11.82 | 0.02 | 3.82 | 0.04 | 0.64 | 16 | 3.09 | 0.03 | |
| 483.141 | $CH_3OH$ | $10_0 - 9_0$ | 483.141 | 127.60 | 5.05e-04 | 11.73 | 0.02 | 3.72 | 0.04 | 0.68 | 15 | 3.48 | 0.03 | |
| 483.389 | $CH_3OH$ | $10_2 - 9_2$ | 483.389 | 165.35 | 4.89e-04 | 12.22 | 0.12 | 5.07 | 0.27 | 0.12 | 20 | 0.81 | 0.04 | |
| 483.472 | $CH_3OH$ | $10_{-4} - 9_{-4}$ | 483.472 | 215.55 | 6.30e-04 | 12.65 | 0.25 | 5.21 | 0.70 | 0.04 | 15 | 0.18 | 0.02 | B: $CH_3OH$ ($10_4 - 9_4$) |
| 483.539 | $CH_3OH$ | $10_4 - 9_4$ | 483.539 | 207.99 | 6.31e-04 | | | | | | | | | B: $CH_3OH$ ($10_4 - 9_4$) |
| 483.539 | $CH_3OH$ | $10_4 - 9_4$ | 483.539 | 207.99 | 6.31e-04 | | | | | | | | | B: $CH_3OH$ ($10_4 - 9_4$) |
| 483.551 | $CH_3OH$ | $10_3 - 9_3$ | 483.551 | 177.46 | 6.82e-04 | | | | | | | | | B: $CH_3OH$ ($10_4 - 9_4$) |
| 483.556 | $CH_3OH$ | $10_3 - 9_3$ | 483.556 | 190.37 | 6.87e-04 | | | | | | | | | B: $CH_3OH$ ($10_4 - 9_4$) |
| 483.566 | $CH_3OH$ | $10_{-3} - 9_{-3}$ | 483.566 | 177.46 | 6.82e-04 | | | | | | | | | B: $CH_3OH$ ($10_4 - 9_3$) |
| 483.686 | $CH_3OH$ | $10_1 - 9_1$ | 483.686 | 148.73 | 7.60e-04 | | | | | | | | | B: $CH_3OH$ ($10_1 - 9_1$) |
| 483.697 | $CH_3OH$ | $10_1 - 9_1$ | 483.697 | 175.39 | 6.85e-04 | | | | | | | | | B: $CH_3OH$ ($10_1 - 9_1$) |
| 483.761 | $CH_3OH$ | $10_2 - 9_2$ | 483.761 | 165.40 | 4.90e-04 | 12.16 | 0.11 | 4.09 | 0.26 | 0.11 | 16 | 0.51 | 0.02 | |
| 484.005 | $CH_3OH$ | $2_2 - 2_1$ | 484.005 | 44.67 | 4.01e-04 | | | | | | | | | B: $CH_3OH$ ($10_2 - 9_2$) |
| 484.023 | $CH_3OH$ | $10_2 - 9_2$ | 484.023 | 149.97 | 4.83e-04 | | | | | | | | | B: $CH_3OH$ ($2_2 - 2_1$) |
| 484.072 | $CH_3OH$ | $10_{-2} - 9_{-2}$ | 484.072 | 153.63 | 4.89e-04 | 11.92 | 0.06 | 4.25 | 0.14 | 0.18 | 14 | 0.94 | 0.02 | |
| 485.263 | $CH_3OH$ | $3_2 - 3_1$ | 485.263 | 51.64 | 5.05e-04 | 11.74 | 0.03 | 4.33 | 0.06 | 0.20 | 16 | 1.10 | 0.03 | |
| 485.418 | $H_2Cl^+$ | $1_{1,1,5/2} - 0_{0,0,3/2}$ | 485.418 | 23.30 | 1.59e-03 | 8.26 | 0.16 | 2.82 | 0.38 | -0.03 | 11 | -0.18 | 0.01 | |
| 486.941 | $CH_3OH$ | $4_2 - 4_1$ | 486.941 | 60.92 | 5.51e-04 | 11.57 | 0.03 | 4.40 | 0.06 | 0.24 | 16 | 1.27 | 0.03 | |
| 487.532 | $CH_3OH$ | $5_2 - 5_1$ | 487.532 | 72.53 | 5.15e-04 | 12.02 | 0.02 | 3.84 | 0.05 | 0.38 | 17 | 2.00 | 0.03 | |
| 489.037 | $CH_3OH$ | $6_2 - 6_1$ | 489.037 | 72.53 | 5.79e-04 | 11.90 | 0.03 | 3.88 | 0.08 | 0.23 | 17 | 1.27 | 0.04 | |
| 489.751 | CS | $10 - 9$ | 489.751 | 129.29 | 2.51e-03 | 11.85 | 0.09 | 8.28 | 0.21 | 0.51 | 17 | 4.34 | 0.04 | |
| 491.551 | $CH_3OH$ | $7_2 - 7_1$ | 491.551 | 86.46 | 6.00e-04 | 11.88 | 0.06 | 4.65 | 0.15 | 0.21 | 18 | 1.11 | 0.04 | |
| 491.935 | $SO_2$ | $7_{4,4} - 6_{3,3}$ | 491.935 | 65.01 | 9.49e-04 | | | | | | 17 | | | B: $H_2CO$ ($7_{1,7} - 6_{1,6}$) |
| 491.968 | $H_2CO$ | $7_{1,7} - 6_{1,6}$ | 491.968 | 106.31 | 3.44e-03 | | | | | | 18 | | | B: $SO_2$ ($7_{4,4} - 6_{3,3}$) |
| 492.161 | C | $1 - 0$ | 492.161 | 23.62 | 7.99e-08 | 11.90 | 0.02 | 1.81 | 0.05 | 1.30 | 18 | 4.74 | 0.04 | |
| 492.279 | $CH_3OH$ | $4_1 - 3_0$ | 492.279 | 37.55 | 3.83e-04 | 11.69 | 0.02 | 4.15 | 0.04 | 0.66 | 17 | 3.47 | 0.04 | |
| 493.699 | $CH_3OH$ | $5_3 - 4_2$ | 493.699 | 84.62 | 6.62e-04 | 11.67 | 0.02 | 3.72 | 0.05 | 0.49 | 14 | 2.44 | 0.03 | |
| 493.734 | $CH_3OH$ | $5_3 - 4_2$ | 493.734 | 84.62 | 6.62e-04 | 11.59 | 0.02 | 3.71 | 0.05 | 0.50 | 15 | 2.56 | 0.03 | |
| 494.455 | $NH_2D$ | $1_{1,0,1} - 0_{0,0,0}$ | 494.455 | 23.73 | 9.95e-04 | 11.19 | 0.03 | 2.21 | 0.08 | 0.14 | 14 | 0.42 | 0.02 | |
| 494.482 | $CH_3OH$ | $7_2 - 7_1$ | 494.482 | 102.70 | 6.18e-04 | 11.79 | 0.02 | 5.06 | 0.06 | 0.21 | 14 | 1.30 | 0.03 | |
| 495.173 | $CH_3OH$ | $7_0 - 6_1$ | 495.173 | 78.08 | 3.21e-04 | 11.71 | 0.02 | 3.68 | 0.05 | 0.40 | 13 | 1.94 | 0.03 | |
| 496.924 | $CH_3OH$ | $14_0 - 13_1$ | 496.924 | 256.14 | 3.26e-04 | 12.34 | 0.14 | 4.62 | 0.32 | 0.08 | 13 | 0.35 | 0.03 | |
| 497.828 | $CH_3OH$ | $8_2 - 8_1$ | 497.828 | 121.27 | 6.36e-04 | 11.90 | 0.04 | 4.61 | 0.08 | 0.20 | 14 | 1.29 | 0.03 | |
| 501.589 | $CH_3OH$ | $9_2 - 9_1$ | 501.589 | 142.15 | 6.54e-04 | 11.93 | 0.07 | 3.98 | 0.15 | 0.16 | 14 | 0.87 | 0.03 | |
| 504.294 | $CH_3OH$ | $7_1 - 6_0$ | 504.294 | 86.05 | 3.53e-04 | 11.72 | 0.02 | 4.01 | 0.06 | 0.39 | 15 | 2.09 | 0.03 | |
| 505.565 | $H_2S$ | $2_{2,1} - 2_{1,2}$ | 505.565 | 79.37 | 2.40e-04 | 11.29 | 0.04 | 3.83 | 0.10 | 0.11 | 16 | 0.48 | 0.03 | |
| 505.762 | $CH_3OH$ | $10_2 - 10_1$ | 505.762 | 165.35 | 6.73e-04 | 11.95 | 0.03 | 4.16 | 0.07 | 0.14 | 18 | 0.62 | 0.03 | |
| 505.834 | $H_2CO$ | $7_{0,7} - 6_{0,6}$ | 505.834 | 97.44 | 3.82e-03 | 11.75 | 0.02 | 4.42 | 0.04 | 0.74 | 18 | 4.03 | 0.04 | |
| 506.153 | $CH_3OH$ | $11_1 - 10_2$ | 506.153 | 174.27 | 2.34e-04 | 11.05 | 0.14 | 4.16 | 0.32 | 0.09 | 18 | 0.32 | 0.03 | |
| 506.825 | DCN | $7 - 6$ | 506.825 | 97.30 | 6.33e-03 | 12.04 | 0.37 | 6.78 | 0.87 | 0.04 | 15 | 0.24 | 0.03 | |

Columns (7)–(11) are grouped under **Gaussian fit**; columns (12)–(14) are grouped under **from moments**.

| (1) ν GHz | (2) Species | (3) Transition | (4) ν_rest GHz | (5) E_u K | (6) A_ul rad/s | (7) v_lsr km/s | (8) δv_lsr km/s | (9) FWHM km/s | (10) δFWHM km/s | (11) T_mb K | (12) RMS mK | (13) Flux K km/s | (14) δFlux K km/s | (15) Blend(B) / Notes |
|---|---|---|---|---|---|---|---|---|---|---|---|---|---|---|
| 509.146 | $H_2CO$ | $7_{2,6} - 6_{2,5}$ | 509.146 | 144.93 | 3.58e-03 | 11.70 | 0.02 | 4.51 | 0.05 | 0.39 | 16 | 2.28 | 0.03 | B: $H_2CO$ $(7_{6,1} - 6_{6,0})$ |
| 509.292 | HDO | $1_{1,0} - 1_{0,1}$ | 509.292 | 46.76 | 2.32e-03 | | | | | | 16 | | | B: HDO $(1_{1,0} - 1_{0,1})$ |
| 509.306 | $H_2CO$ | $7_{6,1} - 6_{6,0}$ | 509.306 | 520.81 | 1.03e-03 | | | | | | 17 | | | B: HDO $(1_{1,0} - 1_{0,1})$ |
| 509.306 | $H_2CO$ | $7_{6,0} - 6_{6,1}$ | 509.306 | 520.81 | 1.03e-03 | | | | | | 16 | | | B: $H_2CO$ $(7_{5,2} - 6_{5,1})$ |
| 509.562 | $H_2CO$ | $7_{5,2} - 6_{5,1}$ | 509.562 | 391.84 | 1.91e-03 | | | | | | 17 | | | B: $H_2CO$ $(7_{5,3} - 6_{5,2})$ |
| 509.562 | $H_2CO$ | $7_{5,3} - 6_{5,2}$ | 509.562 | 391.84 | 1.91e-03 | | | | | | 17 | | | B: $H_2CO$ $(7_{5,2} - 6_{5,1})$ |
| 509.565 | $CH_3OH$ | $10_2 - 9_1$ | 509.565 | 149.97 | 4.07e-04 | | | | | | 19 | | | B: $H_2CO$ $(7_{5,3} - 6_{5,2})$ |
| 509.830 | $H_2CO$ | $7_{4,3} - 6_{4,2}$ | 509.830 | 286.17 | 2.63e-03 | | | | | | 19 | | | B: $H_2CO$ $(7_{4,2} - 6_{4,2})$ |
| 509.830 | $H_2CO$ | $7_{4,4} - 6_{4,3}$ | 509.830 | 286.17 | 2.63e-03 | | | | | | 19 | | | B: $H_2CO$ $(7_{4,3} - 6_{4,3})$ |
| 510.156 | $H_2CO$ | $7_{4,4} - 6_{4,3}$ | 510.156 | 203.89 | 3.20e-03 | 11.88 | 0.02 | 4.08 | 0.04 | 0.55 | 16 | 2.89 | 0.03 | |
| 510.238 | $H_2CO$ | $7_{3,5} - 6_{3,4}$ | 510.238 | 203.90 | 3.20e-03 | 11.93 | 0.02 | 4.19 | 0.04 | 0.54 | 16 | 2.77 | 0.03 | |
| 510.345 | $CH_3OH$ | $7_{3,4} - 6_{3,3}$ | 510.345 | 190.86 | 6.94e-04 | 12.02 | 0.10 | 4.28 | 0.24 | 0.10 | 15 | 0.44 | 0.03 | |
| 512.427 | $NH_2D$ | $2_{0,2,1} - 1_{1,0,0}$ | 512.427 | 47.75 | 1.57e-04 | 11.49 | 0.09 | 3.00 | 0.22 | 0.07 | 17 | 0.19 | 0.03 | |
| 513.076 | $H_2CO$ | $7_{2,5} - 6_{2,4}$ | 513.076 | 145.35 | 3.66e-03 | 11.92 | 0.08 | 4.40 | 0.05 | 0.39 | 17 | 2.26 | 0.03 | |
| 514.853 | SO | $12_{21} - 11_{10}$ | 514.853 | 167.59 | 1.82e-03 | 9.30 | 0.08 | 12.17 | 0.19 | 0.06 | 16 | 0.68 | 0.04 | |
| 515.170 | $CH_3OH$ | $16_0 - 15_1$ | 515.170 | 315.21 | 4.10e-04 | 12.05 | 0.08 | 4.31 | 0.18 | 0.22 | 17 | 1.08 | 0.04 | |
| 515.335 | $CH_3OH$ | $12_2 - 12_1$ | 515.335 | 218.69 | 7.16e-04 | 12.57 | 0.15 | 5.69 | 0.34 | 0.09 | 15 | 0.53 | 0.03 | |
| 516.335 | SO | $12_{12} - 11_{11}$ | 516.335 | 174.22 | 1.83e-03 | 9.32 | 0.06 | 7.38 | 0.15 | 0.06 | 15 | 0.45 | 0.03 | |
| 517.354 | SO | $12_{13} - 11_{12}$ | 517.354 | 165.78 | 1.86e-03 | 9.05 | 0.06 | 12.09 | 0.14 | 0.07 | 17 | 0.73 | 0.04 | |
| 517.970 | $H^{13}CN$ | $6 - 5$ | 517.970 | 87.01 | 6.65e-03 | 12.14 | 0.05 | 8.87 | 0.11 | 0.11 | 16 | 0.96 | 0.03 | |
| 519.188 | $SO_2$ | $29_{2,29} - 28_{0,28}$ | 519.188 | 379.20 | 1.91e-03 | 11.50 | 0.42 | 12.03 | 1.00 | 0.02 | 16 | 0.27 | 0.03 | |
| 520.179 | $CH_3OH$ | $2_2 - 1_1$ | 520.179 | 32.86 | 9.76e-04 | 11.78 | 0.05 | 4.78 | 0.11 | 0.43 | 17 | 2.39 | 0.03 | |
| 520.460 | $H^{13}CO^+$ | $6 - 5$ | 520.460 | 87.43 | 1.14e-02 | 11.38 | 0.04 | 2.42 | 0.09 | 0.12 | 16 | 0.45 | 0.03 | |
| 520.728 | $CH_3OH$ | $13_2 - 13_1$ | 520.728 | 248.84 | 7.40e-04 | 11.73 | 0.11 | 2.85 | 0.31 | 0.08 | 18 | 0.49 | 0.03 | |
| 522.077 | ND | $1_{2,3,4} - 0_{1,2,3}$ | 522.077 | 25.06 | 1.07e-03 | 11.17 | 0.11 | 10.67 | 0.26 | 0.03 | 16 | 0.27 | 0.02 | |
| 523.274 | $CH_3OH$ | $14_{-1} - 13_0$ | 523.274 | 248.93 | 6.27e-04 | 11.91 | 0.02 | 3.70 | 0.04 | 0.35 | 17 | 1.63 | 0.04 | |
| 523.972 | CCH | $6_{13/2,7} - 5_{11/2,6}$ | 523.972 | 88.02 | 4.58e-04 | 13.46 | 0.13 | 2.91 | 0.30 | 0.03 | 16 | 0.13 | 0.02 | B: CCH $(6_{13/2,6} - 5_{11/2,5})$ |
| 523.972 | CCH | $6_{13/2,6} - 5_{11/2,5}$ | 523.972 | 88.02 | 4.53e-04 | | | | | | 17 | | | B: CCH $(6_{13/2,7} - 5_{11/2,6})$ |
| 524.003 | $CH_3OH$ | $15_{-4} - 15_{-3}$ | 524.003 | 366.32 | 1.06e-03 | | | | | | 18 | | | |
| 524.034 | CCH | $6_{11/2,6} - 5_{9/2,5}$ | 524.034 | 88.04 | 4.51e-04 | | | | | | 18 | | | B: CCH $(6_{11/2,5} - 5_{9/2,4})$ |
| 524.035 | CCH | $6_{11/2,5} - 5_{9/2,4}$ | 524.035 | 88.04 | 4.43e-04 | | | | | | 18 | | | B: CCH $(6_{11/2,6} - 5_{9/2,5})$ |
| 524.269 | $CH_3OH$ | $13_{-4} - 13_{-3}$ | 524.269 | 299.06 | 7.10e-04 | 11.27 | 0.21 | 3.18 | 0.50 | 0.04 | 16 | 0.27 | 0.03 | |
| 524.385 | $CH_3OH$ | $12_{-4} - 12_{-3}$ | 524.385 | 268.91 | 7.03e-04 | 11.75 | 0.56 | 9.29 | 1.32 | 0.04 | 17 | 0.30 | 0.03 | |
| 524.489 | $CH_3OH$ | $11_{-4} - 11_{-3}$ | 524.489 | 241.07 | 6.94e-04 | 12.58 | 0.23 | 5.52 | 0.53 | 0.05 | 17 | 0.29 | 0.03 | |
| 524.666 | $CH_3OH$ | $9_{-4} - 9_{-3}$ | 524.666 | 192.34 | 6.64e-04 | 11.60 | 0.12 | 3.35 | 0.29 | 0.06 | 16 | 0.25 | 0.02 | |
| 524.740 | $CH_3OH$ | $8_{-4} - 8_{-3}$ | 524.740 | 171.46 | 6.39e-04 | 12.11 | 0.08 | 5.90 | 0.13 | 0.06 | 16 | 0.46 | 0.03 | |
| 524.805 | $CH_3OH$ | $7_{-4} - 7_{-3}$ | 524.805 | 152.90 | 6.03e-04 | 11.85 | 0.06 | 4.81 | 0.10 | 0.08 | 17 | 0.49 | 0.02 | |
| 524.861 | $CH_3OH$ | $6_{-4} - 6_{-3}$ | 524.861 | 136.65 | 5.49e-04 | 11.79 | 0.13 | 3.96 | 0.32 | 0.08 | 16 | 0.56 | 0.02 | |
| 524.908 | $CH_3OH$ | $5_{-4} - 5_{-3}$ | 524.908 | 122.72 | 4.62e-04 | 12.01 | 0.16 | 3.85 | 0.38 | 0.08 | 15 | 0.40 | 0.02 | |
| 524.947 | $CH_3OH$ | $4_{-4} - 4_{-3}$ | 524.947 | 111.12 | 3.08e-04 | 11.81 | 0.14 | 4.16 | 0.33 | 0.06 | 15 | 0.24 | 0.02 | |
| 525.666 | $H_2CO$ | $7_{1,6} - 6_{1,5}$ | 525.666 | 112.79 | 4.20e-03 | 11.82 | 0.01 | 4.92 | 0.03 | 1.22 | 16 | 7.14 | 0.03 | |
| 526.039 | $SH^+$ | $1_{2,3/2} - 0_{1,1/2}$ | 526.039 | 25.25 | 7.99e-04 | | | | | | 16 | | | B: $SH^+$ $(1_{2,5/2} - 0_{1,3/2})$ |
| 526.048 | $SH^+$ | $1_{2,5/2} - 0_{1,3/2}$ | 526.048 | 25.25 | 9.59e-04 | | | | | | 16 | | | B: $SH^+$ $(1_{2,3/2} - 0_{1,1/2})$ |
| 526.520 | $CH_3OH$ | $14_2 - 14_1$ | 526.520 | 281.29 | 7.66e-04 | 12.51 | 0.07 | 5.67 | 0.16 | 0.06 | 13 | 0.37 | 0.02 | |
| 527.053 | $CH_3OH$ | $11_1 - 10_1$ | 527.053 | 166.37 | 6.54e-04 | 11.97 | 0.04 | 3.86 | 0.10 | 0.31 | 13 | 1.41 | 0.03 | |

| ν GHz (1) | Species (2) | Transition (3) | ν_rest GHz (4) | E_u K (5) | A_ul rad/s (6) | Gaussian fit | | | | | from moments | | | Blend(**B**) / Notes |
|---|---|---|---|---|---|---|---|---|---|---|---|---|---|---|
| | | | | | | v_lsr km/s (7) | δv_lsr km/s (8) | FWHM km/s (9) | δFWHM km/s (10) | T_mb K (11) | RMS mK (12) | Flux K km/s (13) | δFlux K km/s (14) | (15) |
| 527.171 | CH₃OH | 15₃ − 15₂ | 527.171 | 326.28 | 8.92e-04 | 13.07 | 0.38 | 4.81 | 0.89 | 0.03 | 17 | 0.14 | 0.02 | |
| 527.658 | CH₃OH | 14₃ − 14₂ | 527.658 | 291.46 | 8.88e-04 | 11.99 | 0.36 | 8.39 | 0.86 | 0.04 | 17 | 0.34 | 0.03 | |
| 528.177 | CH₃OH | 13₃ − 13₂ | 528.177 | 258.96 | 8.81e-04 | 12.51 | 0.15 | 3.74 | 0.36 | 0.05 | 17 | 0.24 | 0.03 | |
| 528.683 | CH₃OH | 12₃ − 12₂ | 528.683 | 228.78 | 8.74e-04 | 12.10 | 0.21 | 4.01 | 0.50 | 0.05 | 17 | 0.25 | 0.03 | |
| 529.143 | CH₃OH | 11₃ − 11₂ | 529.143 | 200.92 | 8.64e-04 | 12.53 | 0.14 | 4.11 | 0.33 | 0.07 | 15 | 0.37 | 0.03 | |
| 529.540 | CH₃OH | 10₃ − 10₂ | 529.540 | 175.39 | 8.52e-04 | 12.15 | 0.12 | 5.06 | 0.29 | 0.09 | 16 | 0.48 | 0.03 | |
| 529.867 | CH₃OH | 9₃ − 9₂ | 529.867 | 152.17 | 8.37e-04 | 12.37 | 0.04 | 5.20 | 0.10 | 0.11 | 16 | 0.64 | 0.03 | |
| 530.123 | CH₃OH | 8₃ − 8₂ | 530.123 | 131.28 | 8.18e-04 | 11.81 | 0.09 | 4.89 | 0.20 | 0.11 | 16 | 0.63 | 0.02 | |
| 530.184 | CH₃OH | 11₀ − 10₀ | 530.184 | 166.05 | 6.69e-04 | 12.00 | 0.04 | 3.72 | 0.11 | 0.38 | 16 | 1.80 | 0.03 | |
| 530.316 | CH₃OH | 7₃ − 7₂ | 530.316 | 112.71 | 7.93e-04 | 12.33 | 0.08 | 4.76 | 0.20 | 0.16 | 13 | 0.74 | 0.03 | |
| 530.455 | CH₃OH | 6₃ − 6₂ | 530.455 | 96.46 | 7.58e-04 | 12.13 | 0.07 | 4.78 | 0.16 | 0.16 | 13 | 0.78 | 0.02 | |
| 530.549 | CH₃OH | 5₃ − 5₂ | 530.549 | 82.53 | 7.04e-04 | 12.17 | 0.09 | 5.77 | 0.22 | 0.16 | 13 | 0.91 | 0.02 | |
| 530.610 | CH₃OH | 4₃ − 4₂ | 530.610 | 70.93 | 6.14e-04 | 12.18 | 0.03 | 5.04 | 0.07 | 0.14 | 14 | 0.79 | 0.02 | |
| 530.647 | CH₃OH | 3₃ − 3₂ | 530.647 | 61.64 | 4.37e-04 | 12.37 | 0.12 | 6.05 | 0.28 | 0.12 | 15 | 0.78 | 0.03 | |
| 531.079 | CH₃OH | 11₋₁ − 10₋₁ | 531.079 | 158.64 | 6.69e-04 | 11.89 | 0.04 | 3.98 | 0.09 | 0.69 | 14 | 3.48 | 0.03 | |
| 531.319 | CH₃OH | 11₀ − 10₀ | 531.319 | 153.10 | 6.75e-04 | 11.87 | 0.02 | 4.02 | 0.04 | 0.74 | 18 | 3.67 | 0.03 | |
| 531.636 | CH₃OH | 11₂ − 10₂ | 531.636 | 190.87 | 6.58e-04 | 12.25 | 0.02 | 4.23 | 0.06 | 0.14 | 19 | 0.70 | 0.03 | |
| 531.716 | HCN | 6 − 5 | 531.716 | 89.32 | 7.20e-03 | 12.04 | 0.05 | 9.01 | 0.12 | 2.23 | 19 | 21.07 | 0.05 | |
| 531.772 | CH₃OH | 11₋₄ − 10₋₄ | 531.772 | 241.07 | 8.70e-04 | 12.56 | 0.31 | 5.97 | 0.75 | 0.04 | 18 | 0.15 | 0.01 | |
| 531.829 | CH₃OH | 11₄ − 10₄ | 531.829 | 249.18 | 8.75e-04 | 12.27 | 0.56 | 5.90 | 1.33 | 0.03 | 17 | 0.13 | 0.02 | |
| 531.866 | CH₃OH | 11₋₃ − 10₋₃ | 531.866 | 215.90 | 9.34e-04 | | | | | | 17 | | | **B**: CH₃OH (11₃ − 10₃) |
| 531.869 | CH₃OH | 11₃ − 10₃ | 531.869 | 202.98 | 9.27e-04 | | | | | | 17 | | | **B**: CH₃OH (11₋₃ − 10₋₃) |
| 531.870 | CH₃OH | 11₄ − 10₄ | 531.870 | 233.52 | 8.72e-04 | | | | | | 16 | | | **B**: CH₃OH (11₋₃ − 10₋₃) |
| 531.871 | CH₃OH | 11₄ − 10₄ | 531.871 | 233.52 | 8.72e-04 | | | | | | 15 | | | **B**: CH₃OH (11₋₃ − 10₋₃) |
| 531.893 | CH₃OH | 11₃ − 10₃ | 531.893 | 202.98 | 6.26e-04 | | | | | | 18 | | | **B**: CH₃OH (11₋₃ − 10₋₃) |
| 532.031 | CH₃OH | 11₁ − 10₁ | 532.031 | 174.27 | 6.86e-04 | 11.94 | 0.03 | 4.00 | 0.07 | 0.27 | 17 | 1.37 | 0.03 | |
| 532.069 | CH₃OH | 11₃ − 10₃ | 532.069 | 200.92 | 6.28e-04 | 11.98 | 0.08 | 4.38 | 0.20 | 0.09 | 16 | 0.48 | 0.02 | |
| 532.133 | CH₃OH | 11₂ − 10₂ | 532.133 | 190.94 | 6.60e-04 | 11.77 | 0.04 | 3.68 | 0.10 | 0.15 | 15 | 0.45 | 0.02 | |
| 532.466 | CH₃OH | 11₂ − 10₂ | 532.466 | 175.53 | 6.51e-04 | 12.00 | 0.02 | 4.14 | 0.04 | 0.37 | 14 | 1.75 | 0.03 | |
| 532.567 | CH₃OH | 11₋₂ − 10₋₂ | 532.567 | 179.19 | 6.58e-04 | 12.00 | 0.03 | 3.74 | 0.06 | 0.19 | 14 | 0.92 | 0.02 | |
| 532.707 | CH₃OH | 15₂ − 15₁ | 532.707 | 316.05 | 1.18e-03 | | | | | | 17 | | | |
| 532.722 | CH | ²Π 1₁,₀,₃/₂,₁ − 1₋₁,₀,₁/₂,₁ | 532.722 | 25.73 | 2.07e-06 | | | | | | 19 | | | **B**: CH (²Π 1₋₁,₀,₃/₂,₁ − 1₋₁,₀,₁/₂,₁) — |
| 532.724 | CH | ²Π 1₁,₀,₃/₂,₂ − 1₋₁,₀,₁/₂,₁ | 532.724 | 25.73 | 6.21e-04 | | | | | | 19 | | | **B**: CH (²Π 1₋₁,₀,₃/₂,₂ − 1₋₁,₀,₁/₂,₁) — |
| 535.062 | HCO⁺ | 6 − 5 | 535.062 | 89.88 | 1.25e-02 | 11.46 | 0.03 | 3.81 | 0.07 | 4.27 | 14 | 22.76 | 0.04 | |
| 536.191 | CH₃OH | 11₁ − 10₁ | 536.191 | 169.01 | 6.89e-04 | 12.01 | 0.02 | 4.05 | 0.04 | 0.44 | 15 | 2.23 | 0.03 | |
| 536.761 | CH | ²Π 1₋₁,₀,₃/₂,₂ − 1₋₁,₀,₁/₂,₁ | 536.761 | 25.76 | 6.38e-04 | 12.74 | 0.08 | 2.48 | 0.18 | 0.06 | 18 | 0.21 | 0.02 | |
| 538.571 | CH | 5₁ − 4₀ | 538.571 | 49.06 | 4.93e-04 | 11.80 | 0.02 | 4.47 | 0.04 | 0.78 | 14 | 4.19 | 0.03 | |
| 538.689 | CS | 11 − 10 | 538.689 | 155.15 | 3.35e-03 | 12.21 | 0.07 | 9.38 | 0.17 | 0.45 | 14 | 4.07 | 0.03 | |
| 539.282 | CH₃OH | 16₂ − 16₁ | 539.282 | 353.12 | 8.24e-04 | 12.33 | 0.05 | 4.79 | 0.73 | 0.03 | 12 | 0.22 | 0.03 | |
| 540.922 | CH₃OH | 15₀ − 14₁ | 540.922 | 290.74 | 4.21e-04 | 12.63 | 0.05 | 4.12 | 0.12 | 0.07 | 15 | 0.35 | 0.02 | |
| 542.001 | CH₃OH | 6₃ − 5₂ | 542.001 | 98.55 | 7.93e-04 | 11.85 | 0.02 | 3.99 | 0.05 | 0.54 | 13 | 2.73 | 0.03 | |
| 542.082 | CH₃OH | 6₃ − 5₂ | 542.082 | 98.55 | 7.93e-04 | 11.79 | 0.02 | 3.99 | 0.04 | 0.54 | 14 | 2.60 | 0.02 | |

| | | | | | | Gaussian fit | | | | | from moments | | | |
|---|---|---|---|---|---|---|---|---|---|---|---|---|---|---|
| ν GHz | Species | Transition | $\nu_{rest}$ GHz | $E_u$ K | $A_{ul}$ rad/s | $v_{lsr}$ km/s | $\delta v_{lsr}$ km/s | FWHM km/s | $\delta$FWHM km/s | $T_{mb}$ K | RMS mK | Flux K km/s | $\delta$Flux K km/s | Blend(**B**) / Notes |
| (1) | (2) | (3) | (4) | (5) | (6) | (7) | (8) | (9) | (10) | (11) | (12) | (13) | (14) | (15) |
| 543.076 | CH$_3$OH | $8_0 - 7_{-1}$ | 543.076 | 96.61 | 4.54e-04 | 11.89 | 0.02 | 3.92 | 0.05 | 0.45 | 13 | 2.31 | 0.03 | |
| 543.898 | HNC | $6_{0,0} - 5_{0,0}$ | 543.898 | 91.37 | 8.04e-03 | 11.54 | 0.04 | 2.20 | 0.09 | 0.38 | 14 | 1.62 | 0.02 | |
| 546.239 | CH$_3$OH | $1_2 - 1_1$ | 546.239 | 392.50 | 1.27e-03 | 12.06 | 0.86 | 8.99 | 2.10 | 0.03 | 18 | 0.17 | 0.02 | |
| 547.676 | H$_2^{18}$O | $1_{1,0} - 1_{0,1}$ | 547.676 | 60.46 | 3.29e-03 | 13.69 | 0.20 | 19.19 | 0.20 | 0.07 | 19 | 1.09 | 0.05 | |
| 548.831 | C$^{18}$O | $5 - 4$ | 548.831 | 79.02 | 1.06e-05 | 11.34 | 0.01 | 2.02 | 0.03 | 1.36 | 16 | 5.10 | 0.04 | |
| 550.926 | $^{13}$CO | $5 - 4$ | 550.926 | 79.33 | 1.07e-05 | 11.60 | 0.02 | 3.09 | 0.04 | 6.94 | 16 | 31.33 | 0.03 | |
| 553.146 | CH$_3$OH | $8_1 - 7_0$ | 553.146 | 104.62 | 4.63e-04 | 11.73 | 0.02 | 4.20 | 0.06 | 0.38 | 16 | 1.82 | 0.03 | |
| 554.055 | CH$_3$OH | $12_1 - 11_2$ | 554.055 | 202.12 | 4.63e-04 | 12.09 | 0.10 | 3.28 | 0.23 | 0.08 | 15 | 0.41 | 0.02 | |
| 556.936 | H$_2$O | $1_{1,0,0} - 1_{0,1,0}$ | 556.936 | 60.96 | 3.46e-03 | 11.36 | 0.04 | 20.54 | 0.10 | 2.12 | 19 | 33.72 | 0.06 | |
| 558.087 | SO | $13_{12} - 12_{11}$ | 558.087 | 194.37 | 2.32e-03 | 9.78 | 0.17 | 9.17 | 0.17 | 0.07 | 20 | 0.59 | 0.03 | |
| 558.345 | CH$_3$OH | $11_2 - 10_1$ | 558.345 | 175.53 | 5.106e-04 | 12.10 | 0.01 | 3.66 | 0.05 | 0.39 | 17 | 1.76 | 0.03 | |
| 558.967 | N$_2$H$^+$ | $6 - 5$ | 558.967 | 93.90 | 1.16e-02 | 11.55 | 0.01 | 2.24 | 0.02 | 2.28 | 16 | 7.71 | 0.03 | |
| 559.319 | SO | $13_{13} - 12_{12}$ | 559.319 | 201.07 | 2.34e-03 | 9.98 | 0.08 | 5.49 | 0.19 | 0.05 | 16 | 0.40 | 0.03 | |
| 560.178 | SO | $13_{14} - 12_{13}$ | 560.178 | 192.66 | 2.37e-03 | 10.93 | 0.07 | 11.15 | 0.17 | 0.05 | 17 | 0.60 | 0.04 | |
| 561.712 | C$^{17}$O | $5_{5/2} - 4_{7/2}$ | 561.712 | 80.88 | 1.20e-07 | 11.18 | 0.02 | 2.52 | 0.05 | 0.35 | 15 | 1.58 | 0.03 | |
| 561.899 | H$_2$CO | $8_{1,8} - 7_{1,7}$ | 561.899 | 133.28 | 5.20e-03 | 11.98 | 0.02 | 4.74 | 0.04 | 1.47 | 37 | 8.38 | 0.08 | |
| 566.730 | CN | $5_{9/2,11/2} - 4_{7/2,9/2}$ | 566.730 | 81.59 | 1.98e-03 | | | | | | 16 | | | **B**: CN $(5_{9/2,7/2} - 4_{7/2,5/2})$ |
| 566.731 | CN | $5_{9/2,7/2} - 4_{7/2,5/2}$ | 566.731 | 81.59 | 1.86e-03 | | | | | | 16 | | | **B**: CN $(5_{9/2,11/2} - 4_{7/2,9/2})$ |
| 566.731 | CN | $5_{9/2,9/2} - 4_{7/2,7/2}$ | 566.731 | 81.59 | 1.88e-03 | | | | | | 16 | | | **B**: CN $(5_{9/2,11/2} - 4_{7/2,9/2})$ |
| 566.947 | CN | $5_{11/2,11/2} - 4_{9/2,9/2}$ | 566.947 | 81.64 | 1.96e-03 | | | | | | 18 | | | **B**: CN $(5_{11/2,13/2} - 4_{9/2,11/2})$ |
| 566.947 | CN | $5_{11/2,13/2} - 4_{9/2,11/2}$ | 566.947 | 81.64 | 2.03e-03 | | | | | | 24 | | | **B**: CN $(5_{11/2,11/2} - 4_{9/2,9/2})$ |
| 566.947 | CN | $5_{11/2,9/2} - 4_{9/2,7/2}$ | 566.947 | 81.65 | 1.95e-03 | | | | | | 16 | | | **B**: CN $(5_{11/2,11/2} - 4_{9/2,9/2})$ |
| 568.566 | CH$_3$OH | $3_{-2} - 2_{-1}$ | 568.566 | 39.83 | 1.01e-03 | 11.90 | 0.05 | 4.96 | 0.12 | 0.51 | 16 | 2.87 | 0.03 | |
| 568.783 | CH$_3$OH | $17_0 - 16_1$ | 568.783 | 354.54 | 5.62e-04 | 12.09 | 0.03 | 3.84 | 0.07 | 0.20 | 15 | 1.00 | 0.03 | |
| 572.113 | $^{15}$NH$_3$ | $1_0 - 0_0$ | 572.113 | 28.00 | 1.57e-03 | 12.83 | 0.38 | 5.76 | 0.90 | 0.02 | 13 | 0.15 | 0.02 | |
| 572.498 | o-NH$_3$ | $1_0 - 0_0$ | 572.498 | 27.48 | 1.57e-03 | 11.79 | 0.02 | 5.28 | 0.04 | 1.76 | 14 | 10.80 | 0.03 | |
| 572.899 | CH$_3$OH | $15_{-1} - 14_0$ | 572.899 | 283.64 | 8.61e-04 | 12.02 | 0.02 | 3.92 | 0.04 | 0.31 | 14 | 1.55 | 0.03 | |
| 574.868 | CH$_3$OH | $12_1 - 11_1$ | 574.868 | 193.96 | 8.53e-04 | 12.24 | 0.02 | 4.44 | 0.04 | 0.28 | 13 | 1.48 | 0.03 | |
| 576.268 | CO | $5 - 4$ | 576.268 | 82.98 | 1.22e-05 | 11.37 | 0.14 | 11.91 | 0.32 | 15.36 | 20 | 171.01 | 0.07 | |
| 576.708 | H$_2$CO | $8_{0,8} - 7_{0,7}$ | 576.708 | 125.12 | 5.70e-03 | 11.94 | 0.02 | 4.76 | 0.04 | 0.59 | 13 | 3.32 | 0.03 | |
| 578.006 | CH$_3$OH | $12_0 - 11_0$ | 578.006 | 193.79 | 8.70e-04 | 12.12 | 0.02 | 3.90 | 0.04 | 0.40 | 13 | 1.87 | 0.02 | |
| 579.085 | CH$_3$OH | $2_1 - 1_0$ | 579.085 | 44.67 | 1.23e-03 | 11.80 | 0.02 | 4.71 | 0.04 | 0.39 | 14 | 2.15 | 0.02 | |
| 579.151 | CH$_3$OH | $12_{-1} - 11_{-1}$ | 579.151 | 186.43 | 8.72e-04 | 12.06 | 0.02 | 3.77 | 0.05 | 0.72 | 13 | 3.43 | 0.02 | |
| 579.201 | DCN | $8 - 7$ | 579.201 | 125.10 | 9.52e-03 | 12.04 | 0.02 | 2.93 | 0.48 | 0.03 | 12 | 0.11 | 0.01 | |
| 579.460 | CH$_3$OH | $12_1 - 11_0$ | 579.460 | 180.91 | 8.79e-04 | 11.96 | 0.02 | 3.98 | 0.04 | 0.72 | 16 | 3.47 | 0.02 | |
| 579.858 | CH$_3$OH | $12_2 - 11_2$ | 579.858 | 218.69 | 8.62e-04 | 12.26 | 0.08 | 4.67 | 0.18 | 0.14 | 13 | 0.80 | 0.02 | |
| 579.921 | CH$_3$OH | $2_2 - 1_1$ | 579.921 | 44.67 | 1.24e-03 | 11.71 | 0.07 | 5.07 | 0.15 | 0.39 | 13 | 2.31 | 0.02 | |
| 580.033 | CH$_3$OH | $12_5 - 11_5$ | 580.033 | 305.00 | 1.08e-03 | | | | | | 14 | | | **B**: CH$_3$OH $(12_5 - 11_5)$ |
| 580.058 | CH$_3$OH | $12_{-5} - 11_{-5}$ | 580.058 | 318.90 | 1.09e-03 | | | | | | 14 | | | **B**: CH$_3$OH $(12_5 - 11_5)$ |
| 580.058 | CH$_3$OH | $12_5 - 11_5$ | 580.058 | 318.90 | 1.09e-03 | | | | | | 14 | | | **B**: CH$_3$OH $(12_5 - 11_5)$ |
| 580.058 | CH$_3$OH | $12_4 - 11_4$ | 580.058 | 268.91 | 1.16e-03 | | | | | | 14 | | | **B**: CH$_3$OH $(12_5 - 11_5)$ |
| 580.059 | CH$_3$OH | $12_{-4} - 11_{-4}$ | 580.059 | 268.91 | 1.16e-03 | | | | | | 14 | | | **B**: CH$_3$OH $(12_4 - 11_4)$ |
| 580.126 | CH$_3$OH | $12_4 - 11_4$ | 580.126 | 277.02 | 1.17e-03 | | | | | | 16 | | | **B**: CH$_3$OH $(12_4 - 11_4)$ |
| 580.163 | CH$_3$OH | $12_{-3} - 11_{-3}$ | 580.163 | 243.74 | 1.23e-03 | | | | | | 13 | | | **B**: CH$_3$OH $(12_3 - 11_3)$ |
| 580.176 | CH$_3$OH | $12_3 - 11_3$ | 580.176 | 230.83 | 1.22e-03 | | | | | | 13 | | | **B**: CH$_3$OH $(12_3 - 11_3)$ |
| 580.195 | CH$_3$OH | $12_4 - 11_4$ | 580.195 | 261.37 | 1.16e-03 | | | | | | 17 | | | **B**: CH$_3$OH $(12_3 - 11_3)$ |

| ν GHz (1) | Species (2) | Transition (3) | ν_rest GHz (4) | E_u K (5) | A_ul rad/s (6) | Gaussian fit | | | | | from moments | | | Blend(B) / Notes (15) |
|---|---|---|---|---|---|---|---|---|---|---|---|---|---|---|
| | | | | | | v_lsr km/s (7) | δv_lsr km/s (8) | FWHM km/s (9) | δFWHM km/s (10) | T_mb K (11) | RMS mK (12) | Flux K km/s (13) | δFlux K km/s (14) | |
| 580.196 | CH$_3$OH | $12_4 - 11_4$ | 580.196 | 261.37 | $1.16\times10^{-3}$ | | | | | | 16 | | | B: CH$_3$OH $(12_{-3} - 11_{-3})$ |
| 580.213 | CH$_3$OH | $12_3 - 11_3$ | 580.213 | 230.83 | $1.22\times10^{-3}$ | | | | | | 15 | | | B: CH$_3$OH $(12_4 - 11_4)$ |
| 580.369 | CH$_3$OH | $12_3 - 11_3$ | 580.369 | 202.12 | $8.84\times10^{-4}$ | 12.02 | 0.02 | 4.16 | 0.04 | 0.25 | 15 | 1.33 | 0.03 | |
| 580.442 | CH$_3$OH | $12_1 - 11_1$ | 580.442 | 228.78 | $8.29\times10^{-4}$ | 11.99 | 0.05 | 4.51 | 0.12 | 0.07 | 16 | 0.37 | 0.02 | |
| 580.502 | CH$_3$OH | $12_3 - 11_3$ | 580.502 | 218.80 | $8.64\times10^{-4}$ | 12.10 | 0.03 | 3.95 | 0.08 | 0.13 | 15 | 0.54 | 0.02 | |
| 580.903 | CH$_3$OH | $12_2 - 11_2$ | 580.903 | 203.41 | $8.53\times10^{-4}$ | 11.94 | 0.02 | 4.06 | 0.04 | 0.36 | 15 | 1.84 | 0.03 | |
| 581.092 | CH$_3$OH | $12_{-2} - 11_{-2}$ | 581.092 | 207.08 | $8.63\times10^{-4}$ | 12.02 | 0.03 | 4.77 | 0.06 | 0.19 | 15 | 1.10 | 0.03 | |
| 581.612 | H$_2$CO | $8_{2,7} - 7_{2,6}$ | 581.612 | 172.84 | $5.49\times10^{-3}$ | 11.83 | 0.02 | 4.42 | 0.04 | 0.34 | 17 | 1.74 | 0.03 | |
| 581.750 | H$_2$CO | $8_{7,1} - 7_{7,0}$ | 581.750 | 700.86 | $1.37\times10^{-3}$ | | | | | | 17 | | | B: H$_2$CO $(8_{7,0} - 7_{7,0})$ |
| 581.750 | H$_2$CO | $8_{7,1} - 7_{7,0}$ | 581.750 | 700.86 | $1.37\times10^{-3}$ | | | | | | 17 | | | B: H$_2$CO $(8_{7,1} - 7_{7,1})$ |
| 582.071 | H$_2$CO | $8_{6,3} - 7_{6,2}$ | 582.071 | 548.74 | $2.57\times10^{-3}$ | | | | | | 14 | | | B: H$_2$CO $(8_{6,1} - 7_{6,1})$ |
| 582.071 | H$_2$CO | $8_{6,2} - 7_{6,1}$ | 582.071 | 548.74 | $2.57\times10^{-3}$ | | | | | | 15 | | | B: H$_2$CO $(8_{6,2} - 7_{6,2})$ |
| 582.382 | H$_2$CO | $8_{5,3} - 7_{5,2}$ | 582.382 | 419.79 | $3.58\times10^{-3}$ | | | | | | 15 | | | B: H$_2$CO $(8_{5,2} - 7_{5,2})$ |
| 582.382 | H$_2$CO | $8_{5,2} - 7_{5,3}$ | 582.382 | 419.79 | $3.58\times10^{-3}$ | | | | | | 15 | | | B: H$_2$CO $(8_{5,3} - 7_{5,3})$ |
| 582.723 | H$_2$CO | $8_{4,4} - 7_{4,3}$ | 582.723 | 314.14 | $4.42\times10^{-3}$ | | | | | | 18 | | | B: H$_2$CO $(8_{4,3} - 7_{4,3})$ |
| 582.724 | H$_2$CO | $8_{4,4} - 7_{4,3}$ | 582.724 | 314.14 | $4.42\times10^{-3}$ | | | | | | 18 | | | B: H$_2$CO $(8_{4,4} - 7_{4,4})$ |
| 583.145 | H$_2$CO | $8_{3,6} - 7_{3,5}$ | 583.145 | 231.88 | $5.07\times10^{-3}$ | 12.02 | 0.02 | 4.38 | 0.04 | 0.48 | 18 | 2.60 | 0.03 | B: H$_2$CO $(8_{4,5} - 7_{4,4})$ |
| 583.309 | H$_2$CO | $8_{3,5} - 7_{3,4}$ | 583.309 | 231.89 | $5.08\times10^{-3}$ | 11.96 | 0.02 | 4.60 | 0.04 | 0.48 | 16 | 2.70 | 0.03 | |
| 584.147 | CH$_3$OH | $16_0 - 15_1$ | 584.147 | 327.63 | $7.81\times10^{-4}$ | 12.51 | 0.13 | 4.52 | 0.32 | 0.07 | 17 | 0.32 | 0.03 | |
| 584.450 | CH$_3$OH | $6_1 - 5_0$ | 584.450 | 62.87 | $6.25\times10^{-4}$ | 11.98 | 0.02 | 4.28 | 0.04 | 0.93 | 17 | 4.59 | 0.03 | |
| 584.822 | CH$_3$OH | $12_1 - 11_1$ | 584.822 | 197.08 | $8.98\times10^{-4}$ | 12.03 | 0.02 | 4.09 | 0.04 | 0.44 | 18 | 2.22 | 0.03 | |
| 587.454 | H$_2$CO | $8_{2,6} - 7_{2,5}$ | 587.454 | 173.55 | $5.66\times10^{-3}$ | 11.96 | 0.02 | 4.35 | 0.05 | 0.31 | 17 | 1.68 | 0.03 | |
| 587.616 | CS | $12 - 11$ | 587.616 | 183.35 | $4.36\times10^{-3}$ | 12.19 | 0.08 | 10.32 | 0.20 | 0.36 | 15 | 3.75 | 0.04 | |
| 590.278 | CH$_3$OH | $7_3 - 6_3$ | 590.278 | 114.79 | $9.50\times10^{-4}$ | 11.78 | 0.02 | 4.08 | 0.04 | 0.55 | 15 | 2.52 | 0.03 | |
| 590.440 | CH$_3$OH | $7_3 - 6_3$ | 590.440 | 114.79 | $9.50\times10^{-4}$ | 11.84 | 0.02 | 4.13 | 0.04 | 0.54 | 16 | 2.67 | 0.03 | |
| 590.791 | CH$_3$OH | $9_0 - 8_1$ | 590.791 | 117.46 | $6.23\times10^{-4}$ | 12.01 | 0.02 | 4.15 | 0.04 | 0.54 | 16 | 2.54 | 0.03 | |
| 599.927 | HDO | $2_{1,1} - 2_{0,2}$ | 599.927 | 95.23 | $3.45\times10^{-3}$ | 12.03 | 0.09 | 5.05 | 0.21 | 0.05 | 15 | 0.25 | 0.02 | |
| 600.331 | H$_2$CO | $8_{1,7} - 7_{1,6}$ | 600.331 | 141.61 | $6.34\times10^{-3}$ | 12.00 | 0.01 | 5.14 | 0.03 | 1.05 | 15 | 6.12 | 0.03 | |
| 601.258 | SO | $14_{13} - 13_{12}$ | 601.258 | 223.23 | $2.92\times10^{-3}$ | 7.39 | 0.52 | 11.75 | 1.22 | 0.04 | 17 | 0.36 | 0.03 | |
| 601.849 | CH$_3$OH | $13_1 - 12_2$ | 601.849 | 232.29 | $4.07\times10^{-4}$ | 11.91 | 0.17 | 5.35 | 0.40 | 0.09 | 17 | 0.58 | 0.03 | |
| 602.233 | CH$_3$OH | $9_1 - 8_0$ | 602.233 | 125.52 | $5.95\times10^{-4}$ | 11.86 | 0.02 | 3.84 | 0.05 | 0.37 | 19 | 1.68 | 0.03 | |
| 602.292 | SO | $14_{14} - 13_{13}$ | 602.292 | 229.97 | $2.93\times10^{-3}$ | 9.58 | 0.08 | 8.56 | 0.19 | 0.04 | 17 | 0.27 | 0.02 | |
| 603.021 | SO | $14_{15} - 13_{14}$ | 603.021 | 221.61 | $2.96\times10^{-3}$ | 8.39 | 0.10 | 9.41 | 0.24 | 0.03 | 14 | 0.32 | 0.02 | |
| 604.268 | H$^{13}$CN | $7 - 6$ | 604.268 | 116.01 | $1.07\times10^{-2}$ | 13.24 | 0.10 | 11.10 | 0.24 | 0.08 | 17 | 0.62 | 0.02 | |
| 607.175 | H$^{13}$CO$^+$ | $7 - 6$ | 607.175 | 116.57 | $1.84\times10^{-2}$ | 11.48 | 0.03 | 2.07 | 0.08 | 0.08 | 16 | 0.24 | 0.02 | |
| 607.216 | CH$_3$OH | $12_2 - 11_1$ | 607.216 | 203.41 | $6.42\times10^{-4}$ | 12.07 | 0.02 | 3.75 | 0.05 | 0.32 | 16 | 1.43 | 0.03 | |
| 611.267 | CCH | $7_{15/2,7} - 6_{13/2,6}$ | 611.267 | 117.35 | $7.29\times10^{-4}$ | | | | | | 15 | | | B: CCH $(7_{15/2,8} - 6_{13/2,7})$ |
| 611.267 | CCH | $7_{15/2,8} - 6_{13/2,7}$ | 611.267 | 117.36 | $7.36\times10^{-4}$ | | | | | | 15 | | | B: CCH $(7_{15/2,7} - 6_{13/2,6})$ |
| 611.330 | CCH | $7_{13/2,7} - 6_{11/2,6}$ | 611.330 | 117.38 | $7.27\times10^{-4}$ | | | | | | 16 | | | B: CCH $(7_{13/2,6} - 6_{11/2,6})$ |
| 611.330 | CCH | $7_{13/2,7} - 6_{11/2,5}$ | 611.330 | 117.38 | $7.18\times10^{-4}$ | | | | | | 15 | | | B: CCH $(7_{13/2,7} - 6_{11/2,6})$ |
| 616.980 | CH$_3$OH | $4_{-2} - 3_{-1}$ | 616.980 | 49.12 | $1.11\times10^{-3}$ | 12.00 | 0.05 | 5.09 | 0.13 | 0.54 | 12 | 3.16 | 0.02 | |
| 620.304 | HCN | $7 - 6$ | 620.304 | 119.09 | $1.16\times10^{-2}$ | 12.26 | 0.01 | 10.13 | 0.03 | 2.03 | 12 | 21.23 | 0.03 | |
| 622.568 | CH$_3$OH | $18_0 - 17_1$ | 622.568 | 396.17 | $7.52\times10^{-4}$ | 12.22 | 0.03 | 4.98 | 0.06 | 0.18 | 14 | 1.07 | 0.02 | |
| 622.659 | CH$_3$OH | $13_1 - 12_1$ | 622.659 | 223.85 | $1.09\times10^{-3}$ | 12.33 | 0.02 | 4.62 | 0.05 | 0.29 | 14 | 1.57 | 0.02 | |
| 622.774 | CH$_3$OH | $16_{-1} - 15_0$ | 622.774 | 320.63 | $1.16\times10^{-3}$ | 12.16 | 0.02 | 4.18 | 0.05 | 0.29 | 15 | 1.50 | 0.03 | |

continued from previous page

| $\nu$ GHz | Species | Transition | $\nu_{rest}$ GHz | $E_u$ K | $A_{ul}$ rad/s | Gaussian fit | | | | | from moments | | | Blend(B) / Notes |
|---|---|---|---|---|---|---|---|---|---|---|---|---|---|---|
| (1) | (2) | (3) | (4) | (5) | (6) | $v_{lsr}$ km/s (7) | $\delta v_{lsr}$ km/s (8) | FWHM km/s (9) | $\delta$FWHM km/s (10) | $T_{mb}$ K (11) | RMS mK (12) | Flux K km/s (13) | $\delta$Flux K km/s (14) | (15) |
| 624.208 | $HCO^+$ | $7 - 6$ | 624.208 | 119.84 | 2.01e-02 | 11.36 | 0.01 | 4.32 | 0.03 | 3.72 | 16 | 21.10 | 0.05 | |
| 624.964 | $H^{37}Cl$ | $1_{3/2} - 0_{3/2}$ | 624.964 | 29.99 | 1.16e-03 | | | | | | 15 | | | B: $H^{37}Cl$ $(1_{5/2} - 0_{3/2})$ |
| 624.978 | $H^{37}Cl$ | $1_{5/2} - 0_{3/2}$ | 624.978 | 29.99 | 1.16e-03 | | | | | | 14 | | | B: $H^{37}Cl$ $(1_{3/2} - 0_{3/2})$ |
| 624.988 | $H^{37}Cl$ | $1_{1/2} - 0_{3/2}$ | 624.988 | 30.00 | 1.16e-03 | | | | | | 14 | | | B: $H^{37}Cl$ $(1_{5/2} - 0_{3/2})$ |
| 625.749 | $CH_3OH$ | $13_0 - 12_0$ | 625.749 | 223.82 | 1.11e-03 | 12.14 | 0.03 | 3.79 | 0.07 | 0.37 | 15 | 1.58 | 0.03 | |
| 625.902 | HCl | $1_{3/2} - 0_{3/2}$ | 625.902 | 30.04 | 1.17e-03 | | | | | | 15 | | | B: HCl $(1_{3/2} - 0_{3/2})$ |
| 625.919 | HCl | $1_{5/2} - 0_{3/2}$ | 625.919 | 30.04 | 1.17e-03 | | | | | | 15 | | | B: HCl $(1_{3/2} - 0_{3/2})$ |
| 625.932 | HCl | $1_{1/2} - 0_{3/2}$ | 625.932 | 30.04 | 1.17e-03 | | | | | | 14 | | | B: HCl $(1_{3/2} - 0_{3/2})$ |
| 626.555 | $CH_3OH$ | $17_0 - 16_1$ | 626.555 | 366.79 | 9.51e-04 | | | | | | 15 | | | B: $CH_3OH$ $(13_4 - 12_4)$ |
| 626.555 | $CH_3OH$ | $13_4 - 12_4$ | 626.555 | 573.97 | 1.49e-03 | | | | | | 15 | | | B: $CH_3OH$ $(17_0 - 16_1)$ |
| 626.626 | $CH_3OH$ | $3_2 - 2_1$ | 626.626 | 51.64 | 1.23e-03 | 11.70 | 0.02 | 4.87 | 0.04 | 0.47 | 19 | 2.52 | 0.03 | |
| 627.170 | $CH_3OH$ | $13_{-1} - 12_{-1}$ | 627.170 | 216.53 | 1.11e-03 | 12.04 | 0.04 | 4.14 | 0.09 | 0.64 | 15 | 3.24 | 0.03 | |
| 627.558 | $CH_3OH$ | $13_0 - 12_0$ | 627.558 | 211.03 | 1.12e-03 | 11.97 | 0.03 | 4.14 | 0.08 | 0.65 | 14 | 3.17 | 0.03 | |
| 628.052 | $CH_3OH$ | $13_2 - 12_2$ | 628.052 | 248.84 | 1.10e-03 | 12.15 | 0.09 | 4.73 | 0.21 | 0.13 | 15 | 0.64 | 0.03 | |
| 628.330 | $CH_3OH$ | $13_{-4} - 12_{-4}$ | 628.330 | 299.06 | 1.51e-03 | 11.72 | 0.06 | 2.69 | 0.14 | 0.05 | 13 | 0.40 | 0.02 | |
| 628.409 | $CH_3OH$ | $13_4 - 12_4$ | 628.409 | 307.18 | 1.02e-03 | 12.42 | 0.46 | 3.89 | 1.07 | 0.02 | 14 | 0.10 | 0.02 | |
| 628.445 | $CH_3OH$ | $13_{-3} - 12_{-3}$ | 628.445 | 273.90 | 1.59e-03 | | | | | | 14 | | | B: $CH_3OH$ $(13_3 - 12_3)$ |
| 628.470 | $CH_3OH$ | $13_3 - 12_3$ | 628.470 | 260.99 | 1.06e-03 | | | | | | 13 | | | B: $CH_3OH$ $(13_{-3} - 12_{-3})$ |
| 628.470 | $CH_3OH$ | $13_3 - 12_3$ | 628.470 | 260.99 | 1.58e-03 | | | | | | 14 | | | B: $CH_3OH$ $(13_{-3} - 12_{-3})$ |
| 628.512 | $CH_3OH$ | $13_4 - 12_4$ | 628.512 | 291.53 | 1.51e-03 | | | | | | 14 | | | B: $CH_3OH$ $(13_4 - 12_4)$ |
| 628.513 | $CH_3OH$ | $13_4 - 12_4$ | 628.513 | 291.53 | 1.51e-03 | | | | | | 13 | | | B: $CH_3OH$ $(13_4 - 12_4)$ |
| 628.525 | $CH_3OH$ | $13_3 - 12_3$ | 628.525 | 261.00 | 1.58e-03 | | | | | | 13 | | | B: $CH_3OH$ $(13_4 - 12_4)$ |
| 628.696 | $CH_3OH$ | $13_1 - 12_1$ | 628.696 | 232.29 | 1.14e-03 | 12.12 | 0.03 | 4.12 | 0.06 | 0.24 | 14 | 1.15 | 0.02 | |
| 628.816 | $CH_3OH$ | $13_3 - 12_3$ | 628.816 | 258.96 | 1.07e-03 | 12.39 | 0.17 | 6.48 | 0.41 | 0.08 | 13 | 0.46 | 0.02 | |
| 628.869 | $CH_3OH$ | $13_3 - 12_3$ | 628.869 | 248.98 | 1.11e-03 | 12.04 | 0.11 | 4.66 | 0.25 | 0.12 | 14 | 0.62 | 0.02 | |
| 629.140 | $CH_3OH$ | $3_2 - 2_1$ | 629.140 | 51.64 | 1.25e-03 | 11.73 | 0.05 | 4.26 | 0.11 | 0.14 | 17 | 2.35 | 0.03 | |
| 629.322 | $CH_3OH$ | $13_2 - 12_2$ | 629.322 | 233.61 | 1.09e-03 | 12.05 | 0.05 | 4.25 | 0.12 | 0.34 | 17 | 1.68 | 0.03 | |
| 629.652 | $CH_3OH$ | $13_{-3} - 12_{-2}$ | 629.652 | 237.30 | 1.11e-03 | 12.20 | 0.03 | 4.84 | 0.07 | 0.17 | 16 | 0.90 | 0.02 | |
| 629.921 | $CH_3OH$ | $7_1 - 6_0$ | 629.921 | 78.97 | 7.78e-04 | 12.02 | 0.02 | 4.18 | 0.04 | 1.02 | 16 | 4.96 | 0.03 | |
| 631.703 | $H_2CO$ | $9_{1,9} - 8_{1,8}$ | 631.703 | 163.60 | 7.46e-03 | 11.95 | 0.02 | 4.86 | 0.04 | 1.00 | 44 | 4.96 | 0.08 | |
| 633.423 | $CH_3OH$ | $13_1 - 12_1$ | 633.423 | 227.48 | 1.15e-03 | 12.29 | 0.04 | 3.68 | 0.10 | 0.34 | 43 | 1.66 | 0.07 | |
| 634.511 | HNC | $7_{0,0} - 6_{0,0}$ | 634.511 | 121.82 | 1.29e-02 | 11.66 | 0.06 | 2.95 | 0.15 | 0.11 | 33 | 0.42 | 0.05 | |
| 636.193 | $CH_3OH$ | $16_0 - 16_3$ | 636.193 | 395.93 | 2.23e-03 | | | | | | 31 | | | B: $CH_3OH$ $(13_4 - 13_3)$ |
| 636.197 | $CH_3OH$ | $13_4 - 13_3$ | 636.197 | 291.53 | 2.14e-03 | | | | | | 32 | | | B: $CH_3OH$ $(16_0 - 16_3)$ |
| 636.199 | $CH_3OH$ | $17_4 - 17_3$ | 636.199 | 435.37 | 2.25e-03 | | | | | | 31 | | | B: $CH_3OH$ $(16_0 - 16_3)$ |
| 636.274 | $CH_3OH$ | $9_4 - 9_3$ | 636.274 | 184.79 | 1.95e-03 | | | | | | 35 | | | B: $CH_3OH$ $(10_4 - 10_3)$ |
| 636.279 | $CH_3OH$ | $10_4 - 10_3$ | 636.279 | 207.09 | 2.02e-03 | | | | | | 31 | | | B: $CH_3OH$ $(9_4 - 9_3)$ |
| 636.281 | $CH_3OH$ | $11_4 - 11_3$ | 636.281 | 233.52 | 2.07e-03 | | | | | | 30 | | | B: $CH_3OH$ $(9_4 - 9_3)$ |
| 636.291 | $CH_3OH$ | $9_1 - 9_3$ | 636.291 | 184.79 | 1.96e-03 | | | | | | 31 | | | B: $CH_3OH$ $(9_4 - 9_3)$ |
| 636.299 | $CH_3OH$ | $12_4 - 12_3$ | 636.299 | 261.36 | 2.11e-03 | | | | | | 32 | | | B: $CH_3OH$ $(9_4 - 9_3)$ |
| 636.303 | $CH_3OH$ | $8_4 - 8_3$ | 636.303 | 163.90 | 1.87e-03 | | | | | | 31 | | | B: $CH_3OH$ $(9_4 - 9_3)$ |
| 636.312 | $CH_3OH$ | $8_4 - 8_3$ | 636.312 | 163.90 | 1.87e-03 | | | | | | 31 | | | B: $CH_3OH$ $(10_4 - 10_3)$ |
| 636.334 | $CH_3OH$ | $7_4 - 7_3$ | 636.334 | 145.33 | 1.76e-03 | | | | | | 27 | | | B: $CH_3OH$ $(8_4 - 8_3)$ |
| 636.337 | $CH_3OH$ | $7_4 - 7_3$ | 636.337 | 145.33 | 1.76e-03 | | | | | | 33 | | | B: $CH_3OH$ $(8_4 - 8_3)$ |
| 636.364 | $CH_3OH$ | $6_4 - 6_3$ | 636.364 | 129.09 | 1.60e-03 | | | | | | 37 | | | B: $CH_3OH$ $(6_4 - 6_3)$ |

| ν GHz (1) | Species (2) | Transition (3) | ν_rest GHz (4) | E_u K (5) | A_ul rad/s (6) | *Gaussian fit* v_lsr km/s (7) | δv_lsr km/s (8) | FWHM km/s (9) | δFWHM km/s (10) | T_mb K (11) | *from moments* RMS mK (12) | Flux K km/s (13) | δFlux K km/s (14) | Blend(**B**) / Notes (15) |
|---|---|---|---|---|---|---|---|---|---|---|---|---|---|---|
| 636.366 | CH$_3$OH | $6_4 - 6_3$ | 636.366 | 129.09 | 1.60e-03 | | | | | | 37 | | | **B**: CH$_3$OH ($7_4 - 7_3$) |
| 636.393 | CH$_3$OH | $5_4 - 5_3$ | 636.393 | 115.16 | 1.34e-03 | | | | | | 31 | | | **B**: CH$_3$OH ($6_4 - 6_3$) |
| 636.394 | CH$_3$OH | $5_4 - 5_3$ | 636.394 | 115.16 | 1.34e-03 | | | | | | 34 | | | **B**: CH$_3$OH ($6_4 - 6_3$) |
| 636.523 | CH$_3$OH | $15_4 - 15_3$ | 636.523 | 358.81 | 2.21e-03 | | | | | | 35 | | | **B**: CS ($13 - 12$) |
| 636.532 | CS | $13 - 12$ | 636.532 | 213.90 | 5.56e-03 | | | | | | 35 | | | **B**: CH$_3$OH ($15_4 - 15_3$) |
| 638.280 | CH$_3$OH | $30_0 - 9_{-1}$ | 638.280 | 140.60 | 8.35e-04 | 12.26 | 0.06 | 3.96 | 0.13 | 0.46 | 42 | 2.05 | 0.07 | |
| 638.524 | CH$_3$OH | $8_3 - 7_2$ | 638.524 | 133.36 | 1.13e-03 | 11.97 | 0.05 | 3.93 | 0.12 | 0.51 | 41 | 2.45 | 0.06 | |
| 638.818 | CH$_3$OH | $8_3 - 7_2$ | 638.818 | 133.36 | 1.13e-03 | 11.98 | 0.06 | 3.91 | 0.14 | 0.57 | 35 | 2.80 | 0.06 | |
| 645.254 | SO | $15_{15} - 14_{14}$ | 645.254 | 260.94 | 3.62e-03 | 11.61 | 0.09 | 9.64 | 0.20 | 0.05 | 36 | 0.36 | 0.06 | |
| 645.875 | SO | $15_{16} - 14_{15}$ | 645.875 | 252.60 | 3.65e-03 | 8.79 | 0.08 | 6.64 | 0.18 | 0.09 | 34 | 0.43 | 0.06 | |
| 647.082 | H$_2$CO | $9_{0,9} - 8_{0,8}$ | 647.082 | 156.18 | 8.10e-03 | 11.80 | 0.02 | 4.91 | 0.05 | 0.55 | 28 | 3.10 | 0.05 | |
| 649.540 | CH$_3$OH | $14_1 - 13_2$ | 649.540 | 264.78 | 5.20e-04 | | | | | | 27 | | | **B**: CH$_3$OH ($14_1 - 13_2$) |
| 649.540 | CH$_3$OH | $14_1 - 13_2$ | 649.540 | 264.78 | 7.71e-04 | | | | | | 26 | | | **B**: CH$_3$OH ($14_1 - 13_2$) |
| 651.617 | CH$_3$OH | $10_1 - 9_0$ | 651.617 | 148.73 | 7.51e-04 | 11.87 | 0.03 | 4.46 | 0.06 | 0.37 | 25 | 1.93 | 0.04 | |
| 652.096 | N$_2$H$^+$ | $7 - 6$ | 652.096 | 125.19 | 1.87e-02 | 11.49 | 0.02 | 2.24 | 0.05 | 1.80 | 30 | 5.79 | 0.05 | |
| 653.970 | H$_2$CO | $9_{2,8} - 8_{2,7}$ | 653.970 | 204.23 | 7.96e-03 | 11.95 | 0.03 | 3.88 | 0.07 | 0.30 | 31 | 1.58 | 0.05 | |
| 654.463 | H$_2$CO | $9_{7,3} - 8_{7,2}$ | 654.463 | 732.27 | 3.32e-03 | | | | | | 32 | | | **B**: H$_2$CO ($9_{7,2} - 8_{7,1}$) |
| 654.463 | H$_2$CO | $9_{7,2} - 8_{7,1}$ | 654.463 | 732.27 | 3.32e-03 | | | | | | 32 | | | **B**: H$_2$CO ($9_{7,3} - 8_{7,2}$) |
| 655.212 | H$_2$CO | $9_{5,5} - 8_{5,4}$ | 655.212 | 451.24 | 5.83e-03 | | | | | | 31 | | | **B**: H$_2$CO ($9_{5,4} - 8_{5,3}$) |
| 655.212 | H$_2$CO | $9_{5,4} - 8_{5,3}$ | 655.212 | 451.24 | 5.83e-03 | | | | | | 30 | | | **B**: H$_2$CO ($9_{5,5} - 8_{5,4}$) |
| 655.640 | H$_2$CO | $9_{4,6} - 8_{4,5}$ | 655.640 | 345.60 | 6.77e-03 | | | | | | 30 | | | **B**: H$_2$CO ($9_{4,5} - 8_{4,4}$) |
| 655.644 | H$_2$CO | $9_{4,5} - 8_{4,4}$ | 655.644 | 345.60 | 6.77e-03 | | | | | | 28 | | | **B**: H$_2$CO ($9_{4,6} - 8_{4,5}$) |
| 656.165 | H$_2$CO | $9_{3,7} - 8_{3,6}$ | 656.165 | 263.37 | 7.52e-03 | | | | | | 28 | | | **B**: CH$_3$OH ($13_2 - 12_1$) |
| 656.169 | CH$_3$OH | $13_2 - 12_1$ | 656.169 | 233.61 | 7.87e-04 | | | | | | 30 | | | **B**: H$_2$CO ($9_{3,7} - 8_{3,6}$) |
| 656.465 | H$_2$CO | $9_{3,6} - 8_{3,5}$ | 656.465 | 263.40 | 7.53e-03 | 11.85 | 0.02 | 4.78 | 0.05 | 0.44 | 29 | 2.34 | 0.05 | |
| 658.553 | C$^{18}$O | $6 - 5$ | 658.553 | 110.63 | 1.86e-05 | 11.39 | 0.02 | 2.22 | 0.04 | 0.93 | 27 | 3.46 | 0.05 | |
| 661.067 | $^{13}$CO | $6 - 5$ | 661.067 | 111.05 | 1.88e-05 | 11.54 | 0.01 | 3.22 | 0.04 | 6.18 | 28 | 28.65 | 0.06 | |
| 662.209 | H$_2$CO | $9_{2,7} - 8_{2,6}$ | 662.209 | 205.33 | 8.27e-03 | 12.09 | 0.02 | 5.29 | 0.05 | 0.23 | 26 | 1.39 | 0.04 | |
| 665.442 | CH$_3$OH | $5_2 - 4_1$ | 665.442 | 60.73 | 1.27e-03 | 12.00 | 0.02 | 4.91 | 0.05 | 0.67 | 30 | 3.66 | 0.04 | |
| 668.117 | CH$_3$OH | $18_0 - 17_1$ | 668.117 | 408.23 | 7.64e-04 | 12.54 | 0.30 | 3.67 | 0.71 | 0.05 | 25 | 0.25 | 0.04 | |
| 670.423 | CH$_3$OH | $14_1 - 13_1$ | 670.423 | 256.02 | 1.36e-03 | 12.04 | 0.03 | 4.71 | 0.06 | 0.24 | 29 | 1.35 | 0.05 | |
| 672.903 | CH$_3$OH | $17_{-1} - 16_0$ | 672.903 | 359.92 | 1.53e-03 | 12.10 | 0.07 | 4.25 | 0.17 | 0.26 | 29 | 1.32 | 0.04 | |
| 673.416 | CH$_3$OH | $14_0 - 13_0$ | 673.416 | 256.14 | 1.38e-03 | 12.32 | 0.08 | 4.69 | 0.18 | 0.29 | 28 | 1.52 | 0.04 | |
| 673.746 | CH$_3$OH | $4_2 - 3_1$ | 673.746 | 60.92 | 1.34e-03 | 11.70 | 0.04 | 4.62 | 0.04 | 0.54 | 30 | 2.91 | 0.05 | |
| 674.009 | C$^{17}$O | $6_{7/2} - 5_{9/2}$ | 674.009 | 113.23 | 8.37e-08 | | | | | | 32 | | | **B**: CH$_3$OH ($14_1 - 13_1$) |
| 674.017 | CH$_3$OH | $14_1 - 13_1$ | 674.017 | 568.00 | 2.06e-03 | | | | | | 32 | | | **B**: C$^{17}$O ($6_{7/2} - 5_{9/2}$) |
| 674.791 | CH$_3$OH | $14_2 - 13_2$ | 674.791 | 541.67 | 2.02e-03 | | | | | | 31 | | | **B**: H$_2$CO ($9_{1,8} - 8_{1,7}$) |
| 674.810 | H$_2$CO | $9_{1,8} - 8_{1,7}$ | 674.810 | 173.99 | 9.09e-03 | | | | | | 32 | | | **B**: CH$_3$OH ($14_2 - 13_2$) |
| 674.991 | CH$_3$OH | $8_1 - 7_0$ | 674.991 | 97.38 | 9.56e-04 | 12.09 | 0.02 | 4.69 | 0.04 | 1.06 | 33 | 5.55 | 0.06 | |
| 675.135 | CH$_3$OH | $14_{-1} - 13_{-1}$ | 675.135 | 248.94 | 1.39e-03 | 12.13 | 0.02 | 4.25 | 0.05 | 0.57 | 31 | 2.97 | 0.05 | |
| 675.613 | CH$_3$OH | $14_0 - 13_0$ | 675.613 | 243.45 | 1.40e-03 | 12.13 | 0.04 | 4.23 | 0.09 | 0.61 | 29 | 3.07 | 0.05 | |
| 675.773 | CH$_3$OH | $3_3 - 2_2$ | 675.773 | 61.64 | 2.56e-03 | 12.35 | 0.08 | 5.22 | 0.20 | 0.44 | 33 | 1.54 | 0.06 | |
| 676.215 | CH$_3$OH | $14_2 - 13_2$ | 676.215 | 281.29 | 1.38e-03 | 12.08 | 0.06 | 6.04 | 0.15 | 0.11 | 31 | 0.65 | 0.05 | |
| 676.495 | CH$_3$OH | $14_5 - 13_5$ | 676.495 | 379.68 | 1.23e-03 | | | | | | 32 | | | **B**: CH$_3$OH ($19_0 - 18_1$) |
| 676.499 | CH$_3$OH | $19_0 - 18_1$ | 676.499 | 440.09 | 9.84e-04 | | | | | | 34 | | | **B**: CH$_3$OH ($14_5 - 13_5$) |

| $\nu$ GHz (1) | Species (2) | Transition (3) | $\nu_{rest}$ GHz (4) | $E_u$ K (5) | $A_{ul}$ rad/s (6) | Gaussian fit $v_{lsr}$ km/s (7) | $\delta v_{lsr}$ km/s (8) | FWHM km/s (9) | $\delta$FWHM km/s (10) | $T_{mb}$ K (11) | from moments RMS mK (12) | Flux K km/s (13) | $\delta$Flux K km/s (14) | Blend(B) / Notes (15) |
|---|---|---|---|---|---|---|---|---|---|---|---|---|---|---|
| 676.585 | $CH_3OH$ | $14_{-4} - 13_{-4}$ | 676.585 | 331.54 | 1.29e-03 | 12.42 | 0.08 | 4.24 | 0.18 | 0.07 | 31 | 0.40 | 0.04 | B: $CH_3OH$ $(14_4 - 13_4)$ |
| 676.749 | $CH_3OH$ | $14_3 - 13_3$ | 676.749 | 293.47 | 1.34e-03 | 12.17 | 0.06 | 4.18 | 0.14 | 0.12 | 28 | 0.48 | 0.04 | B: $CH_3OH$ $(14_4 - 13_4)$ |
| 676.822 | $CH_3OH$ | $14_4 - 13_4$ | 676.822 | 324.01 | 1.30e-03 | | | | | | 31 | | | B: $CH_3OH$ $(14_4 - 13_4)$ |
| 676.824 | $CH_3OH$ | $14_4 - 13_4$ | 676.824 | 324.01 | 1.30e-03 | | | | | | 31 | | | |
| 676.830 | $CH_3OH$ | $14_3 - 13_3$ | 676.830 | 293.48 | 1.34e-03 | 12.19 | 0.04 | 4.25 | 0.09 | 0.20 | 30 | 0.93 | 0.05 | |
| 677.013 | $CH_3OH$ | $14_1 - 13_1$ | 677.013 | 264.78 | 1.43e-03 | | | | | | 29 | | | |
| 677.191 | $CH_3OH$ | $14_3 - 13_3$ | 677.191 | 291.46 | 1.35e-03 | | | | | | 26 | | | B: $CH_3OH$ $(14_2 - 13_2)$ |
| 677.233 | $CH_3OH$ | $14_2 - 13_2$ | 677.233 | 281.49 | 1.39e-03 | 12.17 | 0.03 | 4.09 | 0.06 | 0.32 | 46 | 1.53 | 0.07 | B: $CH_3OH$ $(14_3 - 13_3)$ |
| 677.710 | $CH_3OH$ | $14_2 - 13_2$ | 677.710 | 266.14 | 1.37e-03 | 12.09 | 0.05 | 6.15 | 0.11 | 0.20 | 29 | 1.13 | 0.04 | |
| 678.253 | $CH_3OH$ | $14_{-2} - 13_{-2}$ | 678.253 | 269.85 | 1.39e-03 | 11.90 | 0.02 | 4.52 | 0.05 | 0.58 | 27 | 2.78 | 0.03 | |
| 678.785 | $CH_3OH$ | $4_2 - 3_1$ | 678.785 | 60.93 | 1.37e-03 | | | | | | 31 | | | |
| 680.047 | $CN$ | $6_{11/2,13/2} - 5_{9/2,11/2}$ | 680.047 | 114.23 | 3.50e-03 | | | | | | 31 | | | B: $CN$ $(6_{11/2,9/2} - 5_{9/2,7/2})$ |
| 680.047 | $CN$ | $6_{11/2,9/2} - 5_{9/2,7/2}$ | 680.047 | 114.22 | 3.36e-03 | | | | | | 31 | | | B: $CN$ $(6_{11/2,13/2} - 5_{9/2,11/2})$ |
| 680.047 | $CN$ | $6_{11/2,11/2} - 5_{9/2,11/2}$ | 680.047 | 114.22 | 3.38e-03 | | | | | | 31 | | | B: $CN$ $(6_{11/2,13/2} - 5_{9/2,11/2})$ |
| 680.264 | $CN$ | $6_{13/2,15/2} - 5_{11/2,13/2}$ | 680.264 | 114.29 | 3.47e-03 | 12.32 | 0.03 | 4.86 | 0.07 | 0.38 | 30 | 2.06 | 0.04 | — |
| 680.264 | $CN$ | $6_{13/2,13/2} - 5_{11/2,11/2}$ | 680.264 | 114.29 | 3.56e-03 | | | | | | 31 | | | — |
| 680.264 | $CN$ | $6_{13/2,11/2} - 5_{11/2,9/2}$ | 680.264 | 114.29 | 3.46e-03 | | | | | | 31 | | | — |
| 681.990 | $CH_3OH$ | $14_1 - 13_1$ | 681.990 | 260.21 | 1.43e-03 | | | | | | 29 | | | |
| 685.435 | $CS$ | $14 - 13$ | 685.435 | 246.79 | 6.90e-03 | | | | | | 31 | | | B: $CH_3OH$ $(11_0 - 10_{-1})$ |
| 685.505 | $CH_3OH$ | $11_0 - 10_{-1}$ | 685.505 | 166.05 | 1.10e-03 | | | | | | 29 | | | B: $CS$ $(14 - 13)$ |
| 686.732 | $CH_3OH$ | $9_3 - 8_2$ | 686.732 | 154.25 | 1.34e-03 | 12.10 | 0.02 | 4.16 | 0.05 | 0.51 | 30 | 2.44 | 0.05 | |
| 687.225 | $CH_3OH$ | $9_3 - 8_2$ | 687.225 | 154.25 | 1.34e-03 | 11.83 | 0.02 | 4.45 | 0.06 | 0.50 | 31 | 2.42 | 0.05 | |
| 687.303 | $H_2S$ | $2_{0,2} - 1_{1,1}$ | 687.303 | 54.70 | 9.32e-04 | 11.64 | 0.03 | 3.82 | 0.07 | 0.34 | 36 | 1.59 | 0.05 | |
| 691.473 | $CO$ | $6 - 5$ | 691.473 | 116.16 | 2.14e-05 | 11.30 | 0.03 | 13.00 | 0.07 | 19.50 | 37 | 204.50 | 0.10 | |
| 697.146 | $CH_3OH$ | $15_1 - 14_2$ | 697.146 | 299.59 | 6.56e-04 | 11.37 | 0.03 | 1.86 | 0.15 | 0.12 | 33 | 0.25 | 0.05 | |
| 698.545 | $CCH$ | $8_{17/2,8} - 7_{15/2,7}$ | 698.545 | 150.88 | 1.10e-03 | | | | | | 31 | | | B: $CCH$ $(8_{17/2,9} - 7_{15/2,8})$ |
| 698.545 | $CCH$ | $8_{17/2,9} - 7_{15/2,8}$ | 698.545 | 150.88 | 1.11e-03 | | | | | | 34 | | | B: $CCH$ $(8_{17/2,8} - 7_{15/2,7})$ |
| 698.607 | $CCH$ | $8_{15/2,8} - 7_{13/2,7}$ | 698.607 | 150.91 | 1.10e-03 | | | | | | 33 | | | B: $CCH$ $(8_{17/2,8} - 7_{15/2,7})$ |
| 698.607 | $CCH$ | $8_{15/2,7} - 7_{13/2,6}$ | 698.607 | 150.91 | 1.09e-03 | | | | | | 29 | | | B: $CCH$ $(8_{17/2,8} - 7_{15/2,7})$ |
| 701.367 | $CH_3OH$ | $11_1 - 10_0$ | 701.367 | 174.27 | 9.33e-04 | | | | | | 31 | | | B: $H_2CO$ $(10_{0,10} - 9_{1,9})$ |
| 701.370 | $H_2CO$ | $10_{0,10} - 9_{1,9}$ | 701.370 | 197.26 | 1.03e-02 | 12.09 | 0.04 | 4.56 | 0.09 | 0.23 | 33 | 1.18 | 0.05 | B: $CH_3OH$ $(11_1 - 10_0)$ |
| 705.182 | $CH_3OH$ | $14_2 - 13_1$ | 705.182 | 266.14 | 1.74e-02 | | | | | | 42 | | | |
| 708.877 | $HCN$ | $8 - 7$ | 708.877 | 153.11 | 3.02e-02 | 12.37 | 0.04 | 11.01 | 0.09 | 1.67 | 31 | 18.75 | 0.09 | |
| 713.341 | $HCO^+$ | $8 - 7$ | 713.341 | 154.07 | 1.45e-03 | 11.35 | 0.01 | 4.79 | 0.03 | 2.79 | 39 | 16.69 | 0.08 | |
| 713.982 | $CH_3OH$ | $6_2 - 5_{-1}$ | 713.982 | 74.66 | 1.11e-02 | 11.85 | 0.02 | 4.81 | 0.05 | 0.56 | 48 | 2.92 | 0.06 | |
| 716.938 | $H_2CO$ | $10_{0,10} - 9_{0,9}$ | 716.938 | 190.58 | 1.68e-03 | 11.97 | 0.06 | 3.99 | 0.12 | 0.29 | 43 | 1.38 | 0.06 | |
| 718.159 | $CH_3OH$ | $15_1 - 14_1$ | 718.159 | 290.49 | 3.06e-03 | 12.46 | 0.03 | 4.35 | 0.14 | 0.15 | 40 | 0.78 | 0.06 | |
| 718.436 | $CH_3OH$ | $4_{-4} - 3_{-3}$ | 718.436 | 111.12 | 1.16e-03 | 11.75 | 0.03 | 4.56 | 0.07 | 0.38 | 27 | 1.90 | 0.06 | |
| 719.665 | $CH_3OH$ | $9_1 - 8_0$ | 719.665 | 118.08 | 1.50e-03 | 12.17 | 0.02 | 4.46 | 0.04 | 1.04 | 32 | 5.38 | 0.05 | |
| 720.441 | $CH_3OH$ | $5_2 - 4_1$ | 720.441 | 72.53 | 1.70e-03 | 11.92 | 0.02 | 4.90 | 0.06 | 0.60 | 26 | 3.49 | 0.06 | |
| 721.011 | $CH_3OH$ | $15_0 - 14_0$ | 721.011 | 290.74 | 1.70e-03 | 12.60 | 0.03 | 4.61 | 0.07 | 0.25 | 26 | 1.23 | 0.04 | |

| ν | Species | Transition | $\nu_{rest}$ | $E_u$ | $A_{ul}$ | $v_{lsr}$ | $\delta v_{lsr}$ | Gaussian fit FWHM | $\delta$FWHM | $T_{mb}$ | RMS | from moments Flux | $\delta$Flux | Blend(**B**) / Notes |
|---|---|---|---|---|---|---|---|---|---|---|---|---|---|---|
| GHz | | | GHz | K | rad/s | km/s | km/s | km/s | km/s | K | mK | K km/s | K km/s | |
| (1) | (2) | (3) | (4) | (5) | (6) | (7) | (8) | (9) | (10) | (11) | (12) | (13) | (14) | (15) |
| 723.040 | CH$_3$OH | $15_{-1} - 14_{-1}$ | 723.040 | 283.64 | 1.71e-03 | 12.21 | 0.02 | 4.25 | 0.05 | 0.50 | 28 | 2.50 | 0.04 | |
| 723.280 | CH$_3$OH | $18_{-1} - 17_0$ | 723.280 | 401.50 | 1.98e-03 | 12.48 | 0.04 | 5.37 | 0.09 | 0.20 | 30 | 1.08 | 0.05 | |
| 723.619 | CH$_3$OH | $15_0 - 14_0$ | 723.619 | 278.18 | 1.72e-03 | 12.13 | 0.02 | 4.27 | 0.04 | 0.56 | 31 | 2.52 | 0.05 | |
| 724.122 | CH$_3$OH | $4_3 - 3_2$ | 724.122 | 70.93 | 2.56e-03 | 12.26 | 0.02 | 5.38 | 0.05 | 0.48 | 31 | 2.77 | 0.05 | |
| 724.345 | CH$_3$OH | $15_2 - 14_2$ | 724.345 | 316.05 | 1.71e-03 | 12.26 | 0.07 | 5.00 | 0.16 | 0.10 | 34 | 0.51 | 0.05 | **B**: CH$_3$OH $(15_4 - 14_4)$ |
| 725.013 | CH$_3$OH | $15_3 - 14_3$ | 725.013 | 328.26 | 2.47e-03 | 12.67 | 0.13 | 5.12 | 0.30 | 0.08 | 31 | 0.36 | | **B**: CH$_3$OH $(15_4 - 14_4)$ |
| 725.122 | CH$_3$OH | $15_4 - 14_4$ | 725.122 | 358.81 | 2.39e-03 | | | | | | 31 | | | **B**: CH$_3$OH $(15_4 - 14_4)$ |
| 725.126 | CH$_3$OH | $15_4 - 14_4$ | 725.126 | 358.81 | 2.39e-03 | | | | | | 31 | | | |
| 725.126 | CH$_3$OH | $15_3 - 14_3$ | 725.126 | 328.28 | 2.47e-03 | 12.15 | 0.08 | 3.98 | 0.20 | 0.18 | 32 | 0.72 | 0.05 | **B**: CH$_3$OH $(15_2 - 14_2)$ |
| 725.316 | CH$_3$OH | $15_1 - 14_1$ | 725.316 | 299.59 | 1.76e-03 | | | | | | 24 | | | **B**: CH$_3$OH $(15_3 - 14_3)$ |
| 725.565 | CH$_3$OH | $15_3 - 14_3$ | 725.565 | 326.28 | 2.48e-03 | | | | | | 26 | | | |
| 725.594 | CH$_3$OH | $15_2 - 14_2$ | 725.594 | 316.31 | 1.72e-03 | 12.47 | 0.03 | 4.62 | 0.08 | 0.30 | 28 | 1.53 | 0.04 | |
| 726.052 | CH$_3$OH | $15_2 - 14_2$ | 726.052 | 300.98 | 1.70e-03 | 12.23 | 0.04 | 5.16 | 0.10 | 0.17 | 32 | 0.86 | 0.04 | |
| 726.208 | H$_2$CO | $10_{2,9} - 9_{2,8}$ | 726.208 | 239.08 | 1.11e-02 | 12.11 | 0.16 | 4.34 | 0.38 | 0.13 | 32 | 0.44 | 0.05 | |
| 726.899 | CH$_3$OH | $15_{-2} - 14_{-2}$ | 726.899 | 304.74 | 1.72e-03 | | | | | | 25 | | | **B**: H$_2$CO $(10_{5,4} - 9_{5,4})$ |
| 728.054 | H$_2$CO | $10_{5,6} - 9_{5,5}$ | 728.054 | 486.18 | 8.72e-03 | | | | | | 26 | | | **B**: H$_2$CO $(10_{5,5} - 9_{5,5})$ |
| 728.054 | H$_2$CO | $10_{5,5} - 9_{5,4}$ | 728.054 | 486.18 | 8.72e-03 | | | | | | 26 | | | **B**: H$_2$CO $(10_{4,6} - 9_{4,5})$ |
| 728.583 | H$_2$CO | $10_{4,7} - 9_{4,6}$ | 728.583 | 380.57 | 9.78e-03 | | | | | | 25 | | | **B**: H$_2$CO $(10_{4,7} - 9_{4,6})$ |
| 728.592 | H$_2$CO | $10_{4,6} - 9_{4,5}$ | 728.592 | 380.57 | 9.78e-03 | | | | | | 25 | | | |
| 728.862 | H$_2$CO | $5_2 - 4_1$ | 728.862 | 72.53 | 1.55e-03 | 11.82 | 0.02 | 4.85 | 0.04 | 0.63 | 25 | 3.20 | 0.05 | |
| 729.213 | H$_2$CO | $10_{3,8} - 9_{3,7}$ | 729.213 | 298.36 | 1.06e-02 | 12.14 | 0.02 | 5.10 | 0.06 | 0.29 | 24 | 1.48 | 0.04 | |
| 729.725 | H$_2$CO | $10_{3,7} - 9_{3,6}$ | 729.725 | 298.42 | 1.06e-02 | 12.05 | 0.02 | 4.21 | 0.05 | 0.32 | 25 | 1.56 | 0.04 | |
| 730.520 | CH$_3$OH | $15_1 - 14_1$ | 730.520 | 295.27 | 1.77e-03 | | | | | | 28 | | | **B**: CH$_3$OH $(20_0 - 19_1)$ |
| 730.520 | CH$_3$OH | $20_0 - 19_1$ | 730.520 | 486.30 | 1.26e-03 | | | | | | 26 | | | **B**: CH$_3$OH $(15_1 - 14_1)$ |
| 731.596 | SO | $17_{18} - 16_{17}$ | 731.596 | 320.77 | 5.32e-03 | 8.98 | 0.10 | 8.66 | 0.23 | 0.05 | 24 | 0.33 | 0.04 | |
| 732.432 | CH$_3$OH | $10_3 - 9_2$ | 732.432 | 193.79 | 1.42e-03 | 12.21 | 0.04 | 4.38 | 0.05 | 0.24 | 28 | 2.33 | 0.04 | |
| 734.324 | CS | $15 - 14$ | 734.324 | 282.04 | 8.58e-03 | 12.21 | 0.15 | 11.05 | 0.35 | 0.17 | 26 | 1.90 | 0.05 | |
| 734.894 | CH$_3$OH | $10_3 - 9_2$ | 734.894 | 177.46 | 1.58e-03 | 11.88 | 0.06 | 5.03 | 0.14 | 0.46 | 25 | 2.50 | 0.04 | |
| 735.673 | CH$_3$OH | $10_3 - 9_2$ | 735.673 | 177.46 | 1.58e-03 | 12.06 | 0.05 | 4.47 | 0.11 | 0.46 | 25 | 2.22 | 0.04 | |
| 736.034 | H$_2$S | $2_{1,2} - 1_{0,1}$ | 736.034 | 55.10 | 1.33e-03 | 11.65 | 0.07 | 5.27 | 0.16 | 0.22 | 29 | 1.01 | 0.06 | |
| 737.343 | H$_2$CO | $10_{2,8} - 9_{2,7}$ | 737.343 | 240.72 | 1.16e-02 | 11.84 | 0.04 | 4.94 | 0.10 | 0.19 | 26 | 1.01 | 0.04 | |
| 745.210 | N$_2$H$^+$ | $8 - 7$ | 745.210 | 160.96 | 2.81e-02 | 11.51 | 0.02 | 2.60 | 0.06 | 1.18 | 32 | 4.09 | 0.05 | |
| 749.072 | H$_2$CO | $10_{1,9} - 9_{1,8}$ | 749.072 | 209.94 | 1.25e-02 | 12.06 | 0.02 | 5.59 | 0.04 | 0.59 | 36 | 3.49 | 0.06 | |
| 751.551 | CH$_3$OH | $12_1 - 11_0$ | 751.551 | 202.12 | 1.14e-03 | 12.16 | 0.03 | 4.38 | 0.07 | 0.35 | 35 | 1.61 | 0.06 | |
| 752.033 | H$_2$O | $2_{1,1,0} - 2_{0,2,0}$ | 752.033 | 136.94 | 6.98e-03 | 11.91 | 0.06 | 14.88 | 0.14 | 2.19 | 43 | 38.51 | 0.15 | |
| 752.312 | CH$_3$OH | $7_{-5} - 7_{-4}$ | 752.312 | 189.00 | 2.35e-03 | 12.23 | 0.11 | 7.29 | 0.25 | 0.06 | 41 | 0.37 | 0.06 | |
| 754.222 | CH$_3$OH | $15_2 - 14_1$ | 754.222 | 300.98 | 1.15e-03 | 12.14 | 0.04 | 3.83 | 0.11 | 0.21 | 36 | 1.02 | 0.06 | |
| 762.636 | CH$_3$OH | $7_{-2} - 6_{-1}$ | 762.636 | 90.91 | 1.66e-03 | 12.04 | 0.02 | 4.96 | 0.04 | 0.60 | 38 | 2.99 | 0.06 | |
| 762.676 | CH$_3$OH | $13_{-3} - 13_{-2}$ | 762.676 | 273.90 | 2.31e-03 | 12.03 | 0.29 | 4.79 | 0.69 | 0.07 | 39 | 0.31 | 0.04 | |
| 763.883 | CH$_3$OH | $12_{-3} - 12_{-2}$ | 763.883 | 243.74 | 2.32e-03 | 12.67 | 0.08 | 2.80 | 0.19 | 0.11 | 33 | 0.44 | 0.04 | |
| 763.951 | CH$_3$OH | $8_5 - 9_4$ | 763.951 | 221.45 | 4.87e-04 | | | | | | 38 | | | **B**: CH$_3$OH $(8_5 - 9_4)$ |
| 763.951 | CH$_3$OH | $8_5 - 9_4$ | 763.951 | 221.45 | 4.87e-04 | | | | | | 40 | | | **B**: CH$_3$OH $(8_5 - 9_4)$ |
| 763.953 | CH$_3$OH | $10_1 - 9_0$ | 763.953 | 141.08 | 1.39e-03 | | | | | | 40 | | | **B**: CH$_3$OH $(8_5 - 9_4)$ |
| 764.812 | CH$_3$OH | $11_{-3} - 11_{-2}$ | 764.812 | 215.90 | 2.31e-03 | 12.02 | 0.07 | 4.95 | 0.16 | 0.13 | 37 | 0.59 | 0.05 | |
| 765.513 | CH$_3$OH | $10_{-3} - 10_{-2}$ | 765.513 | 190.37 | 2.30e-03 | 12.31 | 0.08 | 10.55 | 0.18 | 0.12 | 34 | 0.95 | 0.06 | |

| ν GHz | Species | Transition | ν$_{rest}$ GHz | E$_u$ K | A$_{ul}$ rad/s | Gaussian fit | | | | | from moments | | | Blend(B) / Notes |
|---|---|---|---|---|---|---|---|---|---|---|---|---|---|---|
| | | | | | | v$_{lsr}$ km/s | δv$_{lsr}$ km/s | FWHM km/s | δFWHM km/s | T$_{mb}$ K | RMS mK | Flux K km/s | δFlux K km/s | |
| (1) | (2) | (3) | (4) | (5) | (6) | (7) | (8) | (9) | (10) | (11) | (12) | (13) | (14) | (15) |
| 765.866 | CH$_3$OH | $16_1 - 15_1$ | 765.866 | 327.24 | 2.04e-03 | 12.54 | 0.05 | 5.19 | 0.12 | 0.15 | 42 | 0.87 | 0.06 | |
| 766.028 | CH$_3$OH | $9_{-3} - 8_{-3}$ | 766.028 | 167.16 | 2.28e-03 | 12.19 | 0.06 | 5.33 | 0.15 | 0.18 | 38 | 0.84 | 0.05 | |
| 766.396 | CH$_3$OH | $8_{-3} - 7_{-2}$ | 766.396 | 146.28 | 2.24e-03 | 12.11 | 0.05 | 5.61 | 0.12 | 0.19 | 42 | 1.16 | 0.07 | |
| 766.648 | CH$_3$OH | $7_{-3} - 7_{-2}$ | 766.648 | 127.71 | 2.18e-03 | 12.28 | 0.05 | 5.61 | 0.13 | 0.21 | 39 | 1.24 | 0.05 | |
| 766.710 | CH$_3$OH | $6_2 - 5_1$ | 766.710 | 86.46 | 1.69e-03 | 11.94 | 0.03 | 4.75 | 0.06 | 0.59 | 42 | 2.69 | 0.06 | |
| 766.761 | CH$_3$OH | $5_{-4} - 4_{-3}$ | 766.761 | 122.72 | 3.13e-03 | 11.67 | 0.03 | 3.99 | 0.07 | 0.43 | 45 | 1.92 | 0.06 | |
| 766.810 | CH$_3$OH | $4_3 - 3_2$ | 766.810 | 111.46 | 2.10e-03 | 11.97 | 0.03 | 5.32 | 0.10 | 0.20 | 43 | 1.03 | 0.06 | B: CH$_3$OH ($3_{-3} - 3_{-2}$) |
| 766.908 | CH$_3$OH | $6_{-3} - 6_{-2}$ | 766.908 | 97.53 | 1.96e-03 | 11.89 | 0.04 | 3.00 | 0.10 | 0.24 | 43 | 0.75 | 0.05 | B: CH$_3$OH ($4_{-3} - 4_{-2}$) |
| 766.960 | CH$_3$OH | $5_{-3} - 5_{-2}$ | 766.960 | 85.92 | 2.54e-03 | | | | | | 39 | | | |
| 766.982 | CH$_3$OH | $4_{-3} - 4_{-2}$ | 766.982 | 76.64 | 1.81e-03 | | | | | | 38 | | | |
| 768.252 | C$^{17}$O | $7 - 6$ | 768.252 | 147.50 | 2.98e-05 | 11.31 | 0.03 | 3.19 | 0.06 | 0.56 | 46 | 2.51 | 0.07 | |
| 768.540 | CH$_3$OH | $16_0 - 15_0$ | 768.540 | 327.63 | 2.06e-03 | 12.31 | 0.04 | 5.03 | 0.10 | 0.20 | 44 | 1.09 | 0.06 | |
| 770.886 | CH$_3$OH | $16_{-1} - 15_{-1}$ | 770.886 | 320.63 | 3.08e-03 | | | | | | 36 | | | B: H$_2$CO ($11_{1,11} - 10_{1,10}$) |
| 770.896 | H$_2$CO | $11_{1,11} - 10_{1,10}$ | 770.896 | 234.26 | 1.37e-02 | | | | | | 36 | | | B: CH$_3$OH ($16_{-1} - 15_{-1}$) |
| 771.184 | $^{13}$CO | $7 - 6$ | 771.184 | 148.06 | 3.01e-05 | 11.37 | 0.02 | 3.58 | 0.04 | 4.91 | 36 | 23.54 | 0.07 | |
| 771.576 | CH$_3$OH | $16_0 - 15_0$ | 771.576 | 315.21 | 2.10e-03 | 12.32 | 0.02 | 4.08 | 0.05 | 0.46 | 36 | 2.06 | 0.05 | |
| 772.454 | CH$_3$OH | $5_3 - 4_2$ | 772.454 | 82.53 | 2.70e-03 | 12.74 | 0.03 | 6.21 | 0.06 | 0.51 | 40 | 3.45 | 0.06 | |
| 773.259 | CH$_3$OH | $16_3 - 15_3$ | 773.259 | 365.37 | 2.03e-03 | 11.95 | 0.02 | 5.15 | 0.16 | 0.12 | 38 | 0.40 | 0.05 | |
| 773.416 | CH$_3$OH | $16_3 - 15_3$ | 773.416 | 365.40 | 3.01e-03 | 11.97 | 0.19 | 5.04 | 0.45 | 0.12 | 38 | 0.65 | 0.06 | |
| 773.602 | CH$_3$OH | $16_1 - 15_1$ | 773.602 | 336.72 | 2.13e-03 | 12.72 | 0.07 | 3.31 | 0.10 | 0.13 | 40 | 0.50 | 0.06 | |
| 773.893 | CH$_3$OH | $19_{-1} - 18_0$ | 773.893 | 445.37 | 2.53e-03 | 12.29 | 0.06 | 5.36 | 0.13 | 0.13 | 35 | 0.66 | 0.05 | |
| 773.941 | CH$_3$OH | $16_3 - 15_3$ | 773.941 | 363.43 | 3.03e-03 | | | | | | 35 | | | B: CH$_3$OH ($16_2 - 15_2$) |
| 773.950 | CH$_3$OH | $16_2 - 15_2$ | 773.950 | 353.45 | 3.10e-03 | | | | | | 35 | | | B: CH$_3$OH ($16_3 - 15_3$) |
| 774.333 | CH$_3$OH | $16_2 - 15_2$ | 774.333 | 338.14 | 2.07e-03 | 12.01 | 0.04 | 3.24 | 0.10 | 0.21 | 39 | 0.71 | 0.06 | |
| 775.596 | CH$_3$OH | $16_{-2} - 15_{-2}$ | 775.596 | 341.96 | 2.10e-03 | 11.74 | 0.09 | 3.82 | 0.20 | 0.11 | 40 | 0.39 | 0.06 | |
| 779.009 | CH$_3$OH | $16_1 - 15_1$ | 779.009 | 332.65 | 2.15e-03 | | | | | | 40 | | | B: CH$_3$OH ($13_0 - 12_{-1}$) |
| 779.031 | CH$_3$OH | $13_0 - 12_{-1}$ | 779.031 | 223.82 | 1.82e-03 | | | | | | 40 | | | B: CH$_3$OH ($16_1 - 15_1$) |
| 779.380 | CH$_3$OH | $6_2 - 5_1$ | 779.380 | 86.46 | 1.78e-03 | 11.85 | 0.02 | 5.08 | 0.06 | 0.55 | 41 | 2.60 | 0.07 | |
| 780.039 | H$_2^{37}$Cl$^+$ | $2_{1,2,5/2} - 1_{0,1,5/2}$ | 780.039 | 57.66 | 1.78e-03 | 9.38 | 1.09 | 1.15 | 1.52 | -0.07 | 35 | -0.38 | 0.04 | |
| 781.609 | H$_2$Cl$^+$ | $2_{1,2,5/2} - 1_{0,1,5/2}$ | 781.609 | 57.75 | 1.79e-03 | | | | | | 32 | | | B: H$_2$Cl$^+$ ($2_{1,2,7/2} - 1_{0,1,5/2}$) |
| 781.622 | H$_2$Cl$^+$ | $2_{1,2,5/2} - 1_{0,1,3/2}$ | 781.622 | 57.75 | 4.17e-03 | | | | | | 30 | | | B: H$_2$Cl$^+$ ($2_{1,2,7/2} - 1_{0,1,5/2}$) |
| 781.627 | H$_2$Cl$^+$ | $2_{1,2,7/2} - 1_{0,1,5/2}$ | 781.627 | 57.75 | 5.96e-03 | | | | | | 31 | | | B: H$_2$Cl$^+$ ($2_{1,2,5/2} - 1_{0,1,5/2}$) |
| 781.628 | H$_2$Cl$^+$ | $2_{1,2,7/2} - 1_{0,1,1/2}$ | 781.628 | 57.75 | 4.96e-03 | | | | | | 30 | | | B: H$_2$Cl$^+$ ($2_{1,2,5/2} - 1_{0,1,5/2}$) |
| 783.002 | CH$_3$OH | $12_{1,12} - 11_{0,2}$ | 783.002 | 202.98 | 1.85e-03 | 12.08 | 0.03 | 4.39 | 0.07 | 0.43 | 33 | 2.13 | 0.05 | |
| 783.199 | CS | $16 - 15$ | 783.199 | 319.62 | 1.04e-02 | 12.80 | 0.27 | 10.51 | 0.62 | 0.13 | 32 | 1.27 | 0.05 | |
| 784.177 | CH$_3$OH | $13_3 - 10_2$ | 784.177 | 202.99 | 1.86e-03 | 12.16 | 0.03 | 4.21 | 0.03 | 0.41 | 30 | 1.89 | 0.05 | |
| 784.693 | CH$_3$OH | $21_0 - 20_1$ | 784.693 | 534.79 | 1.60e-03 | 12.66 | 0.05 | 3.86 | 0.12 | 0.14 | 30 | 0.61 | 0.04 | |
| 785.802 | CCH | $9_{19/2,9} - 8_{17/2,8}$ | 785.802 | 188.59 | 1.58e-03 | | | | | | 35 | | | B: CCH ($9_{19/2,10} - 8_{17/2,9}$) |
| 785.802 | CCH | $9_{19/2,10} - 8_{17/2,9}$ | 785.802 | 188.59 | 1.59e-03 | | | | | | 35 | | | B: CCH ($9_{19/2,10} - 8_{17/2,8}$) |
| 786.280 | H$_2$CO | $11_{0,11} - 10_{0,10}$ | 786.280 | 150.96 | 6.36e-08 | | | | | | 39 | | | B: H$_2$CO ($11_{0,11} - 10_{0,10}$) |
| 786.285 | C$^{17}$O | $7_{9/2} - 6_{11/2}$ | 786.285 | 228.32 | 1.47e-02 | | | | | | 35 | | | B: C$^{17}$O ($7_{9/2} - 6_{11/2}$) |
| 793.336 | CN | $7_{13/2,15/2} - 6_{11/2,13/2}$ | 793.336 | 152.30 | 5.64e-03 | | | | | | 35 | | | B: CN ($7_{13/2,11/2} - 6_{11/2,9/2}$) |
| 793.336 | CN | $7_{13/2,11/2} - 6_{11/2,9/2}$ | 793.336 | 152.30 | 5.48e-03 | | | | | | 35 | | | B: CN ($7_{13/2,15/2} - 6_{11/2,13/2}$) |

| ν GHz (1) | Species (2) | Transition (3) | ν_rest GHz (4) | $E_u$ K (5) | $A_{ul}$ rad/s (6) | $v_{lsr}$ km/s (7) | $\delta v_{lsr}$ km/s (8) | FWHM [Gaussian fit] km/s (9) | δFWHM km/s (10) | $T_{mb}$ K (11) | RMS [from moments] mK (12) | Flux K km/s (13) | δFlux K km/s (14) | Blend(B) / Notes (15) | |
|---|---|---|---|---|---|---|---|---|---|---|---|---|---|---|---|
| 793.337 | CN | $7_{13/2,13/2} - 6_{11/2,11/2}$ | 793.337 | 152.30 | 5.51e-03 | | | | | | 35 | | | **B:** CN $6_{11/2,13/2}$ | $(7_{13/2,15/2})$ — |
| 793.553 | CN | $7_{15/2,15/2} - 6_{13/2,13/2}$ | 793.553 | 152.38 | 5.61e-03 | | | | | | 36 | | | **B:** CN $6_{13/2,15/2}$ | $(7_{15/2,17/2})$ — |
| 793.553 | CN | $7_{15/2,17/2} - 6_{13/2,15/2}$ | 793.553 | 152.38 | 5.71e-03 | | | | | | 37 | | | **B:** CN $6_{13/2,13/2}$ | $(7_{15/2,15/2})$ — |
| 793.554 | CN | $7_{15/2,13/2} - 6_{13/2,11/2}$ | 793.554 | 152.38 | 5.59e-03 | | | | | | 35 | | | **B:** CN $6_{13/2,13/2}$ | $(7_{15/2,15/2})$ — |
| 797.433 | HCN | $9 - 8$ | 797.433 | 191.38 | 2.49e-02 | 12.44 | 0.02 | 11.93 | 0.04 | 1.14 | 47 | 14.60 | 0.11 | | |
| 798.313 | H2CO | $11_{2,10} - 10_{2,9}$ | 798.313 | 277.39 | 1.49e-02 | 9.62 | 0.30 | 6.16 | 0.71 | 0.10 | 49 | 0.49 | 0.07 | | |
| 802.241 | CH3OH | $13_1 - 12_0$ | 802.241 | 232.29 | 1.38e-03 | 12.31 | 0.07 | 2.47 | 0.16 | 0.20 | 58 | 0.52 | 0.07 | | |
| 802.282 | H2CO | $13_{3,9} - 12_{3,8}$ | 802.282 | 336.87 | 1.45e-02 | 11.82 | 0.07 | 3.20 | 0.17 | 0.19 | 57 | 0.51 | 0.07 | | |
| 802.458 | HCO+ | $9 - 8$ | 802.458 | 192.59 | 4.33e-02 | 11.55 | 0.02 | 5.42 | 0.04 | 2.00 | 57 | 13.19 | 0.12 | | |
| 803.112 | H2CO | $13_{3,8} - 12_{3,7}$ | 803.112 | 336.96 | 1.45e-02 | 12.27 | 0.07 | 3.64 | 0.17 | 0.77 | 59 | 0.77 | 0.08 | | |
| 803.239 | CH3OH | $16_2 - 15_1$ | 803.239 | 338.14 | 1.38e-03 | 12.25 | 0.08 | 5.16 | 0.15 | 0.14 | 56 | 0.57 | 0.08 | | |
| 806.652 | CO | $7 - 6$ | 806.652 | 154.88 | 3.42e-05 | 11.27 | 0.03 | 13.27 | 0.06 | 20.42 | 49 | 211.46 | 0.14 | | |
| 807.866 | CH3OH | $11_1 - 10_0$ | 807.866 | 166.37 | 1.65e-03 | | | | | | 51 | | | **B:** CH3OH $(11_1 - 10_0)$ | |
| 807.866 | CH3OH | $11_1 - 10_0$ | 807.866 | 166.37 | 1.65e-03 | | | | | | 50 | | | **B:** CH3OH $(11_1 - 10_0)$ | |
| 809.342 | C | $2 - 1$ | 809.342 | 62.46 | 2.67e-07 | 11.98 | 0.02 | 1.80 | 0.05 | 1.47 | 47 | 4.90 | 0.07 | | |
| 811.445 | CH3OH | $8_{-2} - 7_{-1}$ | 811.445 | 109.49 | 1.88e-03 | 12.38 | 0.03 | 5.11 | 0.06 | 0.57 | 49 | 2.98 | 0.07 | | |
| 812.550 | CH3OH | $7_2 - 6_1$ | 812.550 | 102.70 | 1.92e-03 | 12.16 | 0.03 | 5.05 | 0.06 | 0.54 | 43 | 2.70 | 0.06 | | |
| 812.831 | H2CO | $12_{2,9} - 11_{2,8}$ | 812.831 | 279.73 | 1.57e-02 | 11.93 | 0.10 | 8.71 | 0.24 | 0.09 | 41 | 0.77 | 0.06 | | |
| 813.542 | CH3OH | $17_1 - 16_1$ | 813.542 | 366.29 | 2.45e-03 | 13.96 | 0.33 | 8.69 | 0.82 | 0.14 | 43 | 0.99 | 0.06 | | |
| 815.071 | CH3OH | $6_{-4} - 5_{-3}$ | 815.071 | 136.65 | 3.31e-03 | 11.91 | 0.02 | 4.39 | 0.06 | 0.43 | 41 | 1.76 | 0.06 | | |
| 816.011 | CH3OH | $17_0 - 16_0$ | 816.011 | 366.79 | 2.48e-03 | 12.10 | 0.06 | 5.39 | 0.13 | 0.18 | 46 | 1.20 | 0.07 | | |
| 818.669 | CH3OH | $17_{-1} - 16_{-1}$ | 818.669 | 359.92 | 2.50e-03 | 12.67 | 0.04 | 4.15 | 0.10 | 0.16 | 51 | 1.28 | 0.07 | | |
| 819.479 | CH3OH | $17_0 - 16_0$ | 819.479 | 354.54 | 2.51e-03 | 12.39 | 0.04 | 4.04 | 0.08 | 0.33 | 53 | 1.47 | 0.07 | | |
| 820.499 | CH3OH | $17_2 - 16_2$ | 820.499 | 392.50 | 3.71e-03 | 12.12 | 0.93 | 12.87 | 2.41 | 0.30 | 50 | 0.59 | 0.06 | | |
| 820.762 | CH3OH | $6_3 - 5_2$ | 820.762 | 96.46 | 2.93e-03 | 12.17 | 0.02 | 5.18 | 0.05 | 0.45 | 52 | 2.31 | 0.07 | | |
| 821.698 | CH3OH | $17_3 - 16_3$ | 821.698 | 404.83 | 3.64e-03 | 12.77 | 0.05 | 3.44 | 0.12 | 0.14 | 43 | 0.44 | 0.06 | | |
| 821.869 | CH3OH | $17_1 - 16_1$ | 821.869 | 376.17 | 2.56e-03 | 12.44 | 0.05 | 3.75 | 0.13 | 0.15 | 45 | 0.50 | 0.06 | | |
| 822.301 | CH3OH | $17_2 - 16_2$ | 822.301 | 392.92 | 3.73e-03 | 12.31 | 0.22 | 3.45 | 0.53 | 0.04 | 45 | 0.25 | 0.04 | | |
| 822.540 | CH3OH | $17_2 - 16_2$ | 822.540 | 377.62 | 2.49e-03 | 12.29 | 0.06 | 4.45 | 0.14 | 0.16 | 46 | 0.59 | 0.06 | | |
| 823.083 | H2CO | $11_{1,10} - 10_{1,9}$ | 823.083 | 249.44 | 1.67e-02 | 12.08 | 0.03 | 4.32 | 0.06 | 0.41 | 42 | 1.94 | 0.06 | | |
| 824.343 | CH3OH | $17_{-2} - 16_{-2}$ | 824.343 | 381.52 | 2.53e-03 | 12.73 | 0.06 | 2.85 | 0.14 | 0.14 | 41 | 0.44 | 0.05 | | |
| 825.276 | CH3OH | $14_0 - 13_{-1}$ | 825.276 | 256.14 | 2.28e-03 | 12.17 | 0.04 | 4.45 | 0.09 | 0.31 | 41 | 1.56 | 0.05 | | |
| 827.454 | CH3OH | $17_1 - 16_1$ | 827.454 | 372.37 | 2.58e-03 | 12.68 | 0.06 | 5.47 | 0.14 | 0.19 | 53 | 1.11 | 0.07 | | |
| 829.891 | CH3OH | $4_4 - 3_3$ | 829.891 | 103.56 | 7.63e-03 | | | | | | 50 | | | **B:** CH3OH $(4_4 - 3_3)$ | |
| 829.891 | CH3OH | $4_4 - 3_3$ | 829.891 | 103.56 | 7.63e-03 | | | | | | 50 | | | **B:** CH3OH $(4_4 - 3_3)$ | |
| 830.349 | CH3OH | $7_2 - 6_1$ | 830.349 | 102.72 | 2.05e-03 | 12.04 | 0.02 | 5.30 | 0.06 | 0.60 | 50 | 3.16 | 0.07 | | |
| 831.045 | CH3OH | $12_3 - 11_2$ | 831.045 | 230.83 | 2.14e-03 | 11.99 | 0.04 | 4.97 | 0.09 | 0.36 | 60 | 1.68 | 0.08 | | |
| 832.057 | CS | $17 - 16$ | 832.057 | 359.56 | 1.25e-02 | 12.47 | 0.67 | 12.31 | 1.67 | 0.16 | 56 | 1.15 | 0.10 | | |
| 832.754 | CH3OH | $12_3 - 11_2$ | 832.754 | 230.83 | 2.16e-03 | 12.13 | 0.03 | 3.95 | 0.09 | 0.33 | 48 | 1.11 | 0.06 | | |
| 835.138 | CH3OH | $1_0 - 0_0$ | 835.138 | 40.08 | 6.36e-03 | 9.46 | 0.08 | 6.04 | 0.20 | -0.43 | 40 | -2.81 | 0.05 | | |
| 838.307 | N2H+ | $9 - 8$ | 838.307 | 201.19 | 4.03e-02 | 11.77 | 0.03 | 2.88 | 0.08 | 0.78 | 40 | 2.60 | 0.06 | | |

| ν GHz (1) | Species (2) | Transition (3) | ν_rest GHz (4) | E_u K (5) | A_ul rad/s (6) | Gaussian fit — v_lsr km/s (7) | δv_lsr km/s (8) | FWHM km/s (9) | δFWHM km/s (10) | T_mb K (11) | from moments — RMS mK (12) | Flux K km/s (13) | δFlux K km/s (14) | Blend(B) / Notes (15) |
|---|---|---|---|---|---|---|---|---|---|---|---|---|---|---|
| 838.902 | CH$_3$OH | $22_0 - 21_1$ | 838.902 | 585.57 | $1.99\times10^{-3}$ | 13.44 | 0.07 | 5.05 | 0.16 | 0.09 | 39 | 0.39 | 0.05 | |
| 840.276 | H$_2$CO | $12_{1,12} - 11_{1,11}$ | 840.276 | 274.58 | $1.79\times10^{-2}$ | 12.10 | 0.03 | 5.22 | 0.06 | 0.42 | 36 | 2.40 | 0.05 | |
| 851.415 | CH$_3$OH | $12_1 - 11_0$ | 851.415 | 193.96 | $1.95\times10^{-3}$ | 12.48 | 0.08 | 4.59 | 0.04 | 0.69 | 42 | 3.50 | 0.06 | |
| 852.177 | CH$_3$OH | $17_2 - 16_1$ | 852.177 | 377.62 | $1.65\times10^{-3}$ | 11.88 | 0.07 | 2.85 | 0.18 | 0.11 | 42 | 0.39 | 0.05 | |
| 853.504 | CH$_3$OH | $14_1 - 13_0$ | 853.504 | 264.78 | $1.65\times10^{-3}$ | 12.23 | 0.05 | 5.38 | 0.12 | 0.17 | 42 | 0.79 | 0.06 | |
| 857.959 | CH$_3$OH | $8_2 - 7_1$ | 857.959 | 121.27 | $2.18\times10^{-3}$ | 12.14 | 0.04 | 4.72 | 0.08 | 0.57 | 57 | 2.83 | 0.07 | |
| 860.459 | CH$_3$OH | $9_{-2} - 8_{-1}$ | 860.459 | 130.40 | $2.13\times10^{-3}$ | 12.20 | 0.03 | 4.89 | 0.06 | 0.53 | 55 | 2.77 | 0.07 | |
| 861.186 | CH$_3$OH | $18_1 - 17_1$ | 861.186 | 407.62 | $2.92\times10^{-3}$ | 12.72 | 0.42 | 9.93 | 1.02 | 0.11 | 46 | 0.99 | 0.07 | |
| 863.365 | CH$_3$OH | $7_{-4} - 6_{-3}$ | 863.365 | 152.89 | $3.57\times10^{-3}$ | 11.81 | 0.03 | 5.52 | 0.08 | 0.38 | 47 | 2.19 | 0.06 | |
| 863.431 | CH$_3$OH | $18_0 - 17_0$ | 863.431 | 408.23 | $2.94\times10^{-3}$ | 11.61 | 0.08 | 4.67 | 0.19 | 0.14 | 47 | 0.58 | 0.05 | |
| 866.388 | CH$_3$OH | $18_{-1} - 17_{-1}$ | 866.388 | 401.50 | $2.97\times10^{-3}$ | 12.98 | 0.21 | 7.75 | 0.49 | 0.25 | 44 | 1.68 | 0.06 | |
| 867.327 | CH$_3$OH | $18_0 - 17_0$ | 867.327 | 396.17 | $2.99\times10^{-3}$ | 12.18 | 0.04 | 4.60 | 0.10 | 0.24 | 46 | 0.94 | 0.07 | |
| 869.038 | CH$_3$OH | $7_3 - 6_2$ | 869.038 | 112.71 | $3.21\times10^{-3}$ | 12.35 | 0.04 | 5.12 | 0.06 | 0.45 | 54 | 2.26 | 0.08 | |
| 869.973 | CH$_3$OH | $18_1 - 17_3$ | 869.973 | 446.59 | $4.34\times10^{-3}$ | 12.23 | 0.02 | 5.05 | 0.05 | 0.10 | 50 | 0.47 | 0.05 | |
| 870.273 | H$_2$CO | $12_{2,11} - 11_{2,10}$ | 870.273 | 319.16 | $1.94\times10^{-2}$ | 10.94 | 0.08 | 3.73 | 0.18 | 0.15 | 50 | 0.78 | 0.06 | |
| 870.664 | CH$_3$OH | $18_2 - 17_2$ | 870.664 | 419.41 | $4.39\times10^{-3}$ | | | | | | 52 | | | B: CH$_3$OH ($18_3 - 17_3$) |
| 870.691 | CH$_3$OH | $18_3 - 17_3$ | 870.691 | 444.68 | $4.36\times10^{-3}$ | | | | | | 52 | | | B: CH$_3$OH ($18_2 - 17_2$) |
| 871.152 | CH$_3$OH | $15_0 - 14_{-1}$ | 871.152 | 290.74 | $2.84\times10^{-3}$ | | | | | | 53 | | | |
| 873.036 | CCH | $10_{21/2,11} - 9_{19/2,9}$ | 873.036 | 230.49 | $2.18\times10^{-3}$ | 12.73 | 0.03 | 4.79 | 0.07 | 0.32 | 49 | 1.75 | 0.07 | B: CCH ($10_{21/2,11} - 9_{19/2,10}$) |
| 873.036 | CCH | $10_{21/2,11} - 9_{19/2,10}$ | 873.036 | 230.49 | $2.19\times10^{-3}$ | | | | | | 38 | | | B: CCH ($10_{21/2,11} - 9_{19/2,9}$) |
| 873.138 | CH$_3$OH | $18_{-2} - 17_{-2}$ | 873.138 | 423.43 | $3.01\times10^{-3}$ | 12.75 | 0.06 | 8.22 | 0.15 | 0.14 | 39 | 0.95 | 0.06 | |
| 873.775 | H$_2$CO | $12_{2,8} - 11_{3,7}$ | 873.775 | 566.55 | $1.67\times10^{-2}$ | | | | | | 34 | | | B: H$_2$CO ($12_{2,7} - 11_{5,6}$) |
| 873.776 | H$_2$CO | $12_{2,7} - 11_{5,6}$ | 873.776 | 566.55 | $1.67\times10^{-2}$ | | | | | | 35 | | | B: H$_2$CO ($12_{2,8} - 11_{5,7}$) |
| 875.366 | CH$_3$OH | $12_{3,10} - 11_{3,9}$ | 875.366 | 378.88 | $1.91\times10^{-2}$ | 12.26 | 0.05 | 4.08 | 0.13 | 0.14 | 37 | 0.55 | 0.05 | |
| 875.735 | CH$_3$OH | $21_{-1} - 20_0$ | 875.735 | 539.96 | $3.97\times10^{-3}$ | 12.58 | 0.07 | 4.12 | 0.17 | 0.08 | 36 | 0.38 | 0.04 | |
| 875.852 | CH$_3$OH | $18_1 - 17_1$ | 875.852 | 414.40 | $3.07\times10^{-3}$ | 12.56 | 0.05 | 3.54 | 0.12 | 0.16 | 38 | 0.56 | 0.05 | |
| 876.649 | H$_2$CO | $12_{2,9} - 11_{3,8}$ | 876.649 | 379.03 | $1.92\times10^{-2}$ | 12.04 | 0.08 | 4.84 | 0.18 | 0.12 | 38 | 0.68 | 0.05 | |
| 877.922 | C$^{18}$O | $8 - 7$ | 877.922 | 189.64 | $4.47\times10^{-5}$ | 11.44 | 0.03 | 3.23 | 0.08 | 0.34 | 37 | 1.38 | 0.05 | |
| 878.226 | CH$_3$OH | $5_4 - 4_3$ | 878.226 | 115.16 | $7.60\times10^{-3}$ | | | | | | 37 | | | B: CH$_3$OH ($5_4 - 4_3$) |
| 878.227 | CH$_3$OH | $5_4 - 4_3$ | 878.227 | 115.16 | $7.60\times10^{-3}$ | | | | | | 40 | | | B: CH$_3$OH ($5_4 - 4_3$) |
| 879.013 | CH$_3$OH | $13_1 - 12_2$ | 879.013 | 260.99 | $2.48\times10^{-3}$ | 12.19 | 0.04 | 4.77 | 0.08 | 0.23 | 37 | 1.05 | 0.05 | |
| 880.899 | CS | $18 - 17$ | 880.899 | 401.83 | $1.49\times10^{-2}$ | 12.37 | 0.06 | 12.56 | 1.00 | 0.14 | 39 | 1.25 | 0.06 | |
| 881.273 | $^{13}$CO | $8 - 7$ | 881.273 | 190.36 | $4.52\times10^{-5}$ | 11.29 | 0.02 | 5.07 | 0.04 | 2.95 | 42 | 15.76 | 0.08 | |
| 881.421 | CH$_3$OH | $13_3 - 12_2$ | 881.421 | 261.00 | $2.50\times10^{-3}$ | 11.96 | 0.03 | 3.72 | 0.07 | 0.28 | 38 | 1.21 | 0.05 | |
| 881.782 | CH$_3$OH | $8_2 - 7_1$ | 881.782 | 121.29 | $2.37\times10^{-3}$ | 11.94 | 0.02 | 4.76 | 0.28 | 0.63 | 35 | 3.28 | 0.05 | |
| 885.971 | HCN | $10 - 9$ | 885.971 | 233.90 | $3.44\times10^{-2}$ | 12.56 | 0.01 | 12.24 | 0.03 | 1.09 | 40 | 14.07 | 0.09 | |
| 888.629 | H$_2$CO | $12_{2,10} - 11_{2,9}$ | 888.629 | 322.37 | $2.07\times10^{-2}$ | 11.90 | 0.42 | 7.51 | 0.99 | 0.07 | 37 | 0.47 | 0.04 | |
| 891.557 | HCO$^+$ | $10 - 9$ | 891.557 | 235.38 | $5.97\times10^{-2}$ | 11.42 | 0.08 | 5.52 | 0.05 | 1.68 | 38 | 11.66 | 0.07 | |
| 893.639 | HDO | $1_{1,1} - 0_{0,0}$ | 893.639 | 42.89 | $8.35\times10^{-3}$ | 13.42 | 0.08 | 2.47 | 0.18 | 0.10 | 37 | 0.41 | 0.05 | |
| 894.614 | CH$_3$OH | $13_1 - 12_0$ | 894.614 | 223.84 | $2.27\times10^{-3}$ | 12.49 | 0.08 | 4.44 | 0.05 | 0.57 | 36 | 2.76 | 0.05 | |
| 896.805 | H$_2$CO | $12_{2,11} - 11_{11,10}$ | 896.805 | 292.48 | $2.17\times10^{-2}$ | 12.01 | 0.03 | 5.10 | 0.22 | 0.29 | 33 | 1.75 | 0.05 | |
| 900.972 | CH$_3$OH | $12_7 - 11_7$ | 900.972 | 419.41 | $1.96\times10^{-3}$ | 11.65 | 0.02 | 4.10 | 0.08 | 0.08 | 36 | 0.37 | 0.05 | |
| 902.935 | CH$_3$OH | $9_2 - 8_1$ | 902.935 | 142.15 | $2.47\times10^{-3}$ | 12.36 | 0.02 | 5.24 | 0.05 | 0.45 | 36 | 2.41 | 0.06 | |
| 905.404 | CH$_3$OH | $15_1 - 14_0$ | 905.404 | 299.59 | $1.94\times10^{-3}$ | 11.90 | 0.06 | 4.81 | 0.14 | 0.17 | 52 | 0.77 | 0.06 | |

| ν GHz (1) | Species (2) | Transition (3) | ν_rest GHz (4) | $E_u$ K (5) | $A_{ul}$ rad/s (6) | $v_{lsr}$ km/s (7) | $\delta v_{lsr}$ km/s (8) | FWHM km/s (9) | $\delta$FWHM km/s (10) | $T_{mb}$ K (11) | RMS mK (12) | Flux K km/s (13) | $\delta$Flux K km/s (14) | Blend(**B**) / Notes (15) |
|---|---|---|---|---|---|---|---|---|---|---|---|---|---|---|
| 906.593 | CN | $8_{15/2,17/2} - 7_{13/2,15/2}$ | 906.593 | 195.81 | 8.51e-03 | | | | | | 53 | | | **B:** CN $7_{13/2,11/2}$ ($8_{15/2,13/2}$) — |
| 906.593 | CN | $8_{15/2,17/2} - 7_{13/2,11/2}$ | 906.593 | 195.81 | 8.34e-03 | | | | | | 53 | | | **B:** CN $7_{13/2,15/2}$ ($8_{15/2,17/2}$) — |
| 909.045 | OH+ | $1_{0,1/2} - 0_{1,1/2}$ | 909.045 | 43.63 | 5.23e-03 | 8.75 | 0.21 | 3.45 | 0.62 | -0.08 | 40 | -0.44 | 0.05 | |
| 909.159 | OH+ | $1_{0,1/2} - 0_{1,3/2}$ | 909.159 | 43.63 | 1.05e-02 | 9.36 | 0.07 | 2.74 | 0.16 | -0.20 | 40 | -0.72 | 0.05 | |
| 909.508 | H2CO | $13_{1,13} - 12_{1,12}$ | 909.508 | 318.23 | 2.28e-02 | 12.12 | 0.03 | 4.33 | 0.07 | 0.30 | 39 | 0.89 | 0.05 | |
| 909.738 | CH3OH | $10_{-2} - 9_{-1}$ | 909.738 | 153.63 | 2.38e-03 | 12.36 | 0.02 | 4.97 | 0.06 | 0.45 | 36 | 2.16 | 0.05 | |
| 910.808 | CH3OH | $19_0 - 18_0$ | 910.808 | 451.94 | 3.45e-03 | 12.39 | 0.07 | 6.07 | 0.17 | 0.14 | 37 | 1.01 | 0.05 | |
| 911.642 | CH3OH | $8_{-4} - 7_{-3}$ | 911.642 | 171.46 | 3.89e-03 | 11.84 | 0.02 | 5.29 | 0.05 | 0.37 | 44 | 2.10 | 0.06 | |
| 912.109 | CH3OH | $3_{-3} - 2_{-2}$ | 912.109 | 76.64 | 3.49e-03 | 12.18 | 0.02 | 6.01 | 0.05 | 0.45 | 39 | 2.97 | 0.05 | |
| 914.044 | CH3OH | $19_{-1} - 18_{-1}$ | 914.044 | 445.37 | 3.49e-03 | 12.37 | 0.03 | 4.39 | 0.08 | 0.23 | 38 | 1.06 | 0.05 | |
| 915.117 | CH3OH | $19_0 - 18_0$ | 915.117 | 440.09 | 3.51e-03 | 12.34 | 0.05 | 4.04 | 0.12 | 0.16 | 43 | 0.75 | 0.06 | |
| 916.652 | CH3OH | $16_0 - 15_{-1}$ | 916.652 | 327.63 | 3.49e-03 | 12.52 | 0.04 | 3.78 | 0.08 | 0.29 | 47 | 1.01 | 0.06 | |
| 917.270 | CH3OH | $8_3 - 7_2$ | 917.270 | 131.28 | 3.53e-03 | 12.67 | 0.03 | 6.11 | 0.06 | 0.42 | 50 | 2.46 | 0.07 | |
| 918.699 | CH3OH | $19_2 - 18_2$ | 918.699 | 463.50 | 3.49e-03 | 11.83 | 0.10 | 4.01 | 0.23 | 0.49 | 51 | 0.36 | 0.06 | |
| 921.800 | CO | $8 - 7$ | 921.800 | 199.11 | 5.13e-05 | 11.27 | 0.02 | 12.91 | 0.05 | 23.56 | 48 | 248.50 | 0.12 | |
| 923.588 | H2CO | $13_{0,13} - 12_{0,12}$ | 923.588 | 313.69 | 2.39e-02 | 12.58 | 0.02 | 4.23 | 0.06 | 0.37 | 37 | 2.10 | 0.06 | |
| 924.198 | CH3OH | $19_1 - 18_1$ | 924.198 | 458.76 | 3.61e-03 | 12.58 | 0.07 | 5.26 | 0.16 | 0.13 | 39 | 0.63 | 0.05 | |
| 926.555 | CH3OH | $6_4 - 5_3$ | 926.555 | 129.09 | 7.85e-03 | | | | | | 39 | | | **B:** CH3OH ($6_4 - 5_3$) |
| 926.555 | CH3OH | $6_4 - 5_3$ | 926.555 | 129.09 | 7.85e-03 | | | | | | 40 | | | **B:** CH3OH ($6_4 - 5_3$) |
| 926.893 | CH3OH | $14_3 - 13_2$ | 926.893 | 293.47 | 4.20e-03 | 11.38 | 0.04 | 5.90 | 0.10 | 0.25 | 43 | 1.57 | 0.06 | |
| 929.723 | CS | $19 - 18$ | 929.723 | 446.45 | 1.75e-02 | 12.50 | 0.43 | 9.64 | 1.02 | 0.12 | 45 | 0.95 | 0.08 | |
| 930.198 | CH3OH | $14_3 - 13_2$ | 930.198 | 293.48 | 2.87e-03 | 12.25 | 0.05 | 4.19 | 0.12 | 0.12 | 44 | 0.56 | 0.07 | |
| 931.386 | N2H+ | $10 - 9$ | 931.386 | 245.89 | 5.56e-02 | 11.65 | 0.07 | 4.01 | 0.17 | 0.62 | 45 | 3.03 | 0.06 | |
| 933.693 | CH3OH | $9_2 - 8_1$ | 933.693 | 142.19 | 2.73e-03 | 11.84 | 0.02 | 5.21 | 0.06 | 0.48 | 48 | 2.52 | 0.07 | |
| 937.478 | CH3OH | $14_1 - 13_0$ | 937.478 | 256.02 | 2.64e-03 | 12.46 | 0.03 | 5.78 | 0.05 | 0.40 | 40 | 2.94 | 0.06 | |
| 947.476 | CH3OH | $10_2 - 9_1$ | 947.476 | 165.35 | 2.78e-03 | 12.27 | 0.09 | 4.42 | 0.20 | 0.10 | 61 | 1.69 | 0.07 | |
| 948.454 | H2CO | $13_{3,11} - 12_{3,10}$ | 948.454 | 424.40 | 2.46e-02 | 12.28 | 0.18 | 4.27 | 0.42 | 0.10 | 50 | 0.51 | 0.07 | |
| 950.365 | H2CO | $13_{3,10} - 12_{3,9}$ | 950.365 | 424.65 | 2.47e-02 | 11.88 | 0.38 | 4.86 | 0.15 | 0.10 | 51 | 0.41 | 0.06 | |
| 959.346 | CH3OH | $11_{-2} - 10_{-1}$ | 959.346 | 179.19 | 2.65e-03 | 13.42 | 0.06 | 3.57 | 0.11 | 0.35 | 134 | 1.71 | 0.19 | |
| 959.901 | CH3OH | $4_{-3} - 3_{-2}$ | 959.901 | 192.34 | 4.26e-03 | 12.86 | 0.05 | 9.23 | 1.02 | 0.34 | 117 | 1.50 | 0.17 | |
| 960.471 | CH3OH | $4_3 - 3_3$ | 960.471 | 85.92 | 5.57e-03 | 11.70 | 0.05 | 7.30 | 0.11 | 0.44 | 110 | 2.51 | 0.15 | |
| 961.635 | CH3OH | $17_0 - 16_{-1}$ | 961.635 | 491.52 | 4.07e-03 | 13.44 | 0.07 | 3.60 | 0.16 | 0.22 | 74 | 0.50 | 0.17 | |
| 961.777 | CH3OH | $20_0 - 19_0$ | 961.777 | 366.79 | 4.25e-03 | 12.71 | 0.07 | 4.27 | 0.07 | 0.23 | 79 | 0.77 | 0.09 | |
| 962.848 | CH3OH | $20_{-1} - 19_{-1}$ | 962.848 | 486.30 | 4.10e-03 | 11.83 | 0.03 | 3.68 | 0.14 | 0.45 | 57 | 1.14 | 0.09 | |
| 965.448 | CH3OH | $9_3 - 8_2$ | 965.448 | 152.17 | 3.90e-03 | 12.52 | 0.05 | 5.11 | 0.07 | 0.45 | 50 | 2.53 | 0.07 | |
| 966.507 | CH3OH | $20_3 - 19_3$ | 966.507 | 537.04 | 6.00e-03 | | | | | | 50 | | | **B:** CH3OH ($20_1 - 19_1$) |
| 966.515 | CH3OH | $20_1 - 19_1$ | 966.515 | 508.38 | 6.19e-03 | | | | | | 52 | | | **B:** CH3OH ($20_3 - 19_3$) |
| 970.199 | H2CO | $13_{1,12} - 12_{1,11}$ | 970.199 | 339.05 | 2.76e-02 | 10.78 | 0.31 | 4.87 | 0.72 | 0.14 | 82 | 0.67 | 0.11 | |
| 971.804 | OH+ | $2_{5/2} - 0_{1,3/2}$ | 971.804 | 46.64 | 1.82e-02 | | | | | | 108 | | | **B:** OH+ ($2_{3/2} - 0_{1,1/2}$) |
| 971.805 | OH+ | $2_{3/2} - 0_{1,1/2}$ | 971.805 | 46.65 | 1.52e-02 | | | | | | 107 | | | **B:** OH+ ($2_{5/2} - 0_{1,3/2}$) |
| 972.489 | CH3OH | $20_{-1} - 19_1$ | 972.489 | 505.43 | 4.21e-03 | 12.78 | 0.08 | 2.83 | 0.18 | 0.19 | 99 | 0.60 | 0.11 | |
| 974.462 | NH | $1_{2,5/2,3/2} - 0_{1,3/2,1/2}$ | 974.462 | 46.77 | 5.04e-03 | | | | | | 55 | | | **B:** NH ($1_{2,3/2,3/2} - 0_{1,1/2,1/2}$) |
| 974.471 | NH | $1_{2,5/2,3/2} - 0_{1,3/2,3/2}$ | 974.471 | 46.77 | 5.66e-03 | | | | | | 55 | | | **B:** NH ($1_{2,3/2,3/2} - 0_{1,1/2,3/2}$) |

| $\nu$ GHz (1) | Species (2) | Transition (3) | $\nu_{rest}$ GHz (4) | $E_u$ K (5) | $A_{ul}$ rad/s (6) | Gaussian fit $v_{lsr}$ km/s (7) | $\delta v_{lsr}$ km/s (8) | FWHM km/s (9) | $\delta$FWHM km/s (10) | $T_{mb}$ K (11) | from moments RMS mK (12) | Flux K km/s (13) | $\delta$Flux K km/s (14) | Blend(**B**) / Notes (15) |
|---|---|---|---|---|---|---|---|---|---|---|---|---|---|---|
| 974.475 | NH | $1_{2,3/2,3/2} - 0_{1,1/2,1/2}$ | 974.475 | 46.77 | 3.38e-03 | | | | | | 57 | | | **B:** NH ($1_{2,5/2,3/2} - 0_{0,3/2,1/2}$) |
| 974.478 | NH | $1_{2,5/2,7/2} - 0_{1,3/2,5/2}$ | 974.478 | 46.77 | 6.94e-03 | | | | | | 55 | | | **B:** NH ($1_{2,5/2,3/2} - 0_{0,3/2,1/2}$) |
| 974.479 | NH | $1_{2,3/2,5/2} - 0_{1,3/2,3/2}$ | 974.479 | 46.78 | 6.01e-03 | | | | | | 56 | | | **B:** NH ($1_{2,5/2,3/2} - 0_{0,3/2,1/2}$) |
| 974.487 | HCN | $11 - 10$ | 974.487 | 280.67 | 4.59e-02 | | | | | | 55 | | | **B:** NH ($1_{2,5/2,3/2} - 0_{0,3/2,1/2}$) |
| 974.673 | CH$_3$OH | $15_3 - 14_2$ | 974.673 | 328.26 | 3.24e-03 | 11.58 | 0.04 | 2.14 | 0.10 | 0.19 | 57 | 0.70 | 0.06 | **B:** CH$_3$OH ($7_4 - 6_3$) |
| 974.877 | CH$_3$OH | $7_4 - 6_3$ | 974.877 | 145.33 | 8.28e-03 | | | | | | 56 | | | **B:** CH$_3$OH ($7_4 - 6_3$) |
| 974.878 | CH$_3$OH | $7_4 - 6_3$ | 974.878 | 145.33 | 8.28e-03 | | | | | | 57 | | | **B:** H$_2$CO ($14_{1,14} - 13_{1,13}$) |
| 978.529 | CS | $20 - 19$ | 978.529 | 493.42 | 2.04e-02 | | | | | | 51 | | | **B:** CS ($20 - 19$) |
| 978.592 | H$_2$CO | $14_{1,14} - 13_{1,13}$ | 978.592 | 365.20 | 2.84e-02 | | | | | | 56 | | | |
| 979.110 | CH$_3$OH | $15_3 - 14_2$ | 979.110 | 328.28 | 3.28e-03 | 12.45 | 0.05 | 4.62 | 0.13 | 0.20 | 53 | 1.08 | 0.07 | |
| 980.025 | CH$_3$OH | $15_1 - 14_0$ | 980.025 | 290.49 | 3.04e-03 | 12.68 | 0.03 | 4.52 | 0.06 | 0.44 | 61 | 2.15 | 0.09 | |
| 980.636 | HCO$^+$ | $11 - 10$ | 980.636 | 282.44 | 7.98e-02 | 11.44 | 0.02 | 6.14 | 0.06 | 1.23 | 54 | 9.31 | 0.10 | |
| 986.098 | CH$_3$OH | $10_2 - 9_1$ | 986.098 | 165.40 | 3.13e-03 | 12.48 | 0.03 | 4.40 | 0.07 | 0.48 | 63 | 1.87 | 0.09 | |
| 987.560 | C$^{34}$S | $9 - 8$ | 987.560 | 237.03 | 6.38e-05 | 11.01 | 0.14 | 3.37 | 0.32 | 0.23 | 54 | 0.96 | 0.07 | |
| 987.927 | H$_2$O | $2_{0,2,0} - 1_{1,1,0}$ | 987.927 | 100.85 | 5.85e-03 | 12.37 | 0.02 | 16.39 | 0.04 | 4.34 | 72 | 66.59 | 0.18 | |
| 991.329 | $^{13}$CO | $9 - 8$ | 991.329 | 237.94 | 6.45e-05 | 11.32 | 0.02 | 6.92 | 0.04 | 1.82 | 65 | 11.49 | 0.11 | |
| 991.580 | CH$_3$OH | $11_2 - 10_1$ | 991.580 | 190.87 | 3.12e-03 | 12.27 | 0.04 | 5.28 | 0.09 | 0.39 | 59 | 1.90 | 0.07 | |
| 991.660 | H$_2$CO | $14_{0,14} - 13_{0,13}$ | 991.660 | 361.28 | 2.97e-02 | 12.21 | 0.19 | 2.46 | 0.45 | 0.56 | 56 | 0.24 | 0.05 | |
| 993.108 | H$_2$S | $3_{0,3} - 2_{1,2}$ | 993.108 | 102.76 | 3.73e-03 | 11.67 | 0.03 | 5.99 | 0.08 | 0.41 | 64 | 3.41 | 0.11 | |
| 994.217 | CH$_3$OH | $5_{-5} - 4_{-4}$ | 994.217 | 158.83 | 9.26e-03 | 12.18 | 0.06 | 6.32 | 0.14 | 0.17 | 64 | 0.99 | 0.08 | |
| 995.492 | o-NH$_3$ | $4_{0,1} - 4_{3,1}$ | 995.492 | 285.60 | 3.87e-08 | 13.49 | 0.09 | 0.94 | 0.21 | 0.17 | 80 | 1.23 | 0.12 | |
| 997.873 | CH$_3$OH | $20_0 - 19_1$ | 997.873 | 509.89 | 2.76e-03 | 12.74 | 0.10 | 6.56 | 0.23 | 0.17 | 68 | 0.72 | 0.07 | |
| 1002.779 | H$_2$S | $3_{1,3} - 2_{0,2}$ | 1002.779 | 102.82 | 3.87e-03 | 10.88 | 0.24 | 3.42 | 0.56 | 0.16 | 50 | 0.63 | 0.05 | |
| 1005.455 | CH$_3$OH | $21_0 - 20_0$ | 1005.455 | 546.18 | 4.66e-03 | 14.58 | 0.10 | 0.44 | 0.22 | 0.16 | 64 | 0.46 | 0.07 | |
| 1005.642 | CH$_3$OH | $10_4 - 10_3$ | 1005.642 | 223.65 | 5.06e-03 | 12.41 | 0.07 | 7.31 | 0.16 | 0.12 | 60 | 0.78 | 0.07 | |
| 1006.028 | CH$_3$OH | $7_4 - 7_3$ | 1006.028 | 160.99 | 4.45e-03 | | | | | | 63 | | | **B:** CH$_3$OH ($6_4 - 6_3$) |
| 1006.081 | CH$_3$OH | $6_4 - 6_3$ | 1006.081 | 144.75 | 4.04e-03 | | | | | | 63 | | | **B:** CH$_3$OH ($7_4 - 7_3$) |
| 1006.111 | CH$_3$OH | $5_4 - 5_3$ | 1006.111 | 130.82 | 3.40e-03 | | | | | | 64 | | | **B:** CH$_3$OH ($7_4 - 7_3$) |
| 1006.125 | CH$_3$OH | $4_4 - 4_3$ | 1006.125 | 119.21 | 2.26e-03 | | | | | | 60 | | | **B:** CH$_3$OH ($7_4 - 7_3$) |
| 1006.540 | CH$_3$OH | $18_0 - 17_{-1}$ | 1006.540 | 408.23 | 5.12e-03 | 13.49 | 0.06 | 5.14 | 0.13 | 0.25 | 67 | 1.09 | 0.08 | |
| 1008.139 | CH$_3$OH | $10_{-4} - 9_{-3}$ | 1008.139 | 215.55 | 4.69e-03 | 12.16 | 0.07 | 9.13 | 0.15 | 0.26 | 84 | 1.75 | 0.09 | |
| 1008.812 | CH$_3$OH | $5_{-4} - 4_{-3}$ | 1008.812 | 97.53 | 5.63e-03 | 11.93 | 0.04 | 4.85 | 0.09 | 0.47 | 79 | 2.02 | 0.12 | |
| 1009.358 | CH$_3$OH | $12_{-2} - 11_{-1}$ | 1009.358 | 207.08 | 2.91e-03 | 12.22 | 0.05 | 4.80 | 0.12 | 0.29 | 84 | 1.32 | 0.10 | |
| 1011.325 | CH$_3$OH | $17_1 - 16_0$ | 1011.325 | 376.17 | 2.60e-03 | 13.22 | 0.07 | 4.04 | 0.17 | 0.15 | 66 | 0.61 | 0.08 | |
| 1013.563 | CH$_3$OH | $10_3 - 9_2$ | 1013.563 | 175.39 | 4.29e-03 | | | | | | 57 | | | **B:** CH$_3$OH ($10_3 - 9_2$) |
| 1013.563 | CH$_3$OH | $10_3 - 9_2$ | 1013.563 | 175.39 | 4.29e-03 | | | | | | 57 | | | **B:** CH$_3$OH ($10_3 - 9_2$) |
| 1020.720 | CH$_3$OH | $21_1 - 20_1$ | 1020.720 | 554.42 | 4.88e-03 | 12.61 | 0.05 | 4.32 | 0.12 | 0.16 | 62 | 1.00 | 0.08 | |
| 1022.271 | CH$_3$OH | $16_1 - 15_0$ | 1022.271 | 327.24 | 3.49e-03 | 12.53 | 0.04 | 5.29 | 0.08 | 0.30 | 86 | 1.76 | 0.10 | |
| 1022.338 | CH$_3$OH | $16_1 - 15_2$ | 1022.338 | 365.37 | 3.67e-03 | 12.88 | 0.06 | 5.12 | 0.14 | 0.16 | 106 | 0.64 | 0.13 | |
| 1023.193 | CH$_3$OH | $8_4 - 7_3$ | 1023.193 | 163.90 | 8.84e-03 | | | | | | 63 | | | **B:** CH$_3$OH ($8_4 - 7_3$) |
| 1023.197 | CH$_3$OH | $8_4 - 7_3$ | 1023.197 | 163.90 | 8.84e-03 | | | | | | 94 | | | **B:** CH$_3$OH ($8_4 - 7_3$) |
| 1024.443 | N$_2$H$^+$ | $11 - 10$ | 1024.443 | 295.06 | 7.43e-02 | 11.97 | 0.04 | 3.30 | 0.09 | 0.41 | 94 | 1.78 | 0.11 | |
| 1027.314 | CS | $21 - 20$ | 1027.314 | 542.72 | 2.37e-02 | 11.04 | 0.45 | 7.50 | 1.08 | 0.15 | 81 | 0.98 | 0.12 | |
| 1028.180 | CH$_3$OH | $16_5 - 15_2$ | 1028.180 | 365.40 | 3.73e-03 | 13.00 | 0.07 | 8.45 | 0.17 | 0.20 | 79 | 1.42 | 0.11 | |
| 1032.998 | OH$^+$ | $1_{1,1/2} - 0_{1,1/2}$ | 1032.998 | 49.58 | 1.41e-02 | | | | | | 70 | | | **B:** OH$^+$ ($1_{1,3/2} - 0_{1,1/2}$) |

Columns 7–11 are grouped under **Gaussian fit**; columns 12–14 are grouped under **from moments**.

| $\nu$ (GHz) | Species | Transition | $\nu_{rest}$ (GHz) | $E_u$ (K) | $A_{ul}$ (rad/s) | $v_{lsr}$ (km/s) | $\delta v_{lsr}$ (km/s) | FWHM (km/s) | $\delta$FWHM (km/s) | $T_{mb}$ (K) | RMS (mK) | Flux (K km/s) | $\delta$Flux (K km/s) | Blend(B) / Notes |
|---|---|---|---|---|---|---|---|---|---|---|---|---|---|---|
| 1033.004 | OH$^+$ | $1_{1,3/2}-0_{1,1/2}$ | 1033.004 | 49.58 | 3.53e−03 | | | | | | 70 | | | B: OH$^+$ ($1_{1,1/2}-0_{1,1/2}$) — |
| 1033.112 | OH$^+$ | $1_{1,1/2}-0_{1,1/2}$ | 1033.112 | 49.58 | 7.03e−03 | | | | | | 74 | | | B: OH$^+$ ($1_{1,1/2}-0_{1,3/2}$) — |
| 1033.119 | OH$^+$ | $1_{1,3/2}-0_{1,3/2}$ | 1033.119 | 49.58 | 1.76e−02 | | | | | | 75 | | | B: OH$^+$ ($1_{1,1/2}-0_{1,3/2}$) — |
| 1035.247 | CH$_3$OH | $12_2-11_1$ | 1035.247 | 218.69 | 3.48e−03 | 12.07 | 0.06 | 5.07 | 0.15 | 0.28 | 99 | 1.41 | 0.12 | — |
| 1036.912 | CO | $9-8$ | 1036.912 | 248.88 | 7.33e−05 | 11.27 | 0.02 | 12.44 | 0.04 | 22.64 | 188 | 257.05 | 0.48 | |
| 1039.015 | CH$_3$OH | $11_2-10_1$ | 1039.015 | 190.94 | 3.59e−03 | 12.31 | 0.07 | 5.07 | 0.16 | 0.40 | 249 | 1.57 | 0.31 | |
| 1042.522 | CH$_3$OH | $6_{-5}-5_{-4}$ | 1042.522 | 172.76 | 9.19e−03 | 11.24 | 0.06 | 2.30 | 0.14 | 0.30 | 108 | 0.41 | 0.13 | |
| 1047.427 | CCH | $12_{25/2,13}-11_{23/2,12}$ | 1047.427 | 326.85 | 3.81e−03 | | | | | | 105 | | | B: CCH ($12_{25/2,11}$) $11_{23/2,11}$ |
| 1047.427 | CCH | $12_{25/2,12}-11_{23/2,11}$ | 1047.427 | 326.85 | 3.79e−03 | | | | | | 98 | | | B: CCH ($12_{25/2,12}$) $11_{23/2,12}$ |
| 1047.490 | CCH | $12_{23/2,12}-11_{21/2,11}$ | 1047.490 | 326.88 | 3.79e−03 | | | | | | 93 | | | B: CCH ($12_{25/2,12}$) $11_{23/2,12}$ |
| 1047.490 | CCH | $12_{23/2,11}-11_{21/2,10}$ | 1047.490 | 326.88 | 3.78e−03 | | | | | | 92 | | | B: CCH ($12_{25/2,12}$) $11_{23/2,12}$ |
| 1056.355 | CH$_3$OH | $11_{-4}-10_{-3}$ | 1056.355 | 241.07 | 5.17e−03 | 12.57 | 0.04 | 5.51 | 0.11 | 0.28 | 72 | 1.35 | 0.09 | |
| 1057.118 | CH$_3$OH | $6_{-3}-5_{-2}$ | 1057.118 | 111.46 | 5.86e−03 | 11.90 | 0.03 | 5.93 | 0.07 | 0.44 | 75 | 2.57 | 0.10 | |
| 1059.859 | CH$_3$OH | $13_{-2}-12_{-1}$ | 1059.859 | 237.30 | 3.16e−03 | 12.09 | 0.06 | 2.65 | 0.14 | 0.31 | 73 | 1.41 | 0.14 | |
| 1061.609 | CH$_3$OH | $11_3-10_2$ | 1061.609 | 200.92 | 4.72e−03 | 13.11 | 0.23 | 6.04 | 0.54 | 0.27 | 92 | 1.49 | 0.15 | |
| 1062.981 | HCN | $12-11$ | 1062.981 | 331.69 | 5.98e−02 | 12.39 | 0.04 | 14.47 | 0.09 | 0.63 | 124 | 9.19 | 0.22 | |
| 1064.237 | CH$_3$OH | $17_1-16_0$ | 1064.237 | 366.29 | 3.98e−03 | 12.44 | 0.05 | 4.49 | 0.12 | 0.29 | 97 | 1.01 | 0.13 | |
| 1069.694 | HCO$^+$ | $12-11$ | 1069.694 | 333.78 | 1.04e−01 | 11.54 | 0.02 | 5.94 | 0.06 | 0.92 | 56 | 5.86 | 0.07 | |
| 1069.874 | CH$_3$OH | $17_3-16_2$ | 1069.874 | 404.80 | 4.14e−03 | 12.86 | 0.06 | 4.01 | 0.15 | 0.14 | 63 | 0.53 | 0.06 | |
| 1071.505 | CH$_3$OH | $9_4-8_3$ | 1071.505 | 184.79 | 9.51e−03 | | | | | | 65 | | | |
| 1071.514 | CH$_3$OH | $9_4-8_3$ | 1071.514 | 184.79 | 9.52e−03 | | | | | | 66 | | | |
| 1072.837 | H$_2$S | $2_{2,1}-1_{1,0}$ | 1072.837 | 79.37 | 4.13e−03 | 12.33 | 0.14 | 7.48 | 0.33 | 0.31 | 67 | 2.29 | 0.09 | |
| 1076.834 | CH$_3$OH | $5_{-1}-4_4$ | 1076.834 | 170.89 | 1.07e−02 | 12.37 | 0.06 | 3.69 | 0.13 | 0.28 | 87 | 1.01 | 0.12 | |
| 1077.437 | CH$_3$OH | $7_{-3}-6_{-2}$ | 1077.437 | 404.83 | 6.26e−03 | 12.68 | 0.28 | 4.80 | 0.67 | 0.16 | 81 | 0.72 | 0.10 | |
| 1078.477 | CH$_3$OH | $13_3-12_1$ | 1078.477 | 248.84 | 3.88e−03 | 12.31 | 0.04 | 5.31 | 0.10 | 0.28 | 80 | 1.27 | 0.10 | |
| 1090.816 | CH$_3$OH | $7_{-5}-6_{-4}$ | 1090.816 | 189.00 | 9.37e−03 | 12.40 | 0.04 | 6.75 | 0.18 | 0.16 | 84 | 1.18 | 0.09 | B: CH$_3$OH ($9_4-8_3$) |
| 1092.463 | CH$_3$OH | $12_2-11_1$ | 1092.463 | 218.80 | 4.09e−03 | 11.86 | 0.04 | 4.63 | 0.18 | 0.35 | 90 | 1.66 | 0.10 | B: CH$_3$OH ($9_4-8_3$) |
| 1095.063 | CH$_3$OH | $20_6-19_1$ | 1095.063 | 497.33 | 7.23e−03 | 12.40 | 0.02 | 4.12 | 0.18 | 0.15 | 83 | 0.56 | 0.08 | |
| 1097.365 | H$_2$O | $3_{1,2,0}-3_{0,3,0}$ | 1097.365 | 249.44 | 1.64e−02 | 12.66 | 0.02 | 15.63 | 0.04 | 2.17 | 87 | 32.89 | 0.19 | |
| 1101.350 | $^{13}$CO | $10-9$ | 1101.350 | 290.79 | 8.86e−05 | 11.34 | 0.06 | 7.95 | 0.13 | 1.26 | 92 | 9.39 | 0.13 | |
| 1104.547 | CH$_3$OH | $12_{-4}-11_{-3}$ | 1104.547 | 268.91 | 5.69e−03 | 12.42 | 0.06 | 4.69 | 0.13 | 0.20 | 92 | 0.95 | 0.10 | |
| 1105.370 | CH$_3$OH | $18_1-17_0$ | 1105.370 | 127.71 | 6.21e−03 | 12.43 | 0.05 | 5.05 | 0.09 | 0.43 | 85 | 1.94 | 0.10 | |
| 1105.944 | CH$_3$OH | $12_3-11_2$ | 1105.944 | 407.62 | 4.51e−03 | 13.06 | 0.23 | 3.33 | 0.12 | 0.31 | 97 | 1.14 | 0.10 | |
| 1109.585 | CH$_3$OH | $12_3-11_2$ | 1109.585 | 228.78 | 5.18e−03 | 12.74 | 0.23 | 5.21 | 0.55 | 0.28 | 93 | 1.16 | 0.11 | |
| 1110.941 | CH$_3$OH | $14_{-2}-13_{-1}$ | 1110.941 | 269.85 | 3.40e−03 | 12.73 | 0.04 | 4.87 | 0.14 | 0.22 | 93 | 0.96 | 0.10 | |
| 1113.343 | H$_2$O | $1_{1,0}-0_{0,0}$ | 1113.343 | 53.43 | 1.84e−02 | 12.63 | 0.04 | 20.54 | 0.08 | 3.15 | 94 | 48.58 | 0.22 | |
| 1115.204 | H$_2$O$^+$ | $1_{1,1,3/2,5/2}-0_{0,0,1/2,3/2}$ | 1115.204 | 53.52 | 3.10e−02 | 8.40 | 0.04 | 2.50 | 0.00 | −0.22 | 110 | −0.89 | 0.10 | |
| 1116.986 | CH$_3$OH | $23_1-22_2$ | 1116.986 | 659.32 | 6.41e−03 | 13.80 | 0.59 | 5.58 | 1.39 | 0.14 | 119 | 0.65 | 0.11 | |
| 1117.477 | N$_2$H$^+$ | $12-11$ | 1117.477 | 348.69 | 9.68e−02 | 11.78 | 0.06 | 3.48 | 0.13 | 0.35 | 115 | 1.23 | 0.12 | |
| 1119.813 | CH$_3$OH | $10_4-9_3$ | 1119.813 | 207.99 | 1.03e−02 | | | | | | 122 | | | B: CH$_3$OH ($10_4-9_3$) |
| 1119.831 | CH$_3$OH | $10_4-9_3$ | 1119.831 | 207.99 | 1.03e−02 | | | | | | 123 | | | B: CH$_3$OH ($10_4-9_3$) |

Columns 7–11 are grouped under *from Gaussian fit*; columns 12–14 are grouped under *from moments*.

| ν GHz (1) | Species (2) | Transition (3) | ν_rest GHz (4) | E_u K (5) | A_ul rad/s (6) | υ_lsr km/s (7) | δυ_lsr km/s (8) | FWHM km/s (9) | δFWHM km/s (10) | T_mb K (11) | RMS mK (12) | Flux K km/s (13) | δFlux K km/s (14) | Blend(B) / Notes (15) |
|---|---|---|---|---|---|---|---|---|---|---|---|---|---|---|
| 1121.269 | CH$_3$OH | $14_2 - 13_1$ | 1121.269 | 281.29 | 4.30e-03 | 12.23 | 0.07 | 4.55 | 0.17 | 0.15 | 91 | 0.65 | 0.10 | |
| 1125.143 | CH$_3$OH | $6_5 - 5_4$ | 1125.143 | 184.82 | 1.55e-02 | 12.32 | 0.09 | 10.90 | 0.21 | 0.17 | 99 | 1.00 | 0.12 | |
| 1146.464 | CH$_3$OH | $13_2 - 12_1$ | 1146.464 | 248.98 | 4.66e-03 | 12.56 | 0.06 | 4.31 | 0.15 | 0.38 | 153 | 1.74 | 0.17 | |
| 1147.415 | CH$_3$OH | $19_1 - 18_0$ | 1147.415 | 451.23 | 5.10e-03 | 14.11 | 0.07 | 4.21 | 0.17 | 0.29 | 165 | 1.00 | 0.19 | |
| 1151.449 | HCN | $13 - 12$ | 1151.449 | 386.95 | 7.63e-02 | 12.19 | 0.06 | 13.51 | 0.13 | 0.45 | 152 | 5.43 | 0.24 | |
| 1151.985 | CO | $10 - 9$ | 1151.985 | 304.17 | 1.01e-04 | 11.40 | 0.01 | 11.47 | 0.02 | 23.57 | 162 | 267.41 | 0.40 | |
| 1153.127 | H$_2$O | $3_{1,2,0} - 2_{2,1,0}$ | 1153.127 | 249.44 | 2.67e-03 | 12.08 | 0.02 | 16.13 | 0.04 | 2.73 | 172 | 46.14 | 0.39 | |
| 1153.549 | CH$_3$OH | $8_{-3} - 7_{-2}$ | 1153.549 | 146.28 | 6.64e-03 | 12.72 | 0.06 | 5.36 | 0.15 | 0.46 | 140 | 2.07 | 0.17 | |
| 1157.499 | CH$_3$OH | $13_3 - 12_2$ | 1157.499 | 258.96 | 5.66e-03 | 12.52 | 0.46 | 7.96 | 1.14 | 0.35 | 144 | 1.68 | 0.16 | |
| 1158.727 | HCO$^+$ | $13 - 12$ | 1158.727 | 389.39 | 1.33e-01 | 12.16 | 0.18 | 7.63 | 0.43 | 0.64 | 154 | 5.12 | 0.20 | |
| 1162.706 | CH$_3$OH | $15_{-2} - 14_{-1}$ | 1162.706 | 304.74 | 3.61e-03 | 11.58 | 0.39 | 7.87 | 0.92 | 0.28 | 160 | 0.99 | 0.11 | |
| 1162.912 | H$_2$O | $3_{2,1,0} - 3_{1,2,0}$ | 1162.912 | 305.25 | 2.28e-02 | 12.91 | 0.03 | 14.45 | 0.07 | 1.31 | 137 | 18.18 | 0.28 | |
| 1163.624 | CH$_3$OH | $15_2 - 14_1$ | 1163.624 | 316.05 | 4.75e-03 | 13.47 | 0.07 | 3.23 | 0.16 | 0.28 | 136 | 0.80 | 0.14 | |
| 1168.118 | CH$_3$OH | $11_4 - 10_3$ | 1168.118 | 233.52 | 7.51e-03 | | | | | | 161 | | | B: CH$_3$OH ($11_4 - 10_3$) |
| 1168.150 | CH$_3$OH | $11_4 - 10_3$ | 1168.150 | 233.52 | 7.51e-03 | | | | | | 152 | | | B: CH$_3$OH ($11_4 - 10_3$) |
| 1168.452 | p-NH$_3$ | $2_{1,0} - 1_{1,1}$ | 1168.452 | 57.21 | 1.21e-02 | 13.06 | 0.26 | 8.38 | 0.60 | 0.54 | 148 | 3.35 | 0.20 | |
| 1173.435 | CH$_3$OH | $7_5 - 6_4$ | 1173.435 | 201.06 | 1.57e-02 | 10.30 | 0.08 | 7.77 | 0.19 | 0.38 | 165 | 1.45 | 0.18 | |
| 1188.672 | CH$_3$OH | $20_1 - 19_0$ | 1188.672 | 497.13 | 5.75e-03 | 12.69 | 0.09 | 4.05 | 0.21 | 0.20 | 151 | 0.84 | 0.15 | |
| 1199.599 | CH$_3$OH | $4_4 - 3_3$ | 1199.599 | 119.21 | 1.48e-02 | 12.58 | 0.19 | 7.28 | 0.45 | 0.30 | 160 | 2.13 | 0.16 | |
| 1200.855 | CH$_3$OH | $14_{-4} - 13_{-3}$ | 1200.855 | 331.54 | 6.89e-03 | 12.02 | 0.08 | 4.54 | 0.18 | 0.30 | 160 | 0.94 | 0.15 | |
| 1201.630 | CH$_3$OH | $9_{-3} - 8_{-2}$ | 1201.630 | 167.16 | 7.15e-03 | 11.89 | 0.06 | 4.04 | 0.13 | 0.58 | 167 | 1.71 | 0.19 | |
| 1205.368 | CH$_3$OH | $14_3 - 13_2$ | 1205.368 | 291.46 | 6.15e-03 | 13.24 | 0.07 | 3.64 | 0.16 | 0.26 | 153 | 0.96 | 0.14 | |
| 1211.330 | $^{13}$CO | $11 - 10$ | 1211.330 | 348.93 | 1.18e-04 | 11.97 | 0.04 | 9.03 | 0.09 | 1.05 | 176 | 8.77 | 0.28 | |
| 1214.853 | o-NH$_3$ | $2_{0,1} - 1_{0,0}$ | 1214.853 | 85.78 | 1.81e-02 | 13.79 | 0.04 | 7.57 | 0.08 | 0.91 | 183 | 5.54 | 0.27 | |
| 1215.246 | p-NH$_3$ | $2_{1,1} - 1_{1,0}$ | 1215.246 | 58.32 | 1.36e-02 | | | | | | 204 | | | B: CH$_3$OH ($16_{-2} - 15_{-1}$) |
| 1215.261 | CH$_3$OH | $16_{-2} - 15_{-1}$ | 1215.261 | 341.96 | 5.60e-03 | | | | | | 195 | | | B: p-NH$_3$ ($2_{1,1} - 1_{1,0}$) |
| 1216.420 | CH$_3$OH | $12_4 - 11_3$ | 1216.420 | 261.36 | 8.14e-03 | 12.11 | 0.07 | 5.62 | 0.18 | 0.34 | 190 | 1.40 | 0.17 | |
| 1228.789 | H$_2$O | $2_{2,0} - 2_{1,1,0}$ | 1228.789 | 195.91 | 1.88e-02 | 12.93 | 0.04 | 12.80 | 0.09 | 0.77 | 148 | 10.18 | 0.28 | |
| 1229.740 | CH$_3$OH | $21_1 - 20_0$ | 1229.740 | 545.31 | 6.45e-03 | 11.30 | 0.14 | 2.25 | 0.34 | 0.34 | 160 | 0.90 | 0.14 | |
| 1232.476 | HF | $1 - 0$ | 1232.476 | 59.15 | 2.42e-02 | 9.98 | 0.08 | 2.82 | 0.19 | -0.81 | 195 | -2.00 | 0.19 | |
| 1239.890 | HCN | $14 - 13$ | 1239.890 | 446.45 | 9.55e-02 | 12.19 | 0.56 | 14.59 | 1.36 | 0.50 | 191 | 5.02 | 0.29 | |
| 1249.559 | H$^{37}$Cl | $2_{3/2} - 1_{1/2}$ | 1249.559 | 89.97 | 4.65e-03 | | | | | | 83 | | | B: H$^{37}$Cl ($2_{3/2} - 1_{5/2}$) |
| 1249.569 | H$^{37}$Cl | $2_{3/2} - 1_{5/2}$ | 1249.569 | 89.96 | 5.58e-04 | | | | | | 83 | | | B: H$^{37}$Cl ($2_{3/2} - 1_{1/2}$) |
| 1249.571 | H$^{37}$Cl | $2_{3/2} - 1_{3/2}$ | 1249.571 | 89.97 | 9.30e-03 | | | | | | 85 | | | B: H$^{37}$Cl ($2_{3/2} - 1_{3/2}$) |
| 1249.573 | H$^{37}$Cl | $2_{5/2} - 1_{5/2}$ | 1249.573 | 89.97 | 1.12e-02 | | | | | | 83 | | | B: H$^{37}$Cl ($2_{3/2} - 1_{1/2}$) |
| 1249.573 | H$^{37}$Cl | $2_{5/2} - 1_{3/2}$ | 1249.573 | 89.96 | 7.82e-03 | | | | | | 83 | | | B: H$^{37}$Cl ($2_{3/2} - 1_{3/2}$) |
| 1249.582 | H$^{37}$Cl | $2_{5/2} - 1_{3/2}$ | 1249.582 | 89.97 | 5.95e-03 | | | | | | 82 | | | B: H$^{37}$Cl ($2_{3/2} - 1_{1/2}$) |
| 1249.584 | CH$_3$OH | $10_{-3} - 9_{-2}$ | 1249.584 | 190.37 | 1.14e-02 | | | | | | 84 | | | B: H$^{37}$Cl ($2_{3/2} - 1_{1/2}$) |
| 1249.595 | H$^{37}$Cl | $2_{5/2} - 1_{3/2}$ | 1249.595 | 89.97 | 1.86e-03 | | | | | | 84 | | | B: H$^{37}$Cl ($2_{3/2} - 1_{1/2}$) |
| 1251.434 | HCl | $2_{1/2} - 1_{3/2}$ | 1251.434 | 90.10 | 3.36e-03 | | | | | | 96 | | | B: HCl ($2_{3/2} - 1_{1/2}$) |
| 1251.434 | HCl | $2_{5/2} - 1_{5/2}$ | 1251.434 | 90.10 | 4.67e-03 | | | | | | 97 | | | B: HCl ($2_{5/2} - 1_{5/2}$) |
| 1251.447 | HCl | $2_{3/2} - 1_{5/2}$ | 1251.447 | 90.10 | 5.61e-04 | | | | | | 96 | | | B: HCl ($2_{3/2} - 1_{5/2}$) |
| 1251.450 | HCl | $2_{1/2} - 1_{3/2}$ | 1251.450 | 90.10 | 9.34e-03 | | | | | | 96 | | | B: HCl ($2_{5/2} - 1_{3/2}$) |
| 1251.452 | HCl | $2_{5/2} - 1_{3/2}$ | 1251.452 | 90.10 | 7.85e-03 | | | | | | 96 | | | B: HCl ($2_{5/2} - 1_{3/2}$) |

| | | | | | | *Gaussian fit* | | | | | *from moments* | | | |
|---|---|---|---|---|---|---|---|---|---|---|---|---|---|---|
| $\nu$ GHz (1) | Species (2) | Transition (3) | $\nu_{rest}$ GHz (4) | $E_u$ K (5) | $A_{ul}$ rad/s (6) | $v_{lsr}$ km/s (7) | $\delta v_{lsr}$ km/s (8) | FWHM km/s (9) | $\delta$FWHM km/s (10) | $T_{mb}$ K (11) | RMS mK (12) | Flux K km/s (13) | $\delta$Flux K km/s (14) | Blend(**B**) / Notes (15) |
| 1251.464 | HCl | $2_{3/2}-1_{3/2}$ | 1251.464 | 90.10 | 5.98e-03 | | | | | | 97 | | | **B:** HCl ($2_{5/2}-1_{5/2}$) |
| 1251.481 | HCl | $2_{1/2}-1_{3/2}$ | 1251.481 | 90.10 | 1.87e-03 | | | | | | 94 | | | **B:** HCl ($2_{5/2}-1_{5/2}$) |
| 1267.014 | CO | $11-10$ | 1267.014 | 364.97 | 1.34e-04 | 11.66 | 0.04 | 11.29 | 0.11 | 21.62 | 126 | 247.42 | 0.25 | — |
| 1496.923 | CO | $13-12$ | 1496.923 | 503.14 | 2.20e-04 | 12.19 | 0.01 | 12.02 | 0.02 | 15.92 | 209 | 192.13 | 0.44 | — |
| 1541.843 | CH$_3$OH | $20_2-19_1$ | 1541.843 | 525.23 | 1.06e-02 | 14.84 | 0.05 | 2.56 | 0.11 | 0.58 | 241 | 1.14 | 0.19 | — |
| 1611.794 | CO | $14-13$ | 1611.794 | 580.50 | 2.74e-04 | 12.32 | 0.01 | 12.54 | 0.02 | 13.49 | 273 | 165.51 | 0.53 | — |
| 1661.008 | H$_2$O | $2_{2,1,0}-2_{1,2,0}$ | 1661.008 | 194.10 | 3.06e-02 | 14.52 | 0.03 | 10.99 | 0.07 | 2.18 | 210 | 22.30 | 0.33 | — |
| 1669.905 | H$_2$O | $2_{1,2,0}-1_{0,1,0}$ | 1669.905 | 114.38 | 5.60e-02 | 15.24 | 0.03 | 14.58 | 0.07 | 5.32 | 200 | 66.52 | 0.36 | — |
| 1716.770 | H$_2$O | $3_{0,3,0}-2_{1,2,0}$ | 1716.770 | 196.77 | 5.00e-02 | 15.01 | 0.02 | 9.91 | 0.05 | 4.70 | 175 | 53.86 | 0.28 | — |
| 1726.603 | CO | $15-14$ | 1726.603 | 663.36 | 3.35e-04 | 13.05 | 0.01 | 12.34 | 0.02 | 10.34 | 221 | 135.76 | 0.40 | — |
| 1760.486 | $^{13}$CO | $16-15$ | 1760.486 | 718.70 | 3.57e-04 | 16.15 | 0.22 | 3.95 | 0.51 | 0.26 | 172 | 1.15 | 0.17 | — |
| 1763.601 | p-NH$_3$ | $3_{1,0}-2_{1,1}$ | 1763.601 | 142.96 | 5.28e-03 | 13.83 | 0.15 | 2.13 | 0.36 | 0.40 | 166 | 3.22 | 0.20 | — |
| 1763.823 | p-NH$_3$ | $3_{2,0}-2_{2,1}$ | 1763.823 | 126.97 | 3.30e-02 | 13.83 | 0.19 | 2.95 | 0.44 | 0.29 | 169 | 2.05 | 0.18 | — |
| 1834.736 | OH | $^2\Pi\ 2_{-1,0,1}-1_{1,1,0,1}$ | 1834.736 | 269.76 | 2.12e-02 | | | | | | 89 | | | **B:** OH ($^2\Pi\ 2_{-1,0,2}-1_{1,1,0,1}$) |
| 1834.747 | OH | $^2\Pi\ 2_{-1,0,2}-1_{1,1,0,1}$ | 1834.747 | 269.76 | 6.36e-02 | | | | | | | | | |
| 1834.750 | OH | $^2\Pi\ 2_{-1,0,1}-1_{1,1,0,0}$ | 1834.750 | 269.76 | 4.24e-02 | | | | | | 89 | | | **B:** OH ($^2\Pi\ 2_{-1,0,1}-1_{1,1,0,1}$) |
| 1837.747 | OH | $^2\Pi\ 2_{1,1,0,1}-1_{1,-1,0,1}$ | 1837.747 | 270.14 | 2.13e-02 | | | | | | 85 | | | **B:** OH ($^2\Pi\ 2_{1,1,0,2}-1_{1,-1,0,1}$) |
| 1837.817 | OH | $^2\Pi\ 2_{1,1,0,2}-1_{1,-1,0,1}$ | 1837.817 | 270.14 | 6.40e-02 | | | | | | 88 | | | **B:** OH ($^2\Pi\ 2_{1,1,0,1}-1_{1,-1,0,1}$) |
| 1837.837 | OH | $^2\Pi\ 2_{1,1,0,1}-1_{1,-1,0,0}$ | 1837.837 | 270.14 | 4.27e-02 | | | | | | 93 | | | **B:** OH ($^2\Pi\ 2_{1,1,0,1}-1_{1,-1,0,0}$) |
| 1841.346 | CO | $16-15$ | 1841.346 | 751.73 | 4.05e-04 | 12.60 | 0.01 | 12.45 | 0.03 | 5.91 | 83 | 69.82 | 0.14 | — |
| 1900.537 | C$^+$ | $3/2-1/2$ | 1900.537 | 91.21 | 2.32e-06 | 9.07 | 0.01 | 2.08 | 0.01 | 11.58 | 94 | 24.14 | 0.06 | — |



\end{document}